\documentclass[%
reprint,
frontmatterverbose,
preprintnumbers,
nofootinbib,
nobibnotes,
amsmath,amssymb,
aps,
prc,
]{revtex4-2}
\usepackage[pdftex]{graphicx}
\usepackage{dcolumn}
\usepackage{bm}
\usepackage[pdftex]{color}
\usepackage{CJK}
\usepackage[T1]{fontenc}
\usepackage{mathrsfs}
\def\ve#1{{\bm{#1}}}
\def\nuc#1#2#3{{}^{#2}_{#3}\mathrm{#1}}
\def\urm#1{\scriptstyle{\text{\textrm{\textmd{\textup{#1}}}}}}

\def\avr#1{\left\langle{#1}\right\rangle}

\def\ca#1{{\mathcal{#1}}}

\let\temp\epsilon
\let\epsilon\varepsilon
\let\varepsilon\temp
\let\temp\relax
\let\temp\phi
\let\phi\varphi
\let\varphi\temp
\let\temp\relax
\begin{document}
%
\begin{CJK*}{UTF8}{}
  \preprint{RIKEN-iTHEMS-Report-22}
  \title{Comparative study on charge radii and their kinks at magic numbers}
  \author{Tomoya Naito (\CJKfamily{min}{内藤智也})}
  \email{
    tnaito@ribf.riken.jp}
  \affiliation{
    RIKEN Interdisciplinary Theoretical and Mathematical Sciences Program (iTHEMS),
    Wako 351-0198, Japan}
  \affiliation{
    Department of Physics, Graduate School of Science, The University of Tokyo,
    Tokyo 113-0033, Japan}
  \author{Tomohiro Oishi (\CJKfamily{min}{大石知広})}
  \email{
    Current address: RIKEN Nishina Center, Wako 351-0198, Japan;
    tomohiro.oishi@yukawa.kyoto-u.ac.jp}
  \affiliation{
    Yukawa Institute for Theoretical Physics, Kyoto University,
    Kyoto 606-8502, Japan}
  \author{Hiroyuki Sagawa (\CJKfamily{min}{佐川弘幸})}
  \email{
    sagawa@ribf.riken.jp}
  \affiliation{
    Center for Mathematics and Physics, University of Aizu,
    Aizu-Wakamatsu 965-8560, Japan}
  \affiliation{
    RIKEN Nishina Center, Wako 351-0198, Japan}
  \author{Zhiheng Wang (\CJKfamily{gbsn}{王之恒})}
  \email{
    wangzhh2013@lzu.edu.cn}
  \affiliation{
    School of Nuclear Science and Technology, Lanzhou University, 
    Lanzhou 730000, China}
  \affiliation{
    Frontiers Science Center for Rare Isotopes, 
    Lanzhou University,
    Lanzhou 730000, China}
  \date{\today}
  \begin{abstract}
    Isotope dependences of charge radii, i.e., isotope shifts, 
    calculated by
    the Skyrme Hartree-Fock, the relativistic mean-field, and the relativistic Hartree-Fock calculations are compared against the experimental data of magic and semimagic nuclei.
    It is found that the tensor interaction
    plays a role in reproducing the ``kink'' behavior,
    irregularity of isotope shifts at the neutron magic number,
    in the relativistic Hartree-Fock approach.
    With several Skyrme models,
    it is found that the kink behavior can be reproduced with the spin-orbit interaction having nonzero isovector channel.
    The single-particle orbitals near the Fermi energy 
    are crucial to determine the kink size.
    The effects of the symmetry energy and the pairing interaction are also discussed in relation to the kink behavior.
  \end{abstract}
  \maketitle
\end{CJK*}
%
\section{Introduction}
\par
Charge radii of atomic nuclei $ R_{\urm{ch}} $ are one of the important properties of atomic nuclei,
which can be measured precisely via the electron scattering~\cite{
  DeVries1987At.DataNucl.DataTables36_495,
  Suda2017Prog.Part.Nucl.Phys.96_1} 
and the laser spectroscopy~\cite{
  Angeli2013At.DataNucl.DataTables99_69,
  Campbell2016Prog.Part.Nucl.Phys.86_127}.
The isotope dependence of $ R_{\urm{ch}} $, i.e., the isotope shift, has been discussed for a long time theoretically and experimentally.
The slope of $ R_{\urm{ch}} $ as a function of the neutron number $ N $ is known to change at the neutron magic number,
which is called the ``kink'' behavior.
It was discussed in Ref.~\cite{
  Reinhard1995Nucl.Phys.A584_467}
that the kink behavior of $ \mathrm{Pb} $ isotopes can be reproduced 
if the strengths of isoscalar and isovector spin-orbit interaction in the Skyrme energy density functional (EDF) is properly adjusted.
Recently, it was discussed using the M3Y-P6a interaction 
that the kink behavior is also reproduced by the three-body spin-orbit interaction~\cite{
  Nakada2019Phys.Rev.C100_044310}.
The kink behavior was also discussed by using the relativistic mean-field calculation in Ref.~\cite{
  Perera2021Phys.Rev.C104_064313},
in which the importance of the occupation probabilities above the shell gap was pointed out.
The kink behavior of mercury isotopes was also discussed recently in Refs.~\cite{
  Marsh2018Nat.Phys.14_1163,
  DayGoodacre2021Phys.Rev.C104_054322,
  DayGoodacre2021Phys.Rev.Lett.126_032502}.
\par
The aim of this paper is to discuss whether the kink behavior is reproduced even
in the ``normal'' treatment,
i.e., mean-field models with a widely used functional and a standard contact pairing interaction,
without any additional terms or effects.
In this paper, we compare $ R_{\urm{ch}} $ of $ \mathrm{Sn} $ and $ \mathrm{Pb} $ isotopes
obtained by the nonrelativistic Skyrme Hartree-Fock (SHF)~\cite{
  Vautherin1972Phys.Rev.C5_626},
the relativistic mean-field (RMF)~\cite{
  Miller1972Phys.Rev.C5_241,
  Walecka1974Ann.Phys.83_491},
and
the relativistic Hartree-Fock (RHF)~\cite{
  Long2006Phys.Lett.B640_150,
  Long2010Phys.Rev.C81_024308}
calculations.
This comparison enables us to discuss the kink behavior from two perspectives on the spin-orbit splitting. 
In the nonrelativistic Skyrme energy density functional calculation, 
the spin-orbit splitting is phenomenologically determined by the parameters, 
the isoscalar spin-orbit strength $ W_0 $, and the isovector one $ W'_0 $. 
On the other side, in the relativistic mean-field calculations, 
where the single-particle state is described with the Dirac equation obtained self-consistently from the relativistic EDF, 
the spin-orbit splitting is induced by the relativistic effect without introducing any additional parameter.
In the RHF model, 
the tensor interaction can also be included when the Fock terms are explicitly considered. 
Using such comparisons, we will be able to find out how much 
the tensor interaction, and its contribution to the spin-orbit field, is important to reproduce the kink behavior.
In addition, we will also study a correlation between the symmetry energy of the nuclear equation of state and the kink behavior.
\par
This paper is organized as follows.
In Sec.~\ref{sec:theoretical}, the theoretical framework of this paper will be given.
In Sec.~\ref{sec:calculation_Sn}, the calculation results of $ \mathrm{Sn} $ isotopes will be presented, where the detailed discussion of the occurrence mechanism of the kink behaviors will be given.
In Sec.~\ref{sec:calculation_Pb}, the calculation results of $ \mathrm{Pb} $ isotopes will be shown,
where the detailed discussion will be mainly referred to in Sec.~\ref{sec:calculation_Sn}.
In Sec.~\ref{sec:calculation_EoS}, effects of parameters of the nuclear equation of state on the kink behavior will be shown.
In Sec.~\ref{sec:pairing}, effects of the pairing strength and interaction will be discussed.
In Sec.~\ref{sec:calculation_Ca}, the calculation results of $ \mathrm{Ca} $ isotopes will be given,
where it will be found that beyond-mean-field effects are indispensable to reproduce the isotope dependence of $ R_{\urm{ch}} $ of $ \mathrm{Ca} $ isotopes.
In Sec.~\ref{sec:conclusion}, this paper will be summarized.
%
\section{Theoretical Framework}
\label{sec:theoretical}
\par
The methods used in this paper can be classified into two classes:
the nonrelativistic framework and the relativistic one.
We use the SHF calculation for the former
and
the RMF and RHF calculations for the latter.
\par
In the SHF calculation, the following Skyrme EDFs are used:
SAMi~\cite{
  Roca-Maza2012Phys.Rev.C86_031306},
SGII~\cite{
  VanGiai1981Phys.Lett.B106_379},
SLy4~\cite{
  Chabanat1998Nucl.Phys.A635_231},
SLy5~\cite{
  Chabanat1998Nucl.Phys.A635_231},
SkM*~\cite{
  Bartel1982Nucl.Phys.A386_79},
HFB9~\cite{
  Goriely2005Nucl.Phys.A750_425},
UNEDF0~\cite{
  Kortelainen2010Phys.Rev.C82_024313},
UNEDF1~\cite{
  Kortelainen2012Phys.Rev.C85_024304},
and UNEDF2~\cite{
  Kortelainen2014Phys.Rev.C89_054314}.
In addition, the Skyrme EDF with the tensor interaction ``SAMi-T''~\cite{
  Shen2019Phys.Rev.C99_034322}
is used,
in which all the parameters, including the tensor interaction, were simultaneously optimized to fit a set of experimental data.
To see the effect of the tensor interaction, the SAMi-T EDF without the tensor interaction is also used being referred to as ``SAMi-noT''.
A variety of SAMi EDFs with the different value for the effective mass, the symmetry energy, and the nuclear incompressibility,
called ``SAMi-m''~\cite{
  Roca-Maza2013Phys.Rev.C87_034301},
``SAMi-J''~\cite{
  Roca-Maza2013Phys.Rev.C87_034301},
and ``SAMi-K''~\cite{
  Roca-Maza_}
families, respectively, 
are also adopted in order to see whether properties of nuclear equation of state (EoS) affect the kink behavior.
The parameters of all these SAMi EDFs are optimized for the same set of experimental data.
The pairing interaction is taken into account by using the Hartree-Fock-Bogoliubov calculation with the volume-type pairing interaction~\cite{
  Dobaczewski1984Nucl.Phys.A422_103}
\begin{equation}
  \label{eq:pair_volume}
  V_{\urm{pair}} \left( \ve{r}_1, \ve{r}_2 \right)
  =
  -
  V_0
  \delta \left( \ve{r}_1 - \ve{r}_2 \right) 
\end{equation}
with the cutoff energy of $ 60 \, \mathrm{MeV} $,
whose strength is determined to reproduce the neutron pairing gap of $ \nuc{Sn}{120}{} $ as $ \Delta_n = 1.4 \, \mathrm{MeV} $.
The strengths for different EDFs are shown in Table~\ref{tab:pairing}.
All the calculations are performed by using the harmonic oscillator basis~\cite{
  NavarroPerez2017Comput.Phys.Commun.220_363}
under the assumption of spherically symmetric shape.
Since all nuclei studied are magic or semimagic ones, this assumption on the shape is reasonable.
\begin{table}[b]
  \centering
  \caption{
    Adopted pairing strengths for Skyrme EDFs in this study.
    The volume-type pairing interaction is used and the energy cutoff is $ 60 \, \mathrm{MeV} $.}
  \label{tab:pairing}
  \begin{ruledtabular}
    \begin{tabular}{ld}
      \multicolumn{1}{l}{Skyrme EDF} & \multicolumn{1}{c}{$ V_0 $ ($ \mathrm{MeV} \, \mathrm{fm}^3 $)} \\
      \hline
      SLy4      & 194.2 \\
      SLy5      & 188.2 \\
      HFB9      & 164.4 \\
      SkM*      & 156.2 \\
      SGII      & 169.8 \\
      UNEDF0    & 127.6 \\
      UNEDF1    & 138.4 \\
      UNEDF2    & 150.0 \\
      SAMi      & 213.7 \\
      \hline
      SAMi-noT  & 216.4 \\
      SAMi-T    & 225.5 \\
      \hline
      SAMi-K230 & 213.8 \\
      SAMi-K235 & 206.9 \\
      SAMi-K240 & 201.2 \\
      SAMi-K245 & 209.6 \\
      SAMi-K250 & 208.2 \\
      SAMi-K255 & 206.9 \\
      SAMi-K260 & 205.5 \\
      \hline
      SAMi-m60  & 227.6 \\
      SAMi-m65  & 215.2 \\
      SAMi-m70  & 203.9 \\
      SAMi-m75  & 194.3 \\
      SAMi-m80  & 185.4 \\
      SAMi-m85  & 177.3 \\
      \hline
      SAMi-J27  & 205.0 \\
      SAMi-J28  & 208.3 \\
      SAMi-J29  & 212.0 \\
      SAMi-J30  & 215.1 \\
      SAMi-J31  & 217.6 \\
      SAMi-J32  & 219.3 \\
      SAMi-J33  & 220.2 \\
      SAMi-J34  & 220.4 \\
      SAMi-J35  & 220.2 \\
    \end{tabular}
  \end{ruledtabular}
\end{table}
\par
In the RMF calculation, the Fock term is neglected and correspondingly the EDF is optimized in order to reproduce the reference data. 
The RHF calculation, on the other hand, takes the Fock term into account and the parameters of the relativistic EDF are optimized with the reference data as the RMF model.
In the RMF calculation,
the DD-PC1~\cite{
  Niksic2008Phys.Rev.C78_034318},
DD-ME2~\cite{
  Lalazissis2005Phys.Rev.C71_024312},
PKDD~\cite{
  Long2004Phys.Rev.C69_034319},
and DD-LZ1~\cite{
  Wei2020Chin.Phys.C44_074107}
EDFs are used,
while in the RHF calculation, 
the PKO1~\cite{
  Long2008Europhys.Lett.82_12001},
PKO2~\cite{
  Long2008Europhys.Lett.82_12001},
PKO3~\cite{
  Long2008Europhys.Lett.82_12001},
PKA1~\cite{
  Long2007Phys.Rev.C76_034314},
and the modified version of the PKO1, which will be referred to as the PKO1*~\cite{
  Wang2021Phys.Rev.C103_064326},
are used.
The pairing correlation is considered by using the Bardeen-Cooper-Schrieffer (BCS) theory,
the Tian-Ma-Ring (TMR)-type pairing interaction~\cite{
  Tian2009Phys.Lett.B676_44}
is used for DD-PC1 and DD-ME2, 
and
the surface-type pairing interaction~\cite{
  Dobaczewski1996Phys.Rev.C53_2809}
with the pairing strength of $ 500 \, \mathrm{MeV} \, \mathrm{fm}^3 $
is used for DD-ME2, PKDD, DD-LZ1, PKO1, PKO2, PKO3, PKO1*, and PKA1.
All the calculations are performed
assuming the spherically symmetric shape for the harmonic oscillator or Woods-Saxon basis.
\par
In this paper, the same pairing strengths are used for both the proton-proton and the neutron-neutron channels.
We confirmed that the results hardly change even if the Coulomb antipairing effect~\cite{
  Nakada2011Phys.Rev.C83_031302} is considered.
\par
After the proton and neutron root-mean-square radii, $ R_p $ and $ R_n $, are obtained by using SHF, RMF, and RHF,
the charge radius $ R_{\urm{ch}} $ is calculated by using a formula for the finite-size effect of nucleons,
\begin{equation}
  \label{eq:Rch}
  R_{\urm{ch}}^2
  =
  R_p^2
  +
  r_{\urm{E} p}^2
  +
  \frac{N}{Z}
  r_{\urm{E} n}^2
  +
  \avr{r^2}_{\urm{SO} p}
  +
  \frac{N}{Z}
  \avr{r^2}_{\urm{SO} n},
\end{equation}
where $ r_{\urm{E} p} = 0.8409 \, \mathrm{fm} $ is the single proton radius
and
$ r_{\urm{E} n}^2 = -0.1161 \, \mathrm{fm}^2 $ is the single neutron mean-square radius~\cite{
  Zyla2020Prog.Theor.Exp.Phys.2020_083C01}.
The spin-orbit contributions $ \avr{r^2}_{\urm{SO} p} $ and $ \avr{r^2}_{\urm{SO} n} $ obtained in Ref.~\cite{
  Naito2021Phys.Rev.C104_024316}
are calculated with the nucleon magnetic moments
$ \kappa_p = 1.793 $ and $ \kappa_n = -1.913 $~\cite{
  Zyla2020Prog.Theor.Exp.Phys.2020_083C01}.
The importance of the spin-orbit contribution will be discussed in Appendix~\ref{sec:so_contribution}.
%
\section{Calculated Results}
\label{sec:calculation}
\par
In this section, the ``kink'' behaviors of $ \mathrm{Sn} $ and $ \mathrm{Pb} $ isotopes will be presented.
The detailed mechanism will be discussed by using $ \mathrm{Sn} $ isotopes
since $ \nuc{Sn}{132}{} $ or its neighbor nuclei are not used for the fitting criteria of the EDFs, and thus, it is expected that the results may reflect properties of the EDFs better.
Then, the effects of some parameters of the nuclear equation of state on the kink properties will also be discussed,
introducing the symmetry energy $ J $,
the nuclear incompressibility $ K_{\infty} $,
and 
the effective mass $ m^* $.
Dependences of the pairing strength and interaction on the kink behavior will also be discussed.
At last, results of $ \mathrm{Ca} $ isotopes will be presented.
\subsection{Mass-number $ A $ dependence of charge radii in $ \mathrm{Sn} $ and $ \mathrm{Pb} $ isotopes}
\begin{figure*}[tb]
  \centering
  \includegraphics[width=1.0\linewidth]{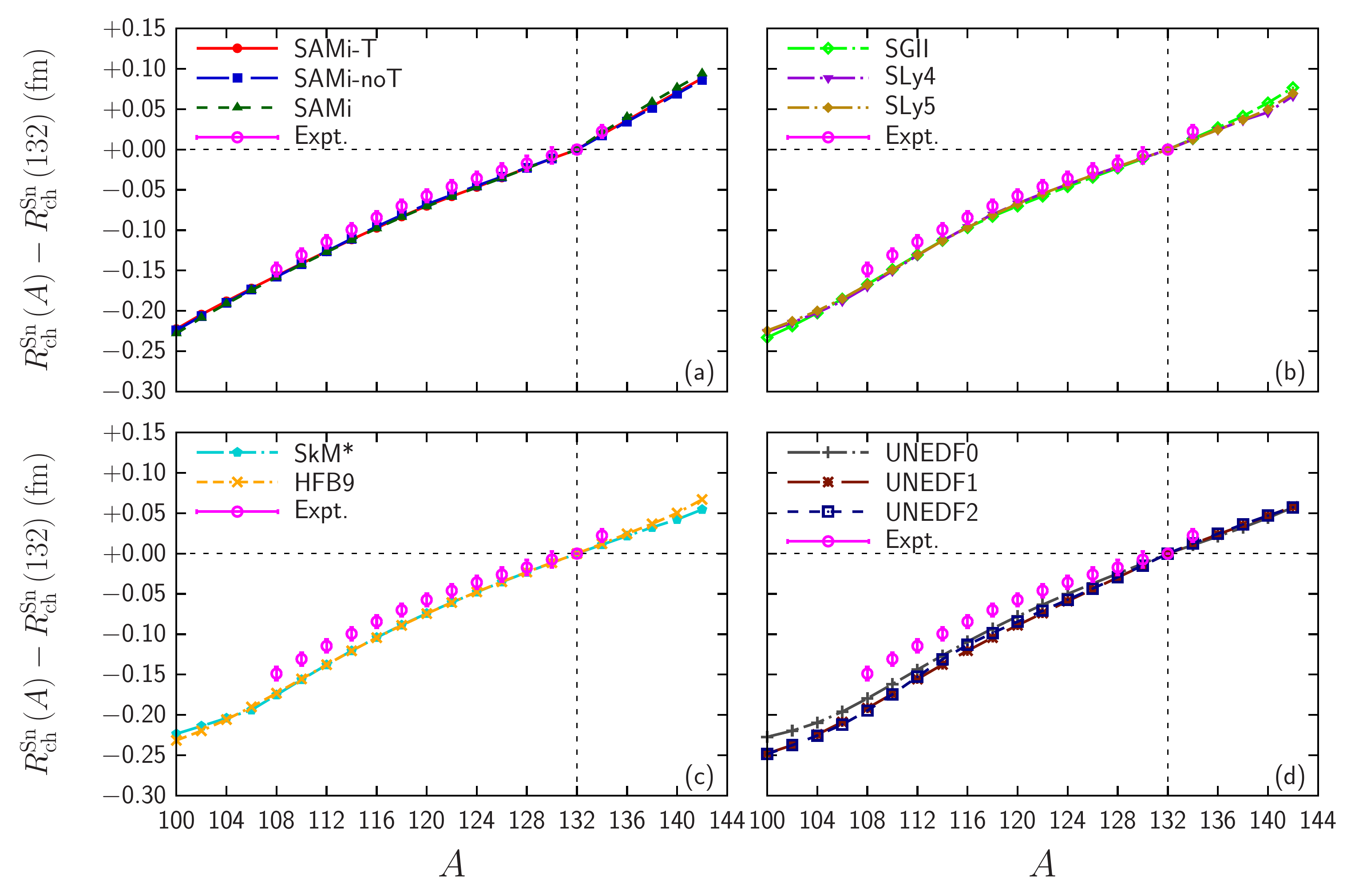}
  \caption{
    Difference of 
    root-mean-square charge radii between $ \nuc{Sn}{A}{} $
    and
    $ \nuc{Sn}{132}{} $, 
    $ R_{\urm{ch}}^{\urm{Sn}} \left( A \right) - R_{\urm{ch}}^{\urm{Sn}} \left( 132 \right) $,
    as a function of $ A $
    calculated by using nonrelativistic EDFs.
    For comparison, experimental data~\cite{
      Angeli2013At.DataNucl.DataTables99_69,
      Gorges2019Phys.Rev.Lett.122_192502}
    are also plotted.}
  \label{fig:Rch_Sn_NR}
\end{figure*}

\begin{figure*}[tb]
  \centering
  \includegraphics[width=1.0\linewidth]{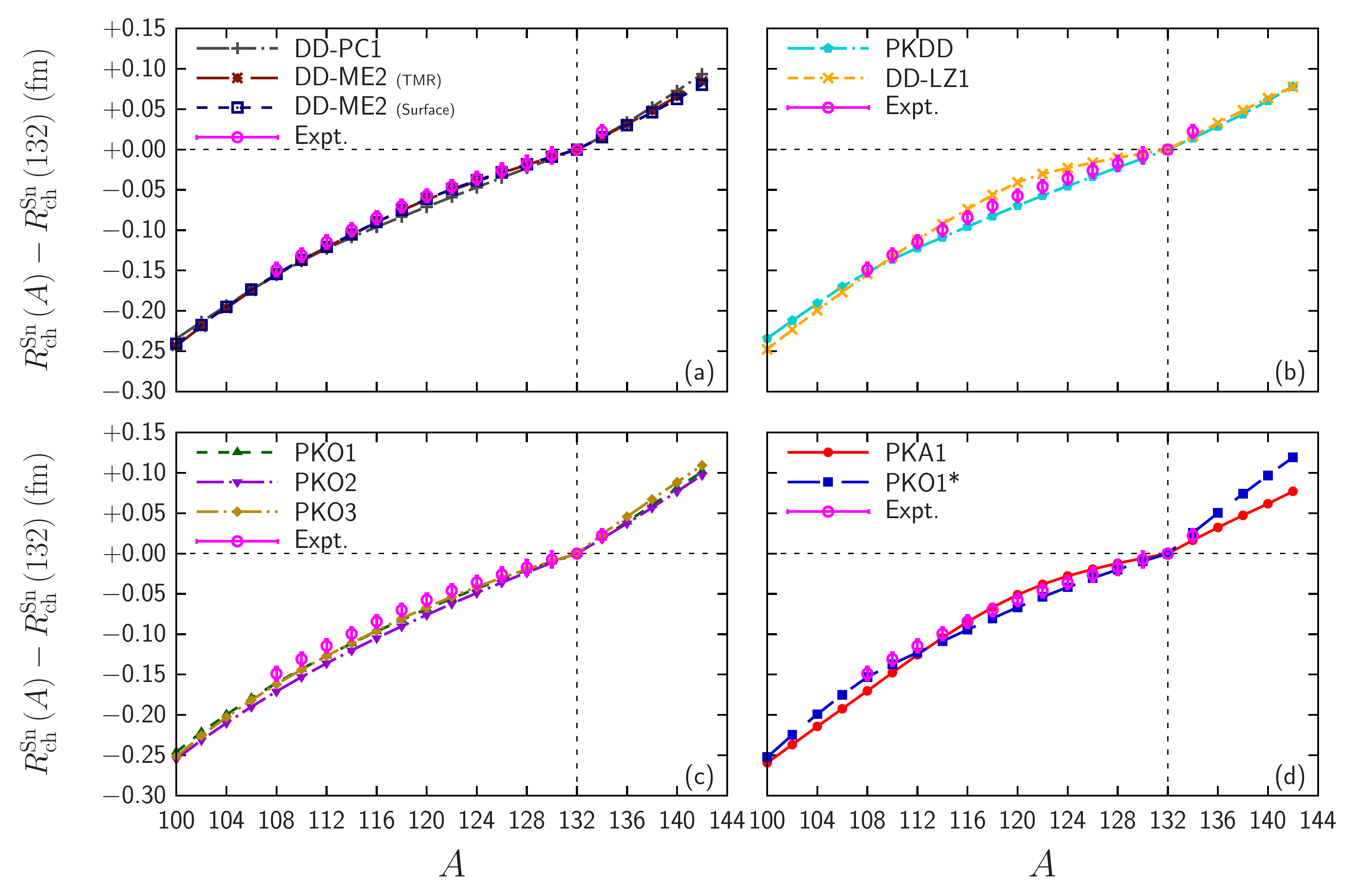}
  \caption{
    Same as Fig.~\ref{fig:Rch_Sn_NR}, but by using relativistic EDFs.}
  \label{fig:Rch_Sn_R}
\end{figure*}

\begin{figure*}[tb]
  \centering
  \includegraphics[width=1.0\linewidth]{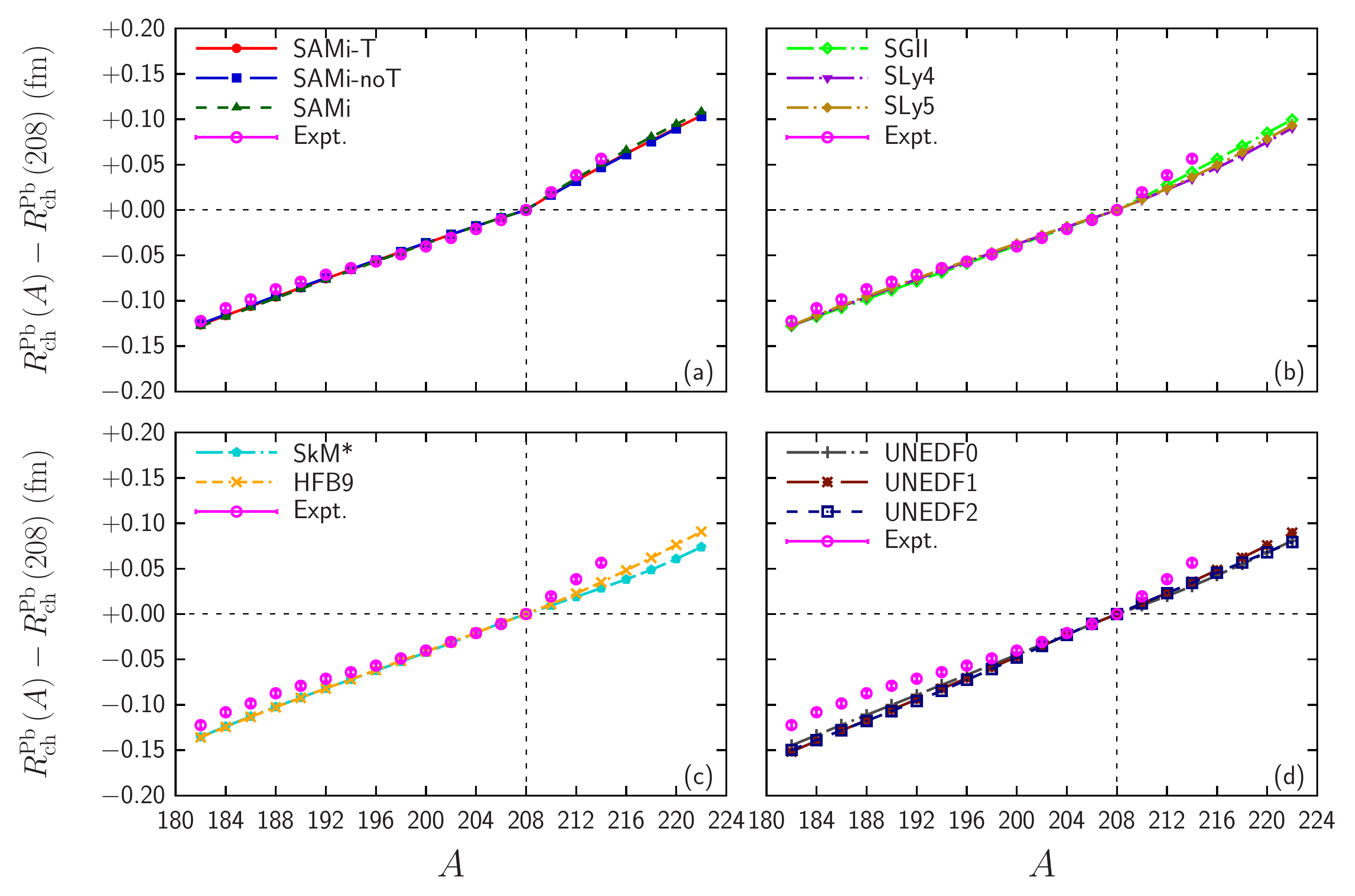}
  \caption{
    Same as Fig.~\ref{fig:Rch_Sn_NR}, but for $ \mathrm{Pb} $ isotopes
    with the reference point at $ \nuc{Pb}{208}{} $,
    $ R_{\urm{ch}}^{\urm{Pb}} \left( A \right) - R_{\urm{ch}}^{\urm{Pb}} \left( 208 \right) $.}
  \label{fig:Rch_Pb_NR}
\end{figure*}

\begin{figure*}[tb]
  \centering
  \includegraphics[width=1.0\linewidth]{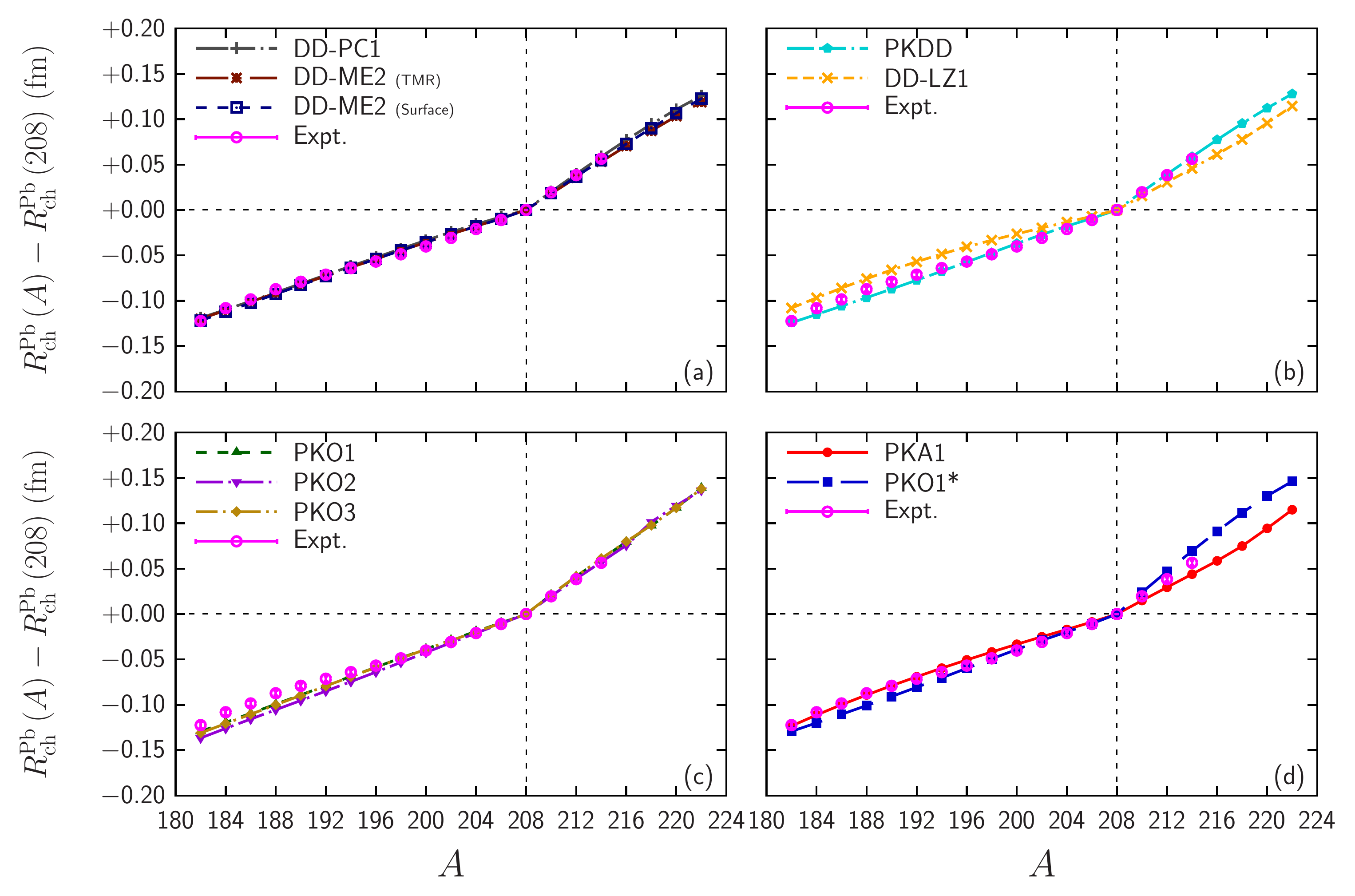}
  \caption{
    Same as Fig.~\ref{fig:Rch_Sn_R}, but for $ \mathrm{Pb} $ isotopes
    with the reference point at $ \nuc{Pb}{208}{} $.}
  \label{fig:Rch_Pb_R}
\end{figure*}

The mass-number $ A $ dependence of the difference between
the root-mean-square charge radius of $ \nuc{Sn}{A}{} $
and
that of $ \nuc{Sn}{132}{} $,
$ R_{\urm{ch}}^{\urm{Sn}} \left( A \right) - R_{\urm{ch}}^{\urm{Sn}} \left( 132 \right) $,
calculated in nonrelativistic (SHF) and relativistic (RMF and RHF) EDFs are shown in
Figs.~\ref{fig:Rch_Sn_NR} and \ref{fig:Rch_Sn_R}, respectively.
For comparison, experimental data~\cite{
  Angeli2013At.DataNucl.DataTables99_69,
  Gorges2019Phys.Rev.Lett.122_192502}
are also plotted.
Results for $ \mathrm{Pb} $ isotopes,
$ R_{\urm{ch}}^{\urm{Pb}} \left( A \right) - R_{\urm{ch}}^{\urm{Pb}} \left( 208 \right) $,
are shown in
Figs.~\ref{fig:Rch_Pb_NR} and \ref{fig:Rch_Pb_R} as well.
\par
It can be seen that most EDFs reproduce well the $ A $ dependence of $ R_{\urm{ch}} $ of stable nuclei of $ \mathrm{Sn} $ isotopes
($ 62 \le N \le 74 $)
except UNEDF2, PKA1, and DD-LZ1.
In Figs.~\ref{fig:Rch_Pb_NR} and \ref{fig:Rch_Pb_R}, 
most EDFs reproduce well the $ A $ dependence of $ \mathrm{Pb} $ isotopes below the magic number $ N = 126 $,
although the UNEDF series underestimate
$ R_{\urm{ch}}^{\urm{Pb}} \left( A \right) - R_{\urm{ch}}^{\urm{Pb}} \left( 208 \right) $
slightly,  
and the DD-LZ1 EDF overestimates it slightly.
Behavior above the magic numbers is the main topic in this paper, and will be presented later.
\subsection{$ \mathrm{Sn} $ isotopes}
\label{sec:calculation_Sn}
\subsubsection{Systematic behavior}
\par
Hereinafter, we focus only on the kink behavior.
To discuss the  size of kink quantitatively, we define the indicator of kink size $ \Delta^2 R_{\urm{ch}}^Z $ as
\begin{align}
  & \Delta^2 R_{\urm{ch}}^Z
    \notag \\
  & = 
    \Delta R_{\urm{ch}}^Z \left( A_{\urm{magic}} + 2 \right)
    -
    \Delta R_{\urm{ch}}^Z \left( A_{\urm{magic}} \right)
    \notag \\
  & =
    R_{\urm{ch}}^Z \left( A_{\urm{magic}} + 2 \right)
    -
    2
    R_{\urm{ch}}^Z \left( A_{\urm{magic}} \right)
    +
    R_{\urm{ch}}^Z \left( A_{\urm{magic}} - 2 \right),
\end{align}
where $ A_{\urm{magic}} $ is the mass number corresponding to doubly magic nuclei,
i.e.,
$ A_{\urm{magic}} = 48 $ for $ Z = 20 $ ($ \mathrm{Ca} $),
$ A_{\urm{magic}} = 132 $ for $ Z = 50 $ ($ \mathrm{Sn} $),
and
$ A_{\urm{magic}} = 208 $ for $ Z = 82 $ ($ \mathrm{Pb} $)
and
\begin{equation}
  \Delta R_{\urm{ch}}^Z \left( A \right)
  =
  R_{\urm{ch}}^Z \left( A \right)
  -
  R_{\urm{ch}}^Z \left( A - 2 \right).
\end{equation}
This indicator corresponds to the discretized second derivative of $ R_{\urm{ch}} $ at $ A = A_{\urm{magic}} $,
which is related to the curvature of the graph in the continuum limit.
The larger value corresponds to the larger (stronger) kink.
The kink indicators for $ \mathrm{Sn} $ isotopes are summarized in Tables~\ref{tab:kink_size_Sn_NR} and \ref{tab:kink_size_Sn_R} for Skyrme EDFs and relativistic EDFs, respectively.
From Tables~\ref{tab:kink_size_Sn_NR} and \ref{tab:kink_size_Sn_R},
one general tendency is found: 
The results of relativistic (RMF and RHF) models give stronger kinks than the nonrelativistic (SHF) ones, 
while the EDF dependence among the members of each model is large. 
In the nonrelativistic SHF, UNEDF series somehow yields a smaller kink commonly than other EDFs.
In the relativistic cases,
the RHF calculations give larger kinks than those of the RMF ones, 
except the DD-LZ1 EDF. 
\begin{table}[tb]
  \centering
  \caption{
    Kink indicator $ \Delta^2 R_{\urm{ch}}^{\urm{Sn}} $ for
    $ \mathrm{Sn} $ calculated by nonrelativistic EDFs.
    The charge radius differences
    $ \Delta R_{\urm{ch}}^{\urm{Sn}} \left( 132 \right) $
    and
    $ \Delta R_{\urm{ch}}^{\urm{Sn}} \left( 134 \right) $
    are also listed.
    All values are given in units of $ 10^{-3} \, \mathrm{fm} $.}
  \label{tab:kink_size_Sn_NR}
  \begin{ruledtabular}
    \begin{tabular}{lddd}
      \multicolumn{1}{l}{EDF} & \multicolumn{1}{c}{$ \Delta R_{\urm{ch}}^{\urm{Sn}} \left( 132 \right) $} & \multicolumn{1}{c}{$ \Delta R_{\urm{ch}}^{\urm{Sn}} \left( 134 \right) $} & \multicolumn{1}{c}{$ \Delta^2 R_{\urm{ch}}^{\urm{Sn}} $} \\
      \hline
      UNEDF1   & +14.913 & +11.886 &  -3.027 \\
      UNEDF2   & +14.850 & +12.571 &  -2.279 \\
      UNEDF0   & +12.495 & +10.479 &  -2.016 \\
      SkM*     & +11.456 & +11.027 &  -0.429 \\
      HFB9     & +11.302 & +12.222 &  +0.920 \\
      SLy4     & +10.665 & +12.449 &  +1.784 \\
      SLy5     & +10.568 & +12.363 &  +1.795 \\
      SGII     & +11.436 & +13.446 &  +2.010 \\
      SAMi-noT & +11.101 & +17.318 &  +6.217 \\
      SAMi-T   & +11.196 & +18.507 &  +7.311 \\
      SAMi     & +11.229 & +21.092 &  +9.863 \\
      \hline
      Expt.    &  +7.4   & +22.4   & +15.0   \\
    \end{tabular}
  \end{ruledtabular}
\end{table}
\begin{table}[tb]
  \centering
  \caption{
    Same as Table~\ref{tab:kink_size_Sn_NR}, but calculated by relativistic EDFs.}
  \label{tab:kink_size_Sn_R}
  \begin{ruledtabular}
    \begin{tabular}{lddd}
      \multicolumn{1}{l}{EDF} & \multicolumn{1}{c}{$ \Delta R_{\urm{ch}}^{\urm{Sn}} \left( 132 \right) $} & \multicolumn{1}{c}{$ \Delta R_{\urm{ch}}^{\urm{Sn}} \left( 134 \right) $} & \multicolumn{1}{c}{$ \Delta^2 R_{\urm{ch}}^{\urm{Sn}} $} \\
      \hline
      PKDD             & +11.088 & +14.166 &  +3.078 \\
      DD-PC1           & +11.411 & +15.862 &  +4.451 \\
      DD-ME2 (Surface) &  +9.055 & +15.159 &  +6.104 \\
      DD-ME2 (TMR)     &  +9.290 & +15.587 &  +6.297 \\
      PKO2             & +11.424 & +18.474 &  +7.050 \\
      PKO1             & +10.172 & +19.566 &  +9.394 \\
      PKA1             &  +5.691 & +16.512 & +10.821 \\
      DD-LZ1           &  +4.836 & +16.631 & +11.795 \\
      PKO3             &  +9.455 & +23.896 & +14.441 \\
      PKO1*            &  +9.629 & +25.748 & +16.119 \\
      \hline
      Expt.            &  +7.4   & +22.4   & +15.0   \\
    \end{tabular}
  \end{ruledtabular}
\end{table}
\begin{table}[tb]
  \centering
  \caption{
    Mesons considered in PKDD, DD-ME2, DD-LZ1, PKO1, PKO2, PKO3, PKO1*, and PKA1 EDFs.
    All the couplings are density dependent.}
  \label{tab:RHF_meson}
  \begin{ruledtabular}
    \begin{tabular}{lllllll}
      \multicolumn{1}{l}{EDF} & \multicolumn{1}{c}{$ \pi $-PV} & \multicolumn{1}{c}{$ \rho $-V} & \multicolumn{1}{c}{$ \rho $-T} & \multicolumn{1}{c}{$ \rho $-VT} & \multicolumn{1}{c}{$ \omega $-V} & \multicolumn{1}{c}{$ \sigma $-S} \\
      \hline
      PKDD   & No  & Yes & No  & No  & Yes & Yes \\
      DD-ME2 & No  & Yes & No  & No  & Yes & Yes \\
      DD-LZ1 & No  & Yes & No  & No  & Yes & Yes \\
      \hline
      PKA1   & Yes & Yes & Yes & Yes & Yes & Yes \\
      PKO1   & Yes & Yes & No  & No  & Yes & Yes \\
      PKO2   & No  & Yes & No  & No  & Yes & Yes \\
      PKO3   & Yes & Yes & No  & No  & Yes & Yes \\
      PKO1*  & Yes & Yes & No  & No  & Yes & Yes \\
    \end{tabular}
  \end{ruledtabular}
\end{table}
\par
One can find that the SHF calculations, except SAMi, SAMi-noT, and SAMi-T, provide a smaller kink than those of the relativistic (RMF and RHF) ones.
We found that, in general, the RHF calculations provide the larger kink than the RMF ones.
This can be understood as follows:
As will be discussed, the spin-orbit interaction is important for the kink behavior~\cite{
  Nakada2019Phys.Rev.C100_044310}.
The spin-orbit mean field appears from both the relativistic effect
and the tensor interaction in RHF calculation,
while only from the former in RMF calculation;
hence, the origin of the spin-orbit interaction of the RMF calculation is different compared to that of the RHF calculation.
The spin-orbit interaction in the Skyrme calculation is introduced phenomenologically as the $ W_0 $ and $ W'_0 $ terms
without any density dependence.
Discussion of the density dependence of the spin-orbit interaction in the Skyrme interaction can be found in Refs.~\cite{
  Pearson1994Phys.Rev.C50_185,
  Pudliner1996Phys.Rev.Lett.76_2416,
  Bender2003Rev.Mod.Phys.75_121,
  Kanada-Enyo:2022dhj}.
\par
Among Skyrme EDFs,
the UNEDF series provide the ``anti-kink'';
the SkM* and HFB9 do not provide visible kink;
and
the SLy4, SLy5, and SGII give the kink, but the magnitude is smaller than the experimental one.
The SAMi, SAMi-noT, and SAMi-T provide the kink larger than the other Skyrme EDF.
The SAMi EDF is constructed to give the better description of spin-isospin properties, such as Gamow-Teller resonances~\cite{
  Roca-Maza2012Phys.Rev.C86_031306},
where the strength of the spin-orbit interaction stems from the spin-isospin transitions involving the spin-orbit partner levels. 
Hence, we conjecture that the modeling with respect to the spin-orbit splitting is essential to describe the kink better.
Note that the SGII is also constructed towards Gamow-Teller resonances~\cite{
  VanGiai1981Phys.Lett.B106_379},
but it does not give the kink as large as the SAMi.
It is worthwhile to mention that, in $ \mathrm{Pb} $ isotopes,
SAMi series and SGII give stronger kinks than other Skyrme EDFs as will be shown later.
\par
The SkI4 EDF
also gives the kink behavior of $ \mathrm{Pb} $ isotopes~\cite{
  Reinhard1995Nucl.Phys.A584_467}.
References~\cite{
  Sharma1995Phys.Rev.Lett.74_3744,
  Reinhard1995Nucl.Phys.A584_467}
claim that the introduction of the isoscalar and isovector spin-orbit interactions,
$ W_0 $ and $ W'_0 $, is essential, while 
the standard Skyrme EDFs have only the isoscalar term ($ W_0 = W'_0 $).
Among better EDFs to describe the kink behavior, 
the two strengths are identical, $ W_0 = W'_0 $ in the SGII EDF;
in the SkI4 EDF, the isoscalar and isovector spin-orbit interactions have opposite signs, namely, 
$ W'_0 < 0 < W_0 $ and $ \left| W'_0 \right| < W_0 $;
in the SAMi EDF, the isoscalar and isovector spin-orbit interactions are of the same direction, but the strengths are different,
$ 0 < W'_0 < W_0 $.
Since their strengths and signs of $ W_0 $ and $ W'_0 $ are different and rather arbitrary,
it is not possible at the moment 
which combination of $ W_0 $ and $ W'_0 $ values are the best for the kink behavior.
It should be noted that the detailed discussion of $ W_0 $ and $ W'_0 $ can be found in Refs.~\cite{
  Reinhard1989Rep.Prog.Phys.52_439,
  Onsi1997Phys.Rev.C55_3166,
  Nayak1998Phys.Rev.C58_878,
  Pearson2001Phys.Lett.B513_319,
  Bender2003Rev.Mod.Phys.75_121}.
\par
Comparing the results of the SAMi-T and the SAMi-noT EDFs,
one can see how the Skyrme tensor interaction affects the kink behavior.
Although the SAMi-T EDF gives the stronger kink than the SAMi-noT EDF
and the Skyrme tensor interaction makes the slope above the shell gap steeper,
its effect is not so significant.
Moreover, the shell structure, which will be discussed later, is not changed much by the tensor terms.
Therefore, the Skyrme tensor interaction introduces additional spin-orbit terms in the mean field,  but its effect 
is not so strong as far as SAMi-T is concerned.
This is because the effect of the tensor interaction may be already included in the original spin-orbit interaction during the fitting procedure.
\par
Next, we shall focus on the relativistic calculation.
Except for the DD-LZ1 EDF, the RMF calculations give a smaller kink than the RHF calculations,
while the DD-LZ1 EDF gives a comparable size of kink with those of the RHF calculations.
Among the RHF calculations, the kink size of PKO2 is the weakest
and PKO3 and PKO1* give appreciable kinks.
It should be noted that although the PKA1 and the DD-LZ1 give the strong kink,
their slopes above $ N = 82 $ are not so steep;
the reason why their kink sizes are large is that the slopes below $ N = 82 $ are loose.
Indeed, the slope above $ N = 82 $ of the PKA1 is the smallest among the RHF calculations,
while the slope above $ N = 82 $ of the DD-LZ1 is still largest among the RMF calculations.
To check the pairing model dependence,
two results of DD-ME2 with different pairing interactions,
denoted by ``DD-ME2 (TMR)'' and ``DD-ME2 (Surface)'',
are shown in Fig.~\ref{fig:Rch_Sn_R},
where the former and the latter, respectively, correspond to
the Tian-Ma-Ring (TMR) type~\cite{  
  Tian2009Phys.Lett.B676_44}
and the surface-type pairing interactions.
One can easily find that these two pairing interactions give almost the same kink size.
\par
The PKO1, PKO2, PKO3, PKO1*, and PKA1 basically start from the same Lagrangian,
while the number of mesons and their meson-nucleon coupling constants are different.
The mesons considered in these EDFs are summarized in Table~\ref{tab:RHF_meson}.
This suggests that $ \pi $-PV, $ \rho $-T, and $ \rho $-VT couplings,
which gives the tensor interaction,
are important to reproduce the kink behavior.
According to Ref.~\cite{
  Wang2021Phys.Rev.C103_064326},
the strength of the tensor interaction of PKO2 is
weakest
and that of PKO1* is the strongest;
that of PKO3 is the second strongest,
and 
those of PKA1 and PKO1 are marginal.
Strengths of the tensor interaction of these interactions are consistent with the kink size:
the stronger the tensor interaction the larger the kink size.
Therefore, the tensor interaction or its outcomes, such as spin-orbit mean field and shell structure, are important to reproduce the kink behavior.
This is in contrast to the nonrelativistic case.
\par
The DD-LZ1 EDF is constructed with the guidance of pseudo-spin symmetry restoration~\cite{
  Geng2019Phys.Rev.C100_051301}
and, as a result, shell evolution was described better than other popular RMF EDFs~\cite{
  Wei2020Chin.Phys.C44_074107}.
It is constructed without an ansatz of density dependences of some meson couplings
and
fitted to $ \nuc{U}{218}{} $, which is related to the sub-shell closure of $ Z = 92 $.
The pseudo-spin symmetry is strongly related to the spin-orbit splitting~\cite{
  Liang2015Phys.Rep.570_1}.
This fact is in agreement with our conjecture that the spin-orbit splitting is important to reproduce the kink size.
\subsubsection{Single-particle energies}
\par
\begin{figure*}[tb]
  \centering
  \includegraphics[width=1.0\linewidth]{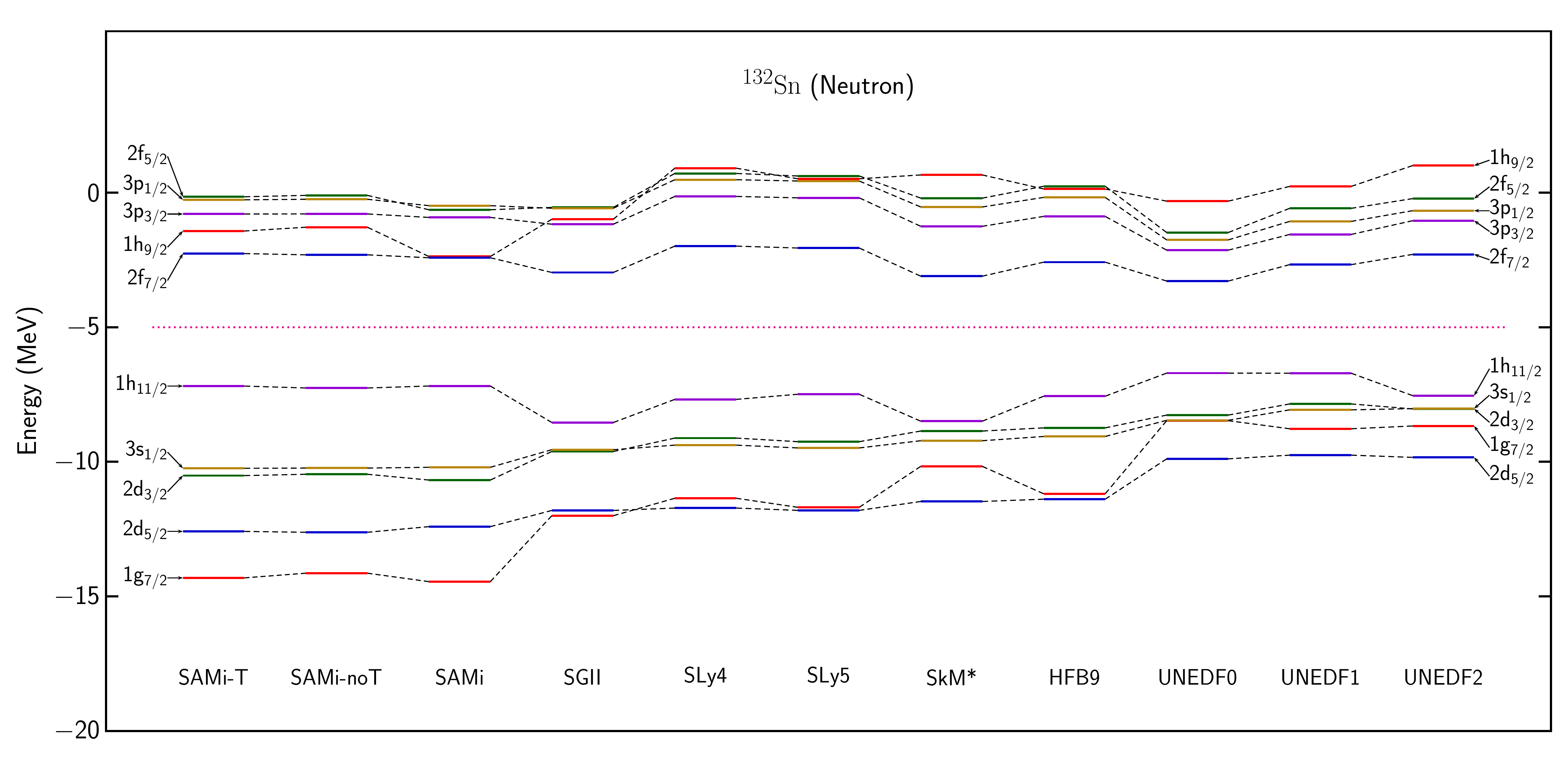}
  \caption{
    Single-particle spectra of $ \nuc{Sn}{132}{} $ calculated by using nonrelativistic EDFs.}
  \label{fig:spectra_Sn_NR}
\end{figure*}
\begin{figure*}[tb]
  \centering
  \includegraphics[width=1.0\linewidth]{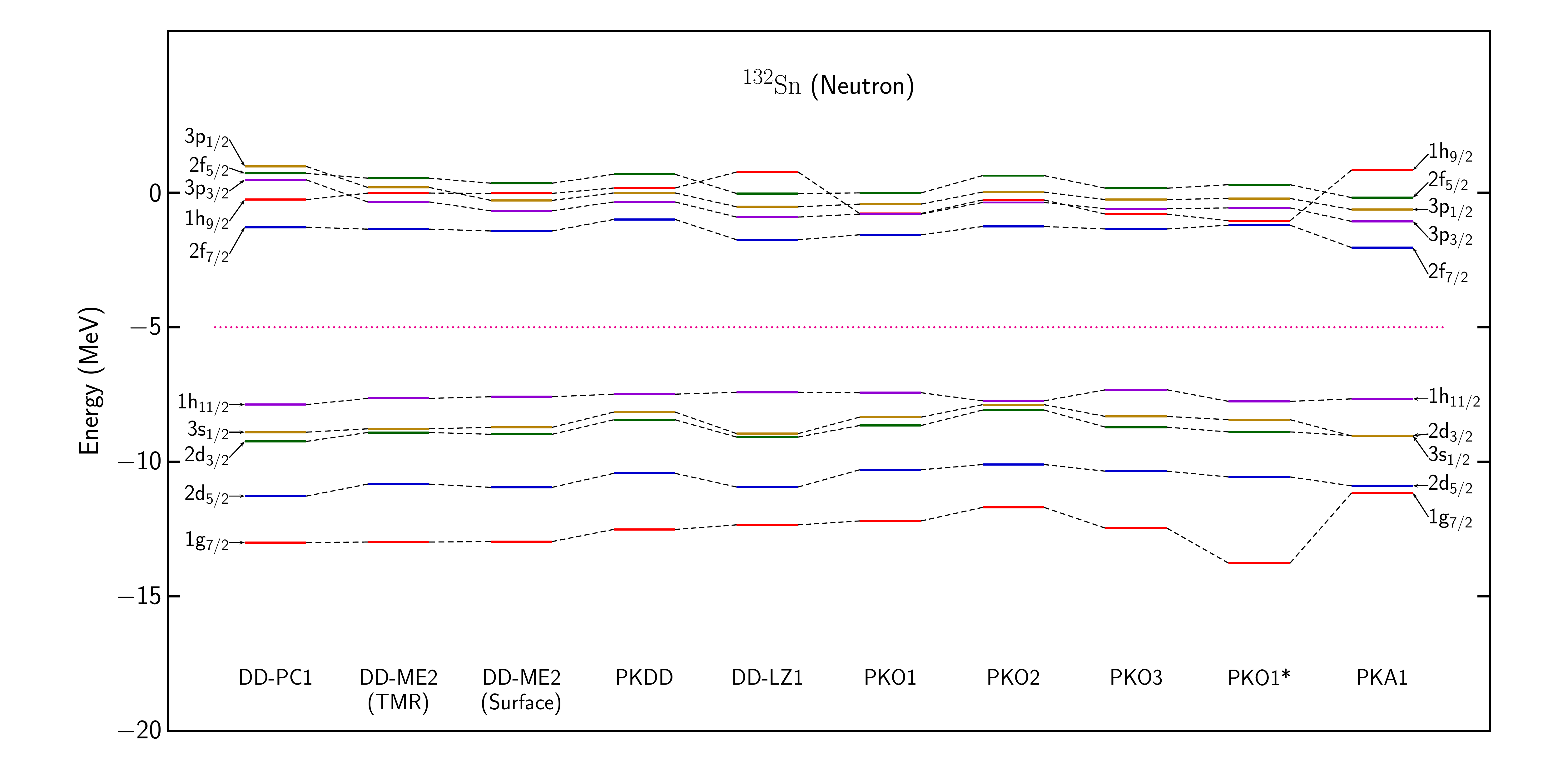}
  \caption{
    Same as Fig.~\ref{fig:spectra_Sn_NR}, but by using relativistic EDFs.}
  \label{fig:spectra_Sn_R}
\end{figure*}
\begin{figure}[tb]
  \centering
  \includegraphics[width=1.0\linewidth]{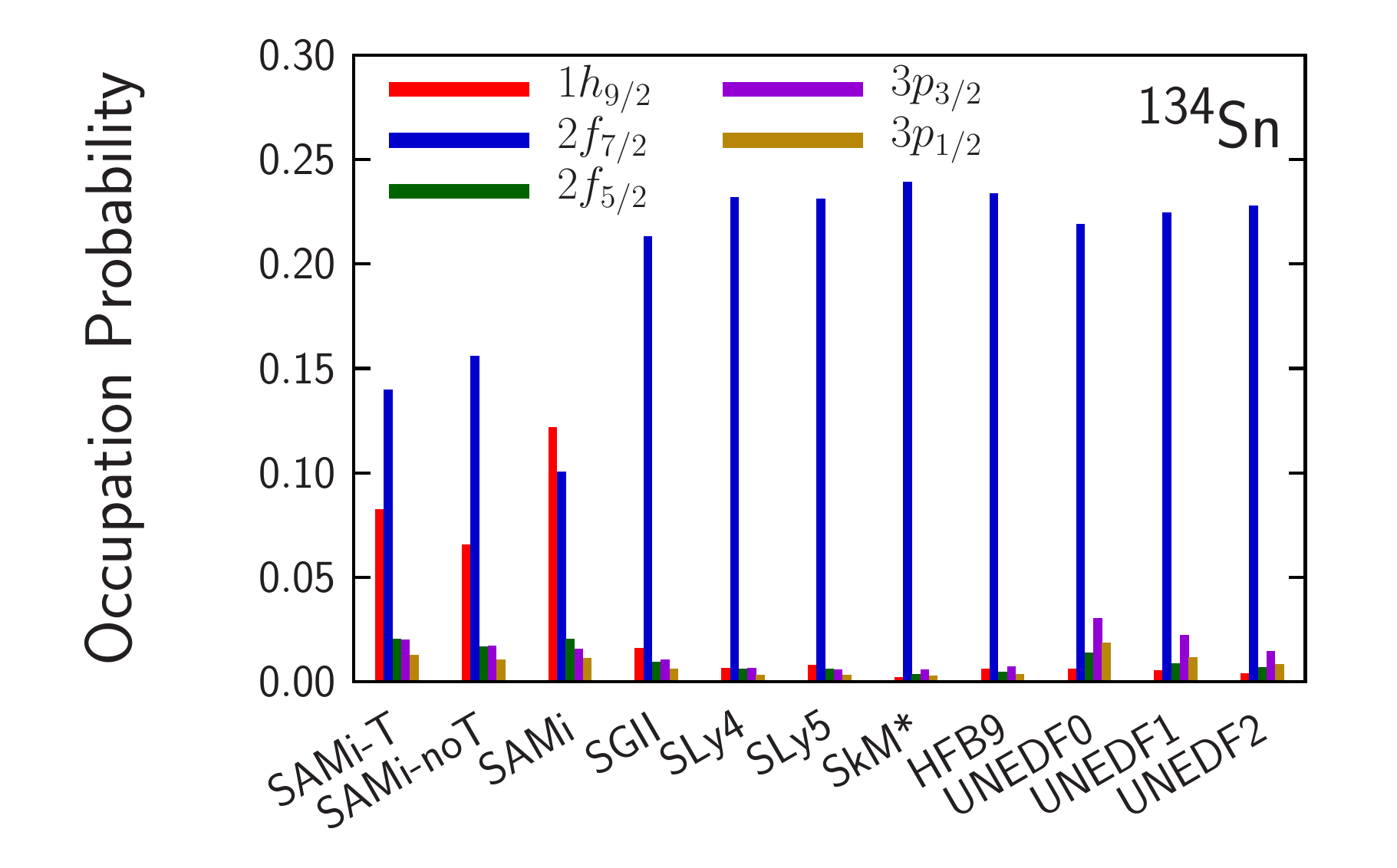}
  \caption{
    Occupation probability of orbitals above the $ N = 82 $ shell gap for $ \nuc{Sn}{134}{} $
    calculated by using nonrelativistic EDFs.}
  \label{fig:occ_Sn_NR}
\end{figure}
\begin{figure}[tb]
  \centering
  \includegraphics[width=1.0\linewidth]{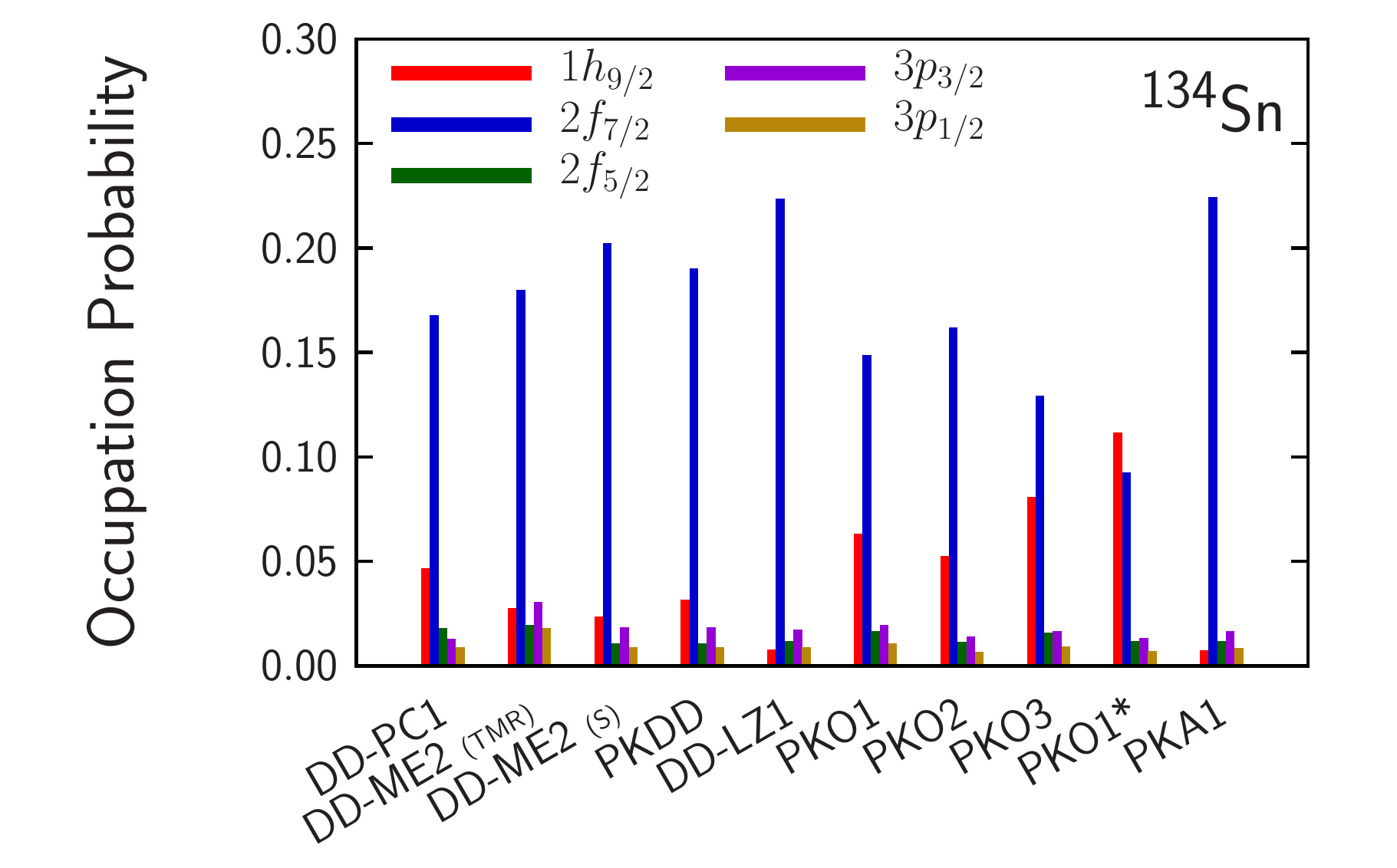}
  \caption{
    Same as Fig.~\ref{fig:occ_Sn_NR}, but by using relativistic EDFs.}
  \label{fig:occ_Sn_R}
\end{figure}
\begin{figure}[tb]
  \centering
  \includegraphics[width=1.0\linewidth]{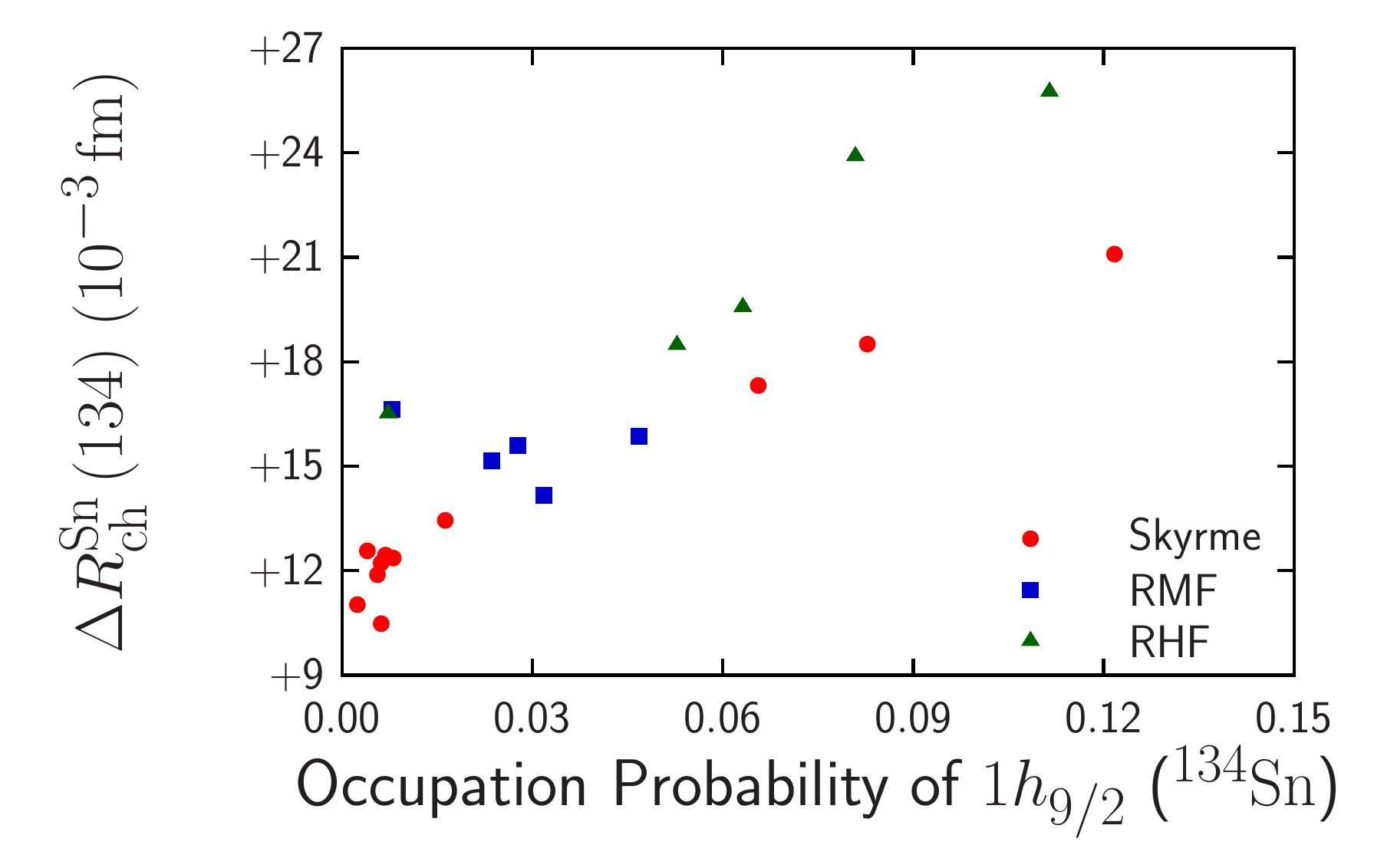}
  \caption{
    Correlation between occupation probability of $ 1f_{9/2} $ orbital of $ \nuc{Sn}{134}{} $ and kink size at $ N = 82 $.}
  \label{fig:occ-kink_Sn}
\end{figure}

To understand the kink behaviors better, 
the single-particle spectra of neutrons of $ \nuc{Sn}{132}{} $ are shown in Figs.~\ref{fig:spectra_Sn_NR} and \ref{fig:spectra_Sn_R}.
The magenta dotted line indicates the $ N = 82 $ shell gap.
All the calculations show that the $ 2f_{7/2} $ orbital is just above the $ N = 82 $ shell gap,
but above the $ 2f_{7/2} $ orbital,
the order of valence-neutron orbitals noticeably depends on the EDF utilized.
Especially, the $ 1h_{9/2} $ orbital appears just above the $ 2f_{7/2} $ orbital in SAMi, SAMi-noT, SAMi-T, DD-PC1, PKO3, and PKO1* calculations,
in which the kink is strong as listed in Tables \ref{tab:kink_size_Sn_NR} and \ref{tab:kink_size_Sn_R}.
Therefore, the order of the single-particle orbitals above the shell gap will play an important role
as mentioned in Ref.~\cite{
  Perera2021Phys.Rev.C104_064313}.
To understand this mechanism more clearly,
the occupation probabilities of the single-particle orbitals just above the shell gap of $ \nuc{Sn}{134}{} $ are plotted
in Figs.~\ref{fig:occ_Sn_NR} and \ref{fig:occ_Sn_R}.
Here, the occupation probability, which ranges between $ 0 $ and $ 1 $, is defined by the occupation number divided by the maximum occupation number.
The correlation between the occupation probability of the $ 1h_{9/2} $ orbital and the slope above the $ N = 82 $ gap [$ \Delta R_{\urm{ch}}^{\urm{Sn}} \left( 134 \right) $] is plotted in Fig.~\ref{fig:occ-kink_Sn}.
It is easily seen that, in general, the larger occupancy of the $ 1h_{9/2} $ orbital gives the steeper slope,
and accordingly the larger kink,
for instance, in SAMi, SAMi-noT, SAMi-T, and the PKO series.
One possible explanation on this kink-evolution effect of $ 1h_{9/2} $ is as follows:
the $ 1h_{9/2} $ orbital does not have the nodal structure,
while the $ 2f_{7/2} $ one has a node.
The former has larger overlap with the protons and thus it extends the charge radius due to the proton-neutron attractive interaction,
as discussed in Ref.~\cite{
  Perera2021Phys.Rev.C104_064313}.
Thus, when the two valence neutrons above the $ N = 82 $ shell occupy this orbital,
the mean radial distribution is suddenly enhanced.
The DD-LZ1 again shows an exceptional behavior: 
Its kink size is noticeably strong, 
although the occupation probability of the $ 1h_{9/2} $ orbital is small.
A similar result is obtained with PKA1.
For the large size of the kink of DD-LZ1 and PKA1, 
an alternative hint is possibly given by considering the single-particle levels.
In Fig.~\ref{fig:spectra_Sn_R},
the energies of $ 1h_{9/2} $ obtained from the DD-LZ1 and PKA1 are located higher than the other relativistic EDFs' results, and thus, their radial distributions can become wider.
Even if its occupation probability is small, 
a finite mixing of this $ 1h_{9/2} $ component can enhance the sudden change of radial distributions between $ \nuc{Sn}{132}{} $ and $ \nuc{Sn}{134}{} $, as well as the size of kink.
\par
It is also shown that the occupation probability of the $ 3p_{3/2} $ orbital is
much larger than the $ 1h_{9/2} $ orbital in the UNEDF0, UNEDF1, and UNEDF2,
which show the ``anti kink'' behavior at $ N = 82 $.
These anti kinks also originate from the steep slope below $ N = 82 $ and the moderate slopes above $ N = 82 $.
\par
\begin{figure}[tb]
  \centering
  \includegraphics[width=1.0\linewidth]{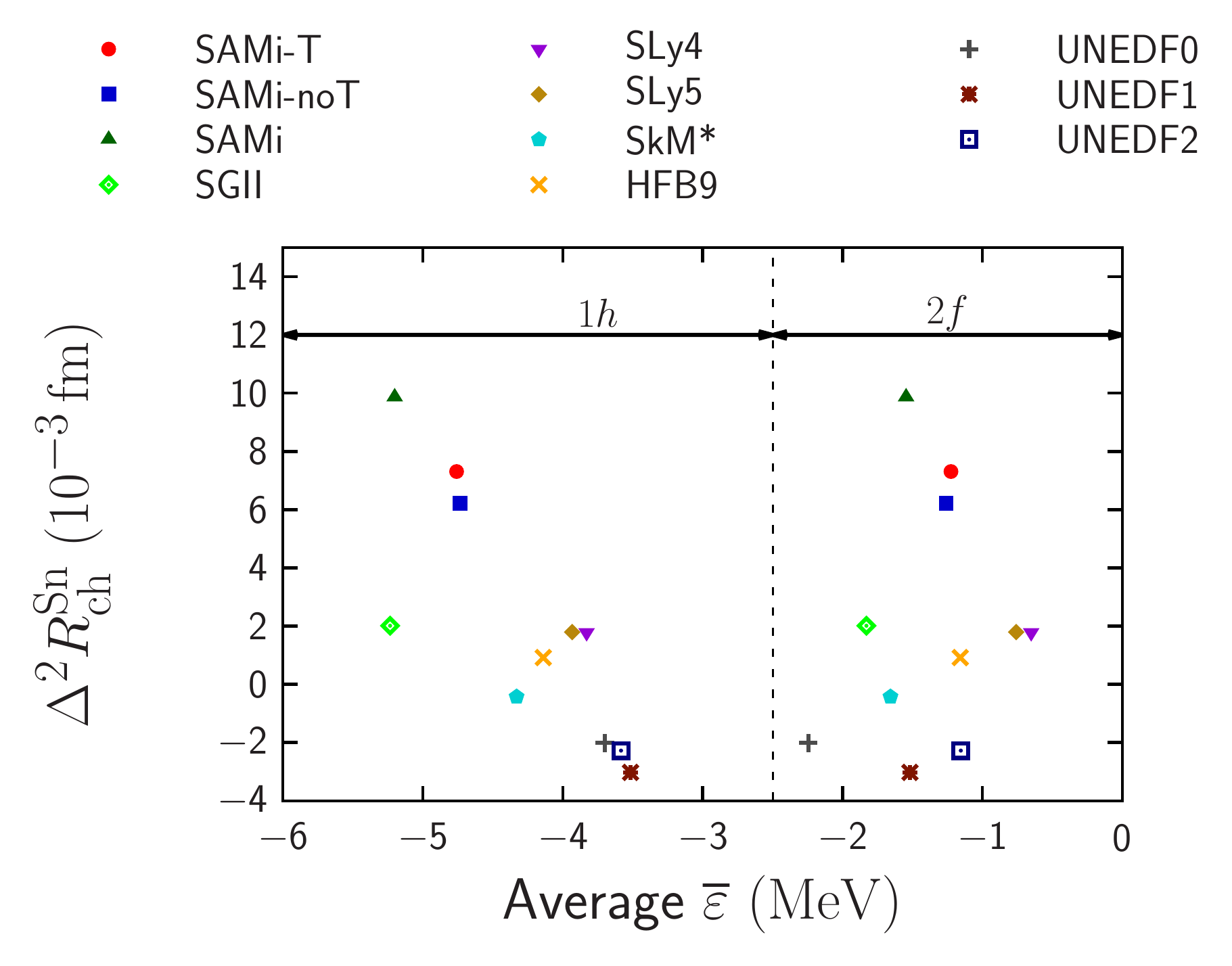}
  \caption{
    Correlations between the averaged single-particle energies of $ 1h $ and $ 2f $ orbitals of $ \nuc{Sn}{134}{} $ and the kink size
    calculated by using nonrelativistic EDFs.}
  \label{fig:average_kink_050_134_nonrel}
\end{figure}
\begin{figure}[tb]
  \centering
  \includegraphics[width=1.0\linewidth]{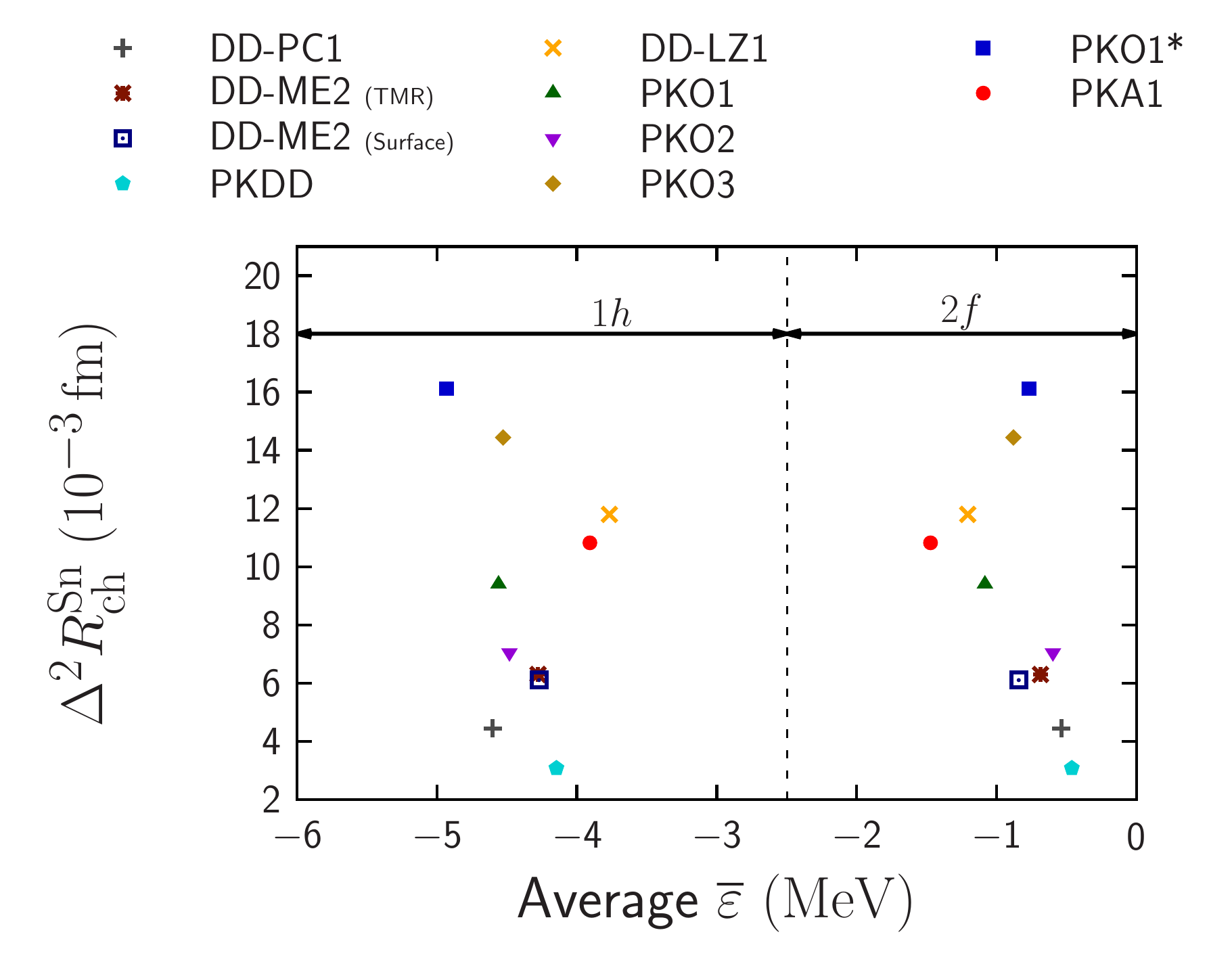}
  \caption{
    Same as Fig.~\ref{fig:average_kink_050_134_nonrel}, but by using relativistic EDFs.}
  \label{fig:average_kink_050_134_rel}
\end{figure}
\begin{figure}[tb]
  \centering
  \includegraphics[width=1.0\linewidth]{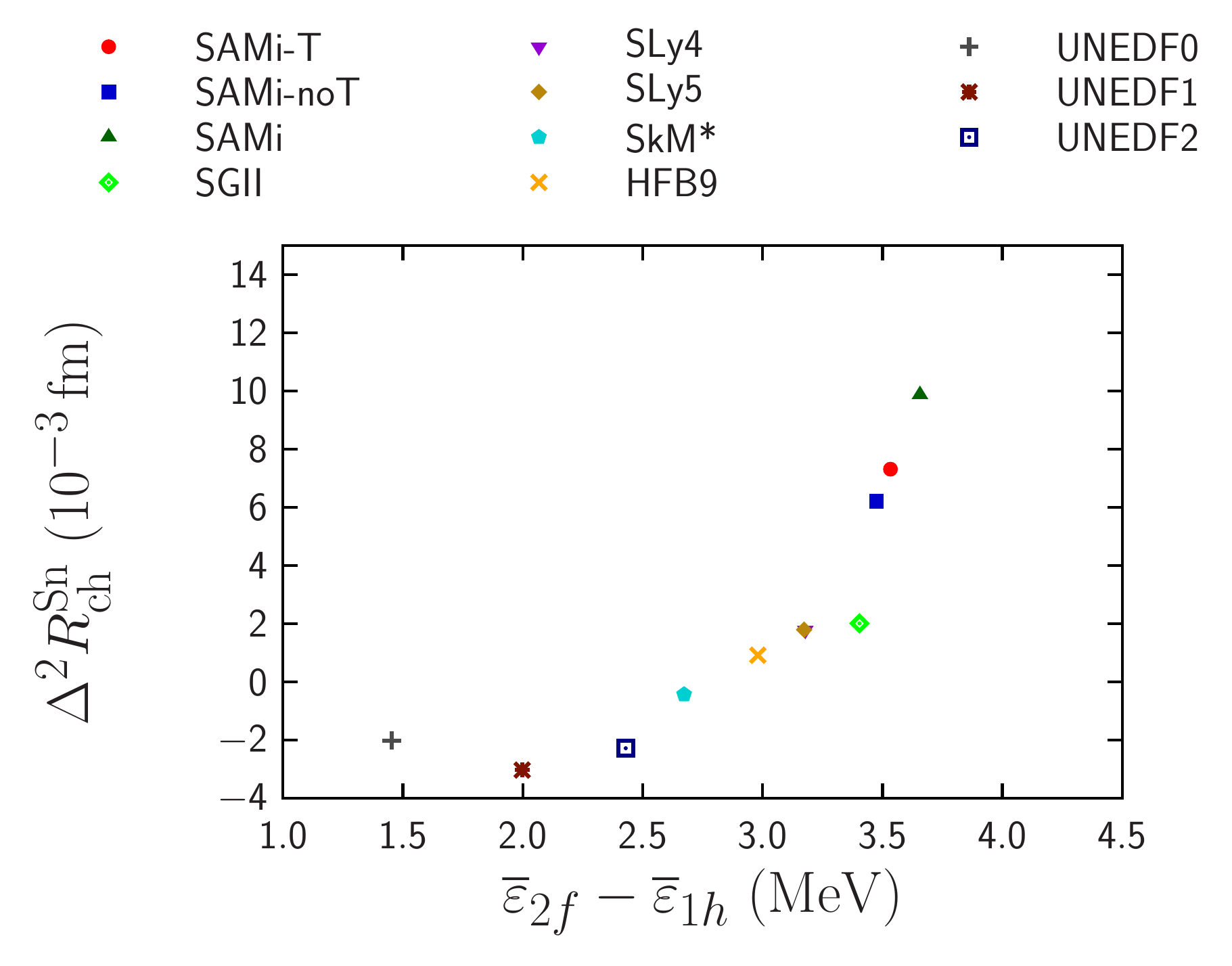}
  \caption{
    Correlations between the difference between the averaged single-particle energies of $ 1h $ and $ 2f $ orbitals of $ \nuc{Sn}{134}{} $ and the kink size
    calculated by using nonrelativistic EDFs.}
  \label{fig:average_diff_050_134_nonrel}
\end{figure}
\begin{figure}[tb]
  \centering
  \includegraphics[width=1.0\linewidth]{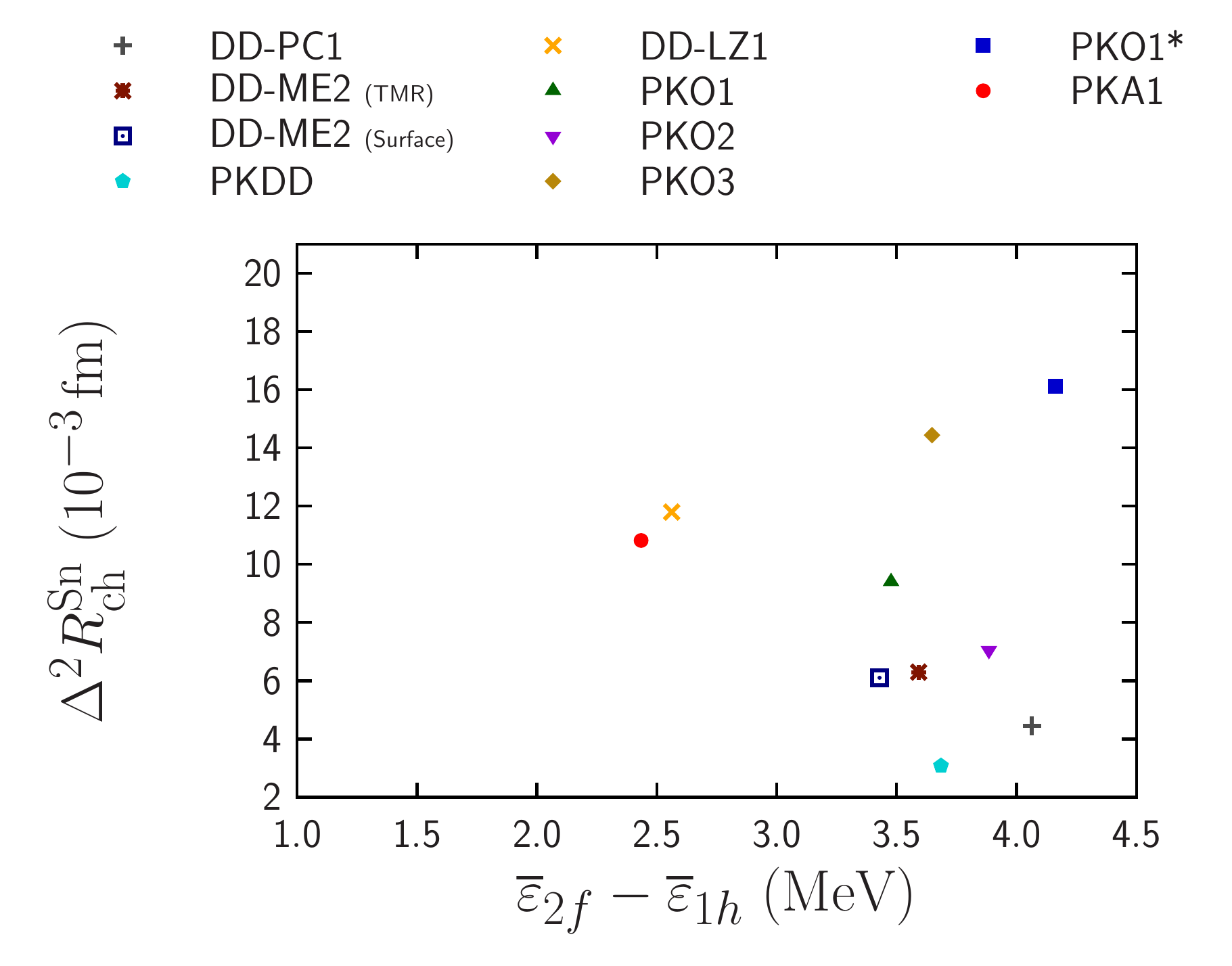}
  \caption{
    Same as Fig.~\ref{fig:average_diff_050_134_nonrel}, but by using relativistic EDFs.}
  \label{fig:average_diff_050_134_rel}
\end{figure}

The occupation probabilities of the orbitals near the Fermi level depend on the single-particle energies.
The occupation probabilities affect the kink size, as well as the spin-orbit interaction.
In order to discuss the effect of the single-particle energies,
the spin-orbit interaction should be \textit{effectively} switched off.
To this end, 
the average energy of the spin-orbit partners $ \overline{\epsilon} $ is considered here.
This $ \overline{\epsilon} $ is not affected by the spin-orbit potential but by the central potential.
The spin-orbit interaction for the orbital with the orbital angular momentum $ l $ is
proportional to
$ l $ for $ j = l + 1/2 $ orbitals
and
$ - \left( l + 1 \right) $ for $ j = l - 1/2 $ orbitals.
Therefore,
the \textit{averaged} single-particle energy for the $ l $ orbital is defined by
\begin{equation}
  \overline{\epsilon}
  =
  \frac{l}{2l + 1}
  \epsilon_{<}
  +
  \frac{l + 1}{2l + 1}
  \epsilon_{>} ,
\end{equation}
where $ \epsilon_{<} $ and $ \epsilon_{>} $ are the single-particle energies for $ j = l - 1/2 $ and $ l + 1/2 $ orbitals, respectively.
Figures \ref{fig:average_kink_050_134_nonrel} and \ref{fig:average_kink_050_134_rel}
show the correlation between the averaged single-particle energies $ \overline{\epsilon} $ of the $ 1h $ and $ 2f $ orbitals in $ \nuc{Sn}{134}{} $ 
and the kink size $ \Delta^2 R_{\urm{ch}} $
in the nonrelativistic and relativistic schemes, respectively.
There is a weak correlation between $ \overline{\epsilon} $ of the $ 1h $ orbital and the kink size.
As $ \overline{\epsilon} $ of the $ 1h $ orbital is small,
the $ 1h_{9/2} $ orbital becomes lower and accordingly the occupation probability of the $ 1h_{9/2} $ orbital becomes larger.
The occupation probability of the $ 1h_{9/2} $ orbitals is correlated to the kink size as discussed above.
We also found that there is no obvious correlation between $ \overline{\epsilon} $ of the $ 2f $ orbital and the kink size.
In Figs.~\ref{fig:average_diff_050_134_nonrel} and \ref{fig:average_diff_050_134_rel},
the correlation between
the difference between two average energies,
$ \overline{\epsilon}_{2f} - \overline{\epsilon}_{1h} $,
and the kink size is shown.
We found that, in contrast to the case of bare $ \overline{\epsilon} $,
there is a clear correlation between
$ \overline{\epsilon}_{2f} - \overline{\epsilon}_{1h} $
and $ \Delta^2 R_{\urm{ch}}^{\urm{Sn}} $ in the nonrelativistic calculation.
We also found that if the pion contribution is included in the RHF calculation;
the smaller $ \overline{\epsilon}_{2f} - \overline{\epsilon}_{1h} $ gives the smaller $ \Delta^2 R_{\urm{ch}}^{\urm{Sn}} $,
while other correlation is not obvious in the relativistic calculation.
Note that similar analysis for the band termination was done in Ref.~\cite{
  Afanasjev2008Phys.Rev.C78_054303}.
\par
By summarizing the above discussions, to evolve the kink behavior of $ R_{\urm{ch}} $ at $ N = 82 $,
the occupation probability of the $ 1h_{9/2} $ orbital,
which is located above the $ 2f_{7/2} $ orbital,
must be large enough.
In order to lower the $ 1h_{9/2} $ orbital,
the spin-orbit interaction should not be too strong.  
Otherwise, the $ 1h_{9/2} $ level becomes higher than the $ 2f_{7/2} $ one in energy
and also the $ 3p_{3/2} $ energy can be too low, which is the origin of the antikink.
The spin-orbit gaps of the $ 1h $ orbitals calculated by the UNDEF0, UNEDF1, and UNEDF2 are almost the same as the other Skyrme EDFs except for the SAMi series.  
However, the occupation probability of the $ 1h_{9/2}$ orbital in the UNEDF0, UNEDF1, and UNEDF2 is smaller.
This may be caused by a larger effective mass of UNEDF EDFs. 
It should be noted that the larger effective mass makes the smaller energy spreading above the shell gap as well,
while the UNEDF series gives as wide energy spreading above the shell gap as the other Skyrme EDFs tested here.
The spin-orbit strength, energy spreading, and the effective masses are related to each other~\cite{
  Satula2008Phys.Rev.C78_011302};
hence, a detailed study of the effect of the effective mass is left for a future investigation.
Another feature is the two parameters of the spin-orbit interaction in some of the Skyrme EDFs.
Even though the strengths of isoscalar and isovector spin-orbit interactions, $ W_0 $ and $ W'_0 $, are the same,
the kink can appear as in the SGII case. 
Thus, the effect of $ W_0 $ and $ W'_0 $ to the kink behavior is still puzzling.
The tensor interactions also induce the spin-orbit mean field in the Skyrme EDF,
SAMi-T, but its effect is tiny.
The average single-particle energy of the $ 1h $ orbital is also correlated with the kink size.
\par
It should be noted that these two orbitals, $ 2f_{7/2} $ and $ 1h_{9/2} $, are pseudospin doublet;
thus if the pseudospin symmetry exactly holds, these two orbitals completely degenerate,
which may make the occupation probability of $ 1h_{9/2} $ larger.
It is known that the relativistic models have the pseudospin symmetry implicitly in the Dirac wave function. 
This may be another reason why the relativistic calculation describes the kink behavior better.
However,
some relativistic EDFs such as the DD-LZ1, which is expected to give better description of the pseudo-spin symmetry, do not give such degeneration.
\subsection{$ \mathrm{Pb} $ isotopes}
\label{sec:calculation_Pb}
\par
In this subsection, results for $ \mathrm{Pb} $ isotopes are shown.
The kink indicators for $ \mathrm{Pb} $ isotopes are summarized in Tables~\ref{tab:kink_size_Pb_NR} and \ref{tab:kink_size_Pb_R}.
Results for $ \mathrm{Pb} $ isotopes are basically similar to those for $ \mathrm{Sn} $ isotopes,
while there are some differences.
For instance,
HFB9 and UNEDF2 give small positive kinks,
SGII and DD-PC1 give significantly larger kinks.
Since properties of $ \nuc{Pb}{208}{} $ are frequently used for fitting criteria of EDFs,
such differences may appear.
\par
\begin{figure*}[tb]
  \centering
  \includegraphics[width=1.0\linewidth]{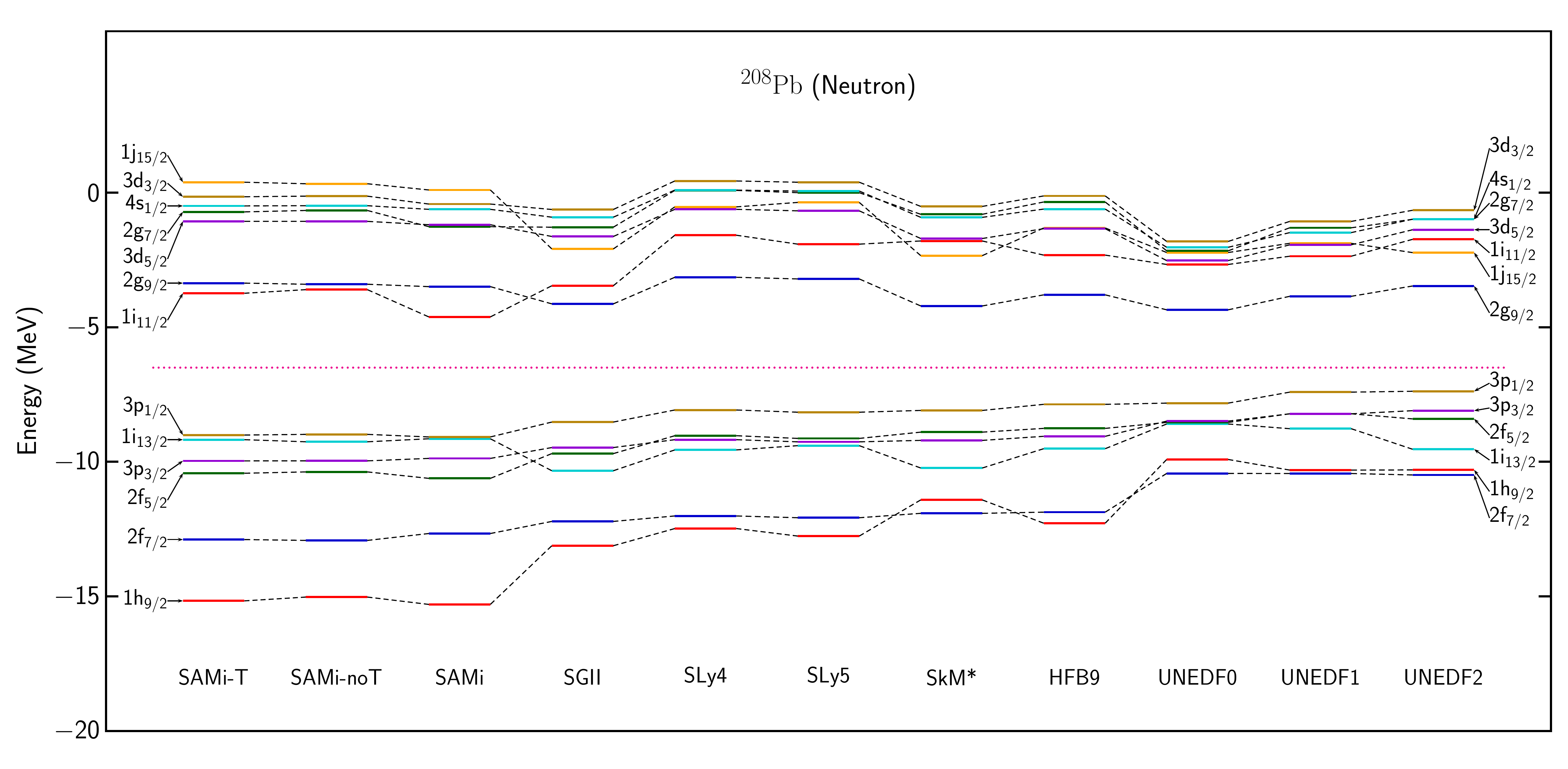}
  \caption{
    Same as Fig.~\ref{fig:spectra_Sn_NR}, but of $ \nuc{Pb}{208}{} $.}
  \label{fig:spectra_Pb_NR}
\end{figure*}
\begin{figure*}[tb]
  \centering
  \includegraphics[width=1.0\linewidth]{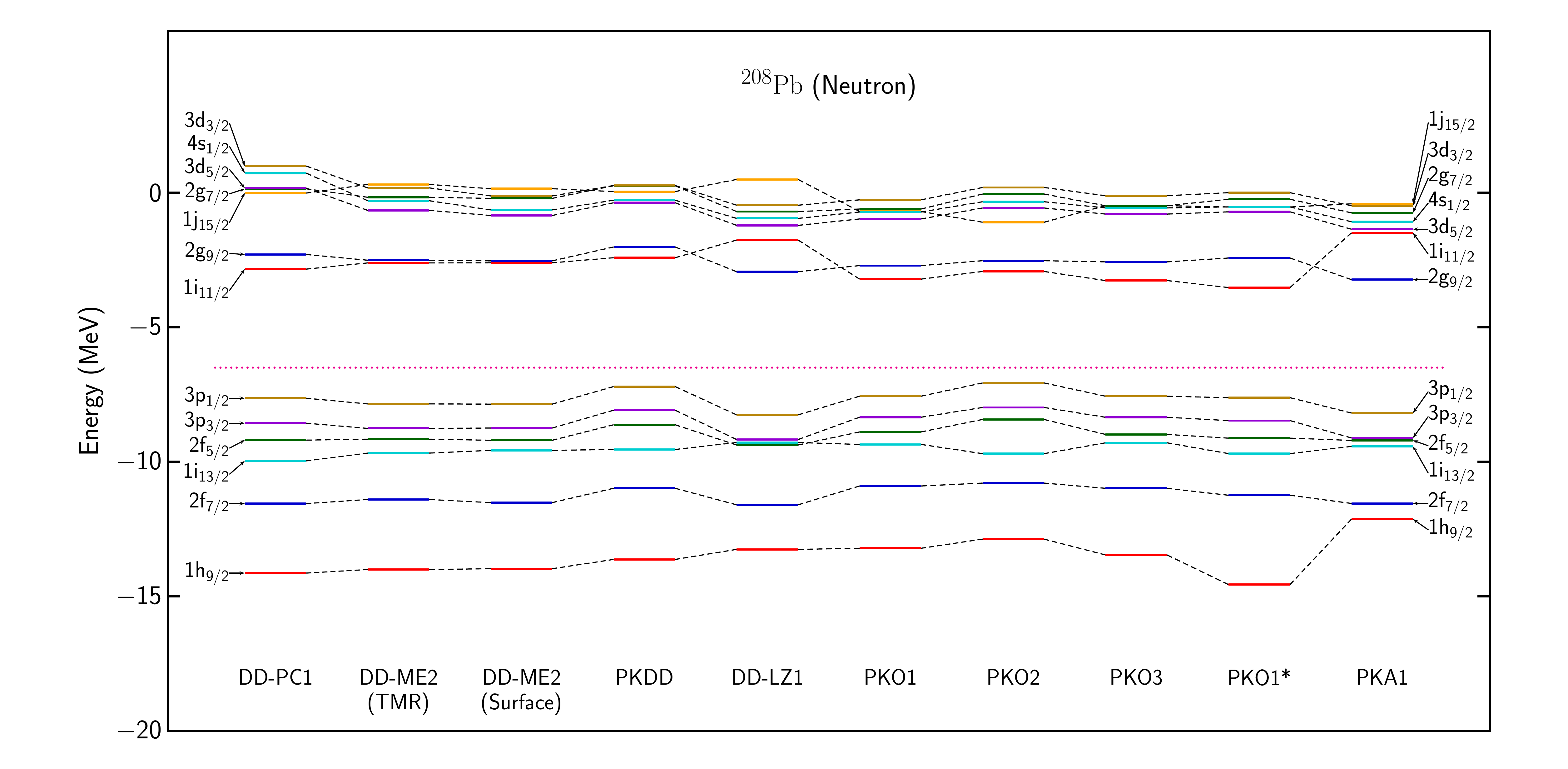}
  \caption{
    Same as Fig.~\ref{fig:spectra_Pb_NR}, but by using relativistic EDFs.}
  \label{fig:spectra_Pb_R}
\end{figure*}
\begin{figure}[tb]
  \centering
  \includegraphics[width=1.0\linewidth]{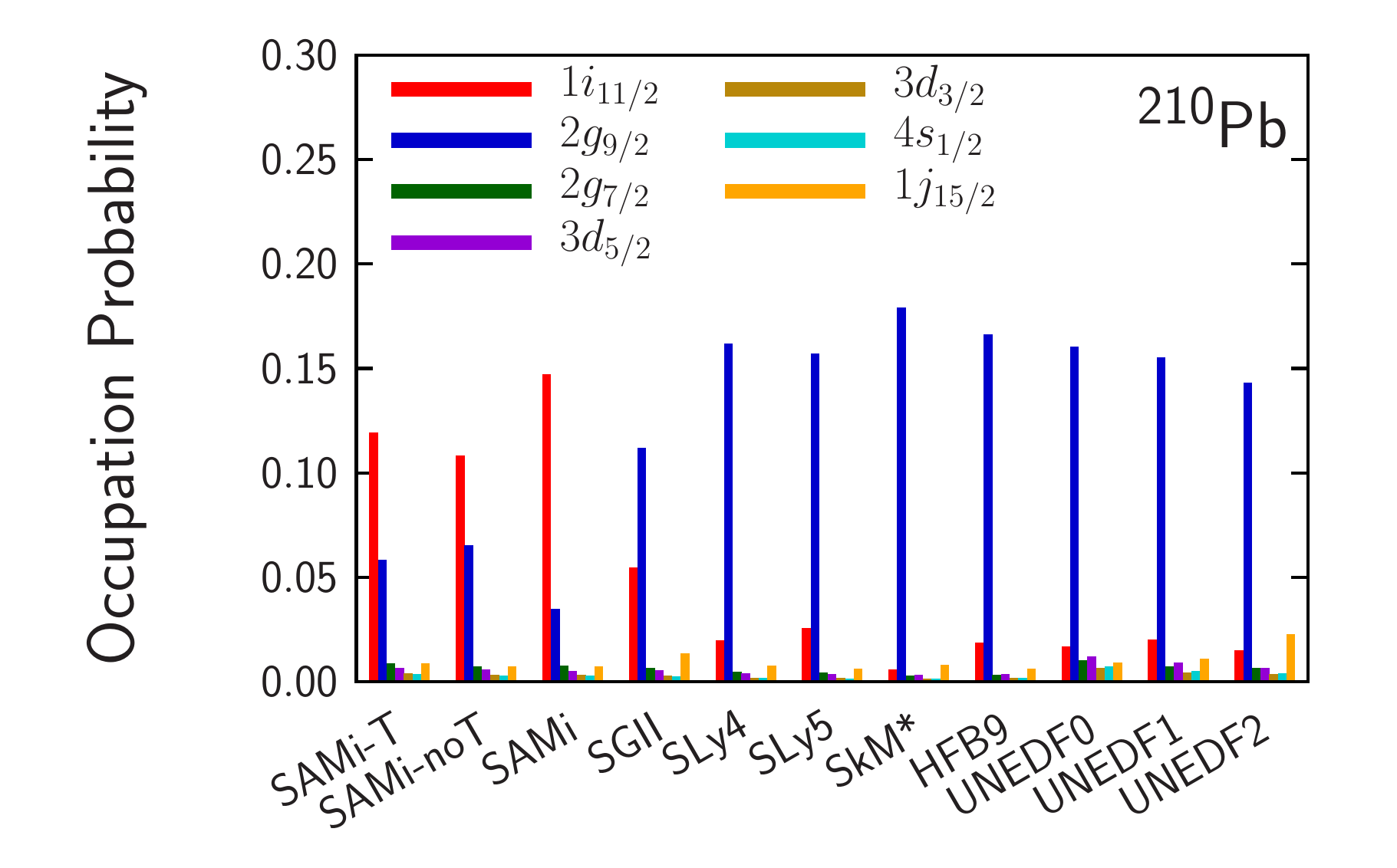}
  \caption{
    Occupation probability of orbitals above the $ N = 126 $ shell gap for $ \nuc{Pb}{210}{} $
    calculated by using nonrelativistic EDFs.}
  \label{fig:occ_Pb_NR}
\end{figure}
\begin{figure}[tb]
  \centering
  \includegraphics[width=1.0\linewidth]{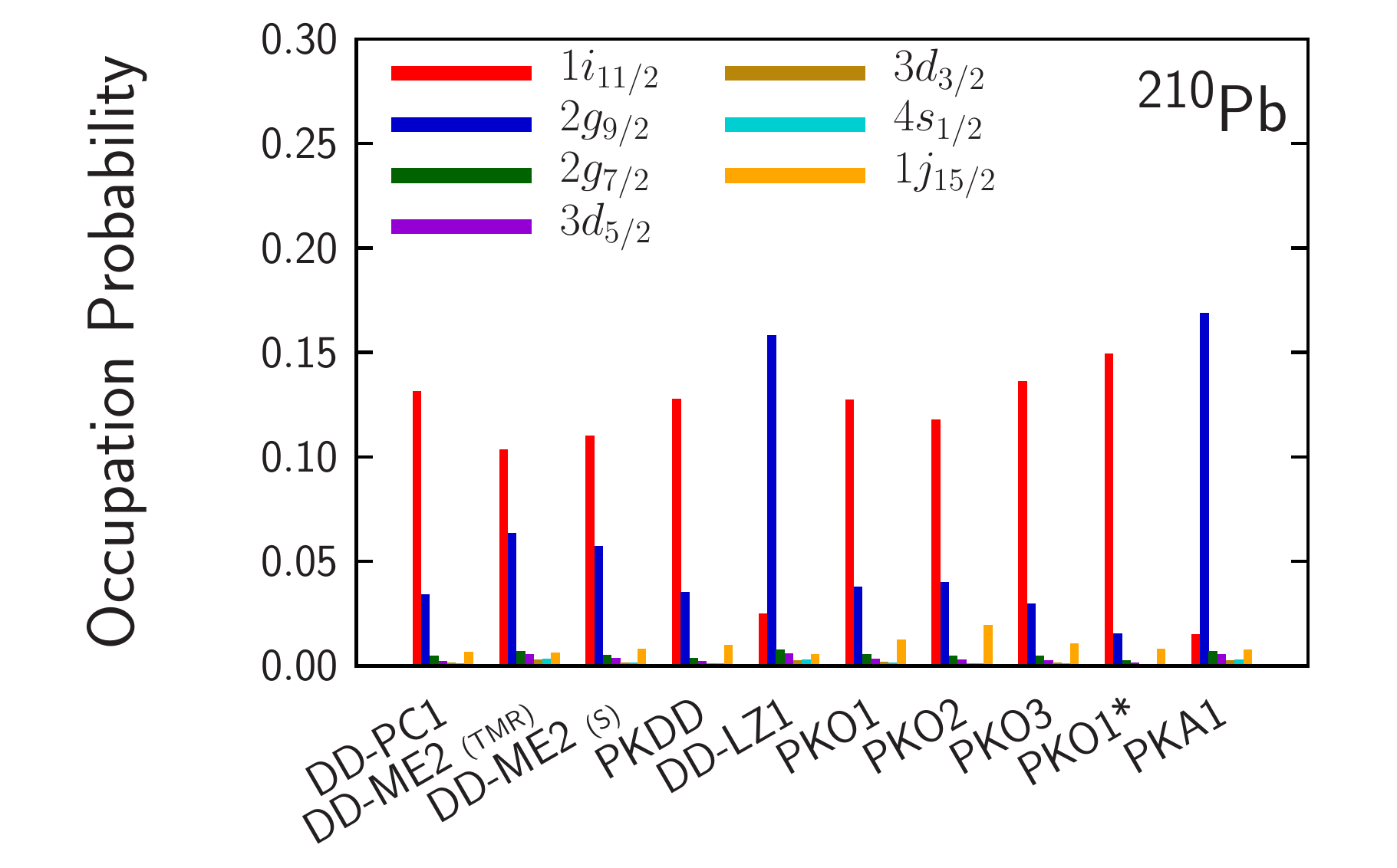}
  \caption{
    Same as Fig.~\ref{fig:occ_Pb_NR}, but by using relativistic EDFs.}
  \label{fig:occ_Pb_R}
\end{figure}
\begin{figure}[tb]
  \centering
  \includegraphics[width=1.0\linewidth]{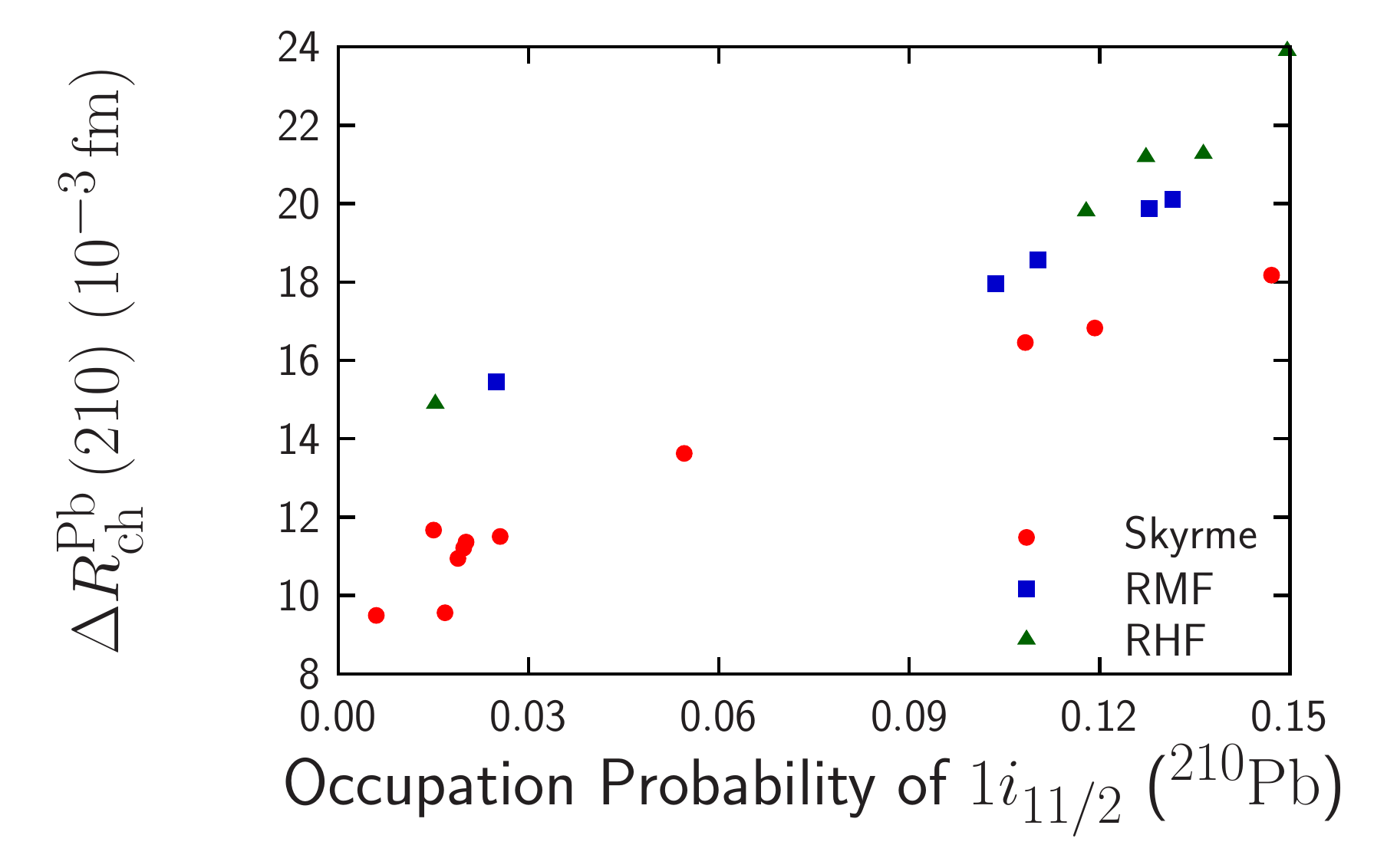}
  \caption{
    Correlation between occupation probability of $ 1i_{11/2} $ orbital of $ \nuc{Pb}{210}{} $ and kink size at $ N = 126 $.}
  \label{fig:occ-kink_Pb}
\end{figure}

To understand these behaviors,
the single-particle spectra of neutrons of $ \nuc{Pb}{208}{} $ are shown in Figs.~\ref{fig:spectra_Pb_NR} and \ref{fig:spectra_Pb_R}.
The magenta dotted line represents the $ N = 126 $ shell gap.
All the calculations show that the $ 3p_{1/2} $ orbital is just below the $ N = 126 $ shell gap,
while above the shell gap, either the $ 2g_{9/2} $ or $ 1i_{11/2} $ orbital appears:
the SAMi-T, SAMi-noT, SAMi, DD-PC1, DD-ME2, PKDD, PKO1, PKO2, PKO3, and PKO1* EDFs give the $ 1i_{11/2} $ orbital lower than the $ 2g_{9/2} $ one,
while the other calculations give the opposite.
The exceptions are the SkM* and UNEDF2 EDFs.
In these two EDFs,
the $ 2g_{9/2} $ orbital appears just above the shell gap, the same as most Skyrme EDFs,
but the $ 1j_{15/2} $ orbital appears instead of the $ 1i_{11/2} $ orbital above the $ 2g_{9/2} $ orbital. 
Referring to Tables \ref{tab:kink_size_Pb_NR} and \ref{tab:kink_size_Pb_R},
one can find that the EDFs,
having the $ 1i_{11/2} $ orbital just above the shell gap,
give a notable kink.
Therefore, the order of the single-particle orbitals above the shell gap may play an important role
as mentioned in Ref.~\cite{
  Perera2021Phys.Rev.C104_064313}.
To understand the mechanism more clearly,
the occupation probabilities of the single-particle orbitals just above the shell gap of $ \nuc{Pb}{210}{} $ are plotted
in Figs.~\ref{fig:occ_Pb_NR} and \ref{fig:occ_Pb_R}.
The correlation between the occupation probability of the $ 1i_{11/2} $ orbital and the slope above the $ N = 126 $ gap [$ \Delta R_{\urm{ch}}^{\urm{Pb}} \left( 210 \right) $] is plotted in Fig.~\ref{fig:occ-kink_Pb}.
It is easily seen that, in general, the larger occupancy of the $ 1i_{11/2} $ orbital gives the steeper slope,
and accordingly the larger kink.
This mechanism is the same as discussed in $ \mathrm{Sn} $ isotopes;
the radial distribution of this single-particle orbital with a smaller $ n $ value can be extended and give a larger kink. 
We especially mention the DD-LZ1 and PKA1, which have  relatively small kinks in Table~\ref{tab:kink_size_Pb_R}.
In the present case of $ \mathrm{Pb} $,
this small kink is attributable to the occupation probabilities shown in Fig.~\ref{fig:occ-kink_Pb}.
Namely, the dominant component is of $ 2g_{9/2} $ with DD-LZ1 and PKA1, 
whereas the other relativistic EDFs conclude the dominance of $ 1i_{11/2} $.
Then, the dominant $ 1i_{11/2} $ ($ 2g_{9/2} $) component leads to a large (small) change of radii for the corresponding kink behavior.
However, as we mentioned in the previous $ \mathrm{Sn} $ case, 
this explanation was less persuading for the $ \mathrm{Sn} $ isotopes, 
where the DD-LZ1 and PKA1 conclude large kinks,
 even though their occupation probabilities of $ 1h_{9/2} $ for extended radii are minor, 
as shown in Fig.~\ref{fig:occ_Sn_R}.
\par
The SGII EDF gives the notable kink,
while the $ 2g_{9/2} $ orbital appears just above the $ N = 126 $ shell gap. 
Nevertheless, this can be understood as follows:
the energy difference between the $ 2g_{9/2} $ orbital and the $ 1i_{11/2} $ one is small for the SGII case 
compared with the other calculations.  
Accordingly, as seen in Fig.~\ref{fig:occ_Pb_NR}, the occupation probability of $ 1i_{11/2} $ is substantially large,
even though that of $ 2g_{9/2} $ is larger than $ 1i_{11/2} $.
Hence, the effect of the $ 1i_{11/2} $ orbital is appreciable.
Then, one puzzle appears.  
Even though the strengths of isoscalar and isovector spin-orbit interactions, $ W_0 $ and $ W'_0 $, are the same in the SGII EDF,
the kink can appear in almost the same size with the SAMi EDF,
in which $ W_0 \ne W'_0 $.
This is in contrast to the claim in Refs.~\cite{
  Sharma1995Phys.Rev.Lett.74_3744,
  Reinhard1995Nucl.Phys.A584_467}.
Thus, we cannot find any strong relation between the isoscalar and isovector spin-orbit strengths and the kink behavior.
Lastly, we should notice that
the Skyrme tensor interaction also introduces the spin-orbit mean field of SAMi-T,
but its effect is tiny, as in the $ \mathrm{Sn} $ case.
\par
\begin{figure}[tb]
  \centering
  \includegraphics[width=1.0\linewidth]{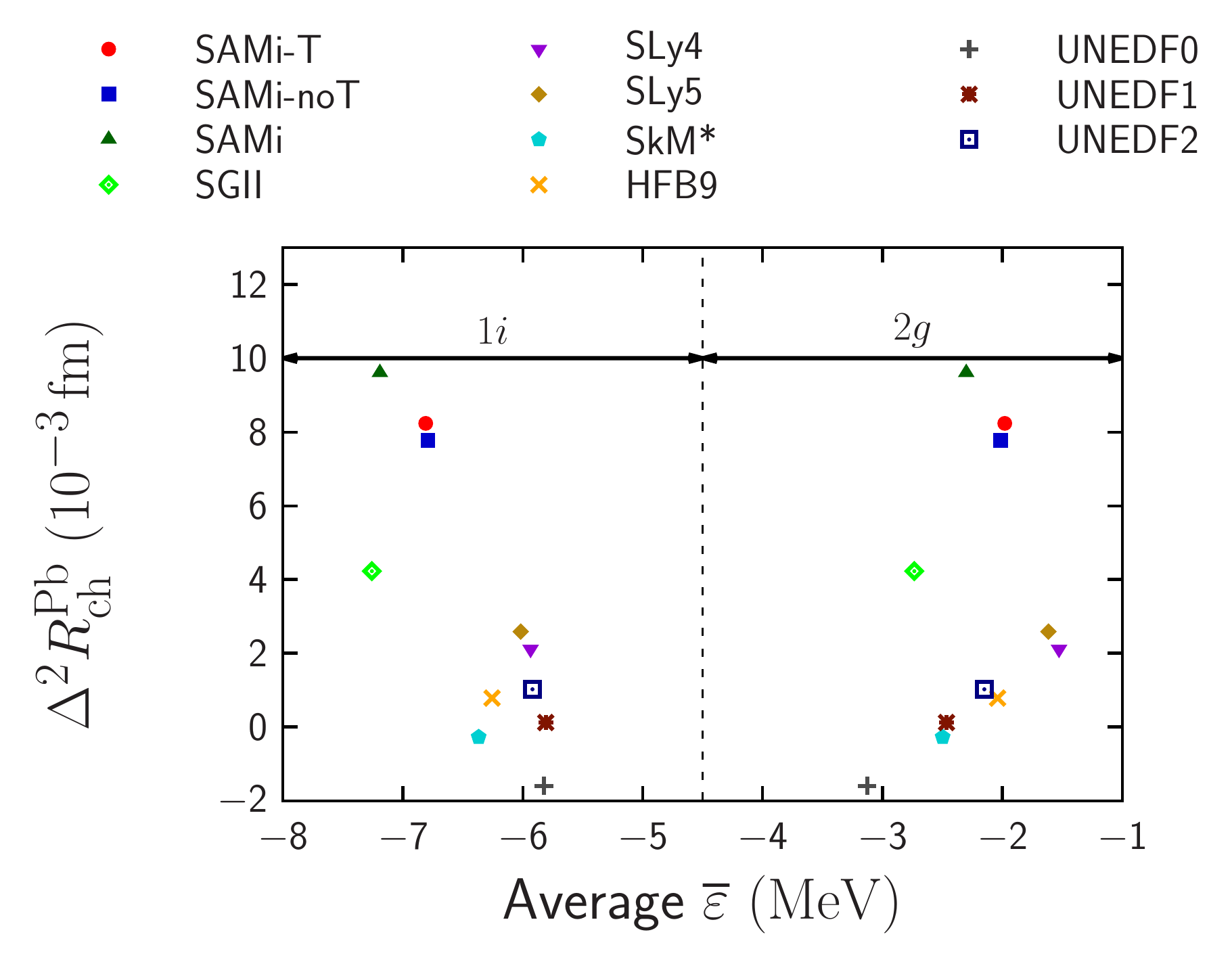}
  \caption{
    Correlation between the averaged single-particle energies of $ 1i $ and $ 2g $ orbitals of $ \nuc{Pb}{210}{} $ and the kink size
    calculated by using nonrelativistic EDFs.}
  \label{fig:average_kink_082_210_nonrel}
\end{figure}
\begin{figure}[tb]
  \centering
  \includegraphics[width=1.0\linewidth]{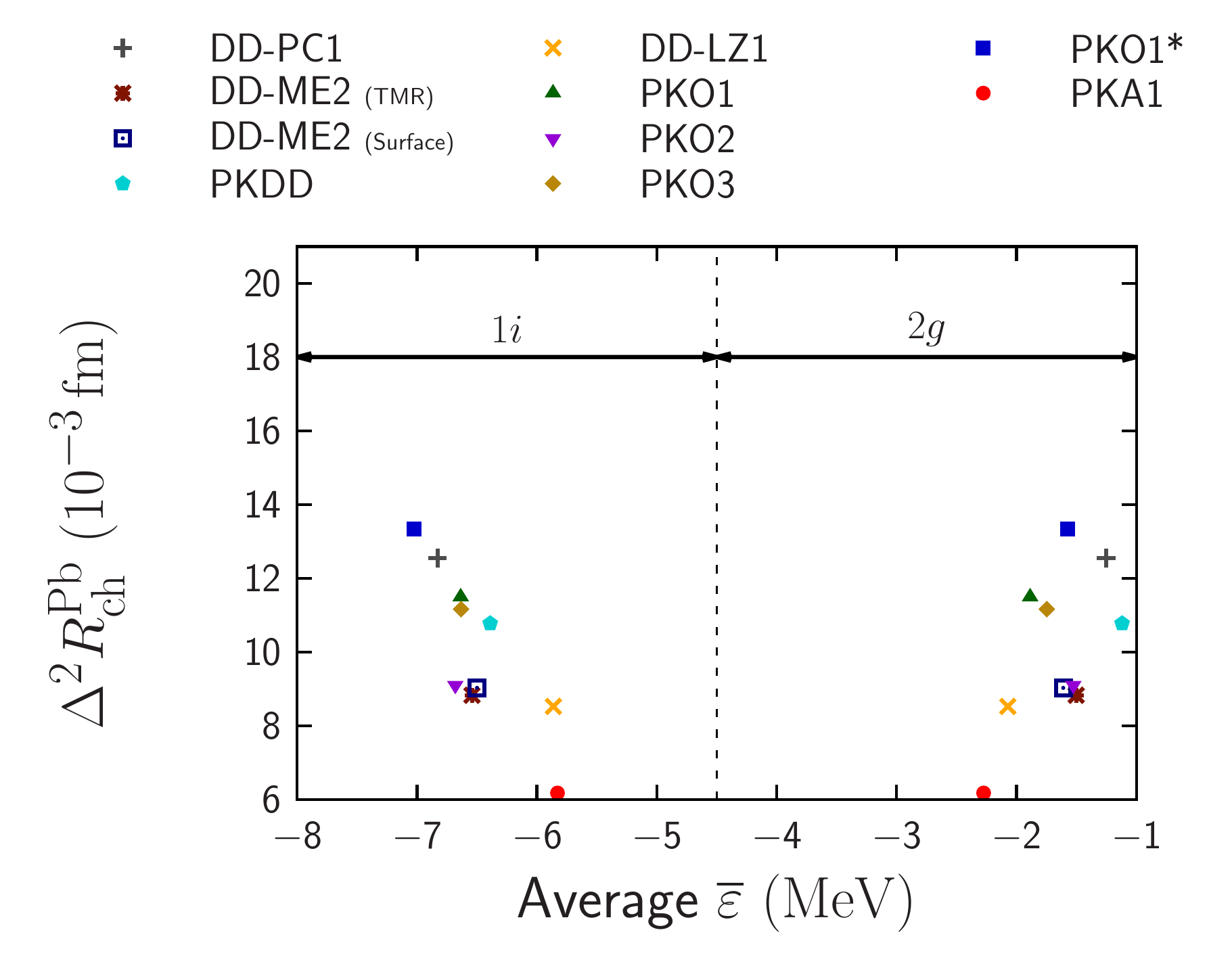}
  \caption{
    Same as Fig.~\ref{fig:average_kink_082_210_nonrel}, but by using relativistic EDFs.}
  \label{fig:average_kink_082_210_rel}
\end{figure}
\begin{figure}[tb]
  \centering
  \includegraphics[width=1.0\linewidth]{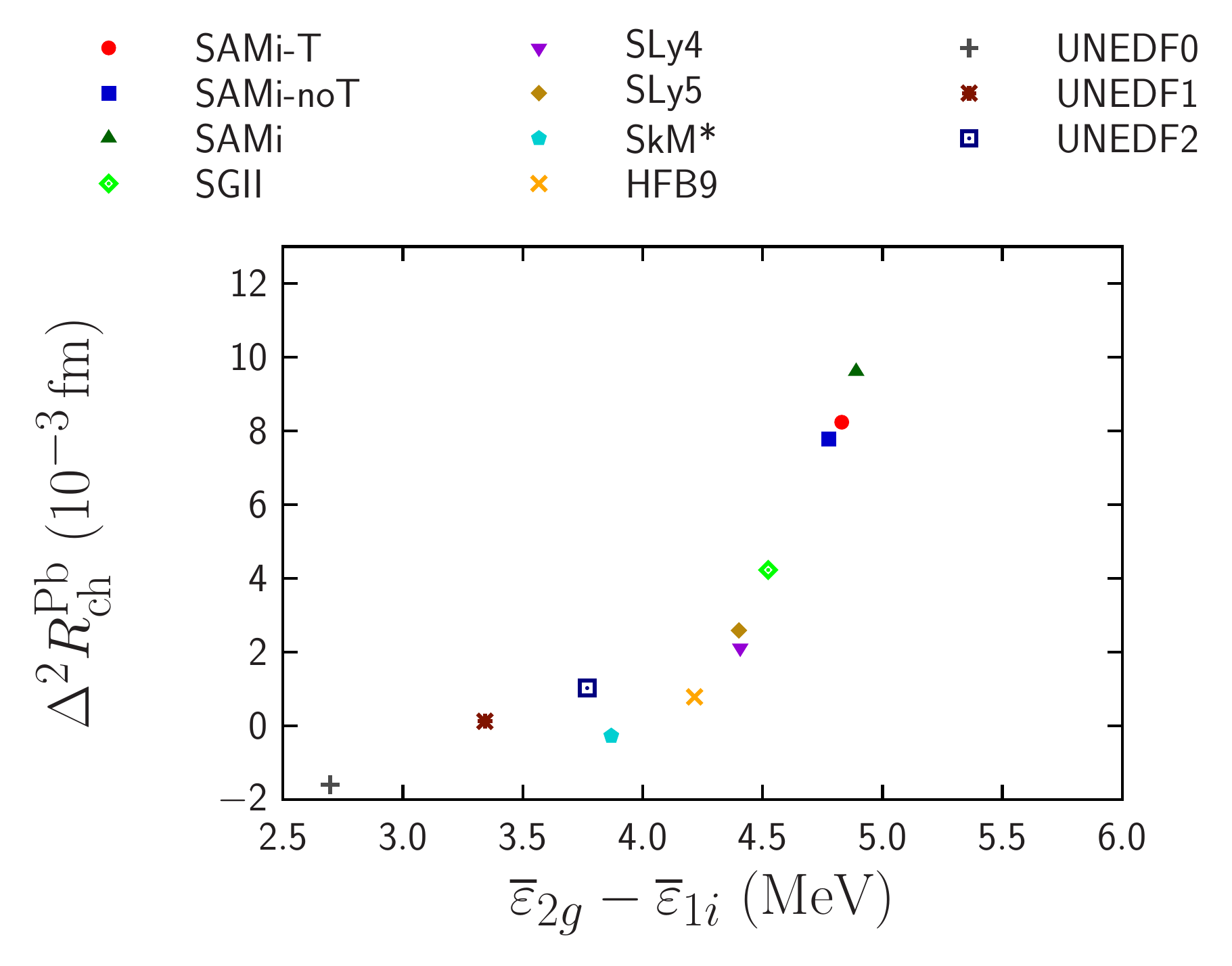}
  \caption{
    Correlations between the difference between the averaged single-particle energies of $ 1i $ and $ 2g $ orbitals of $ \nuc{Pb}{210}{} $ and the kink size
    calculated by using nonrelativistic EDFs.}
  \label{fig:average_diff_082_210_nonrel}
\end{figure}
\begin{figure}[tb]
  \centering
  \includegraphics[width=1.0\linewidth]{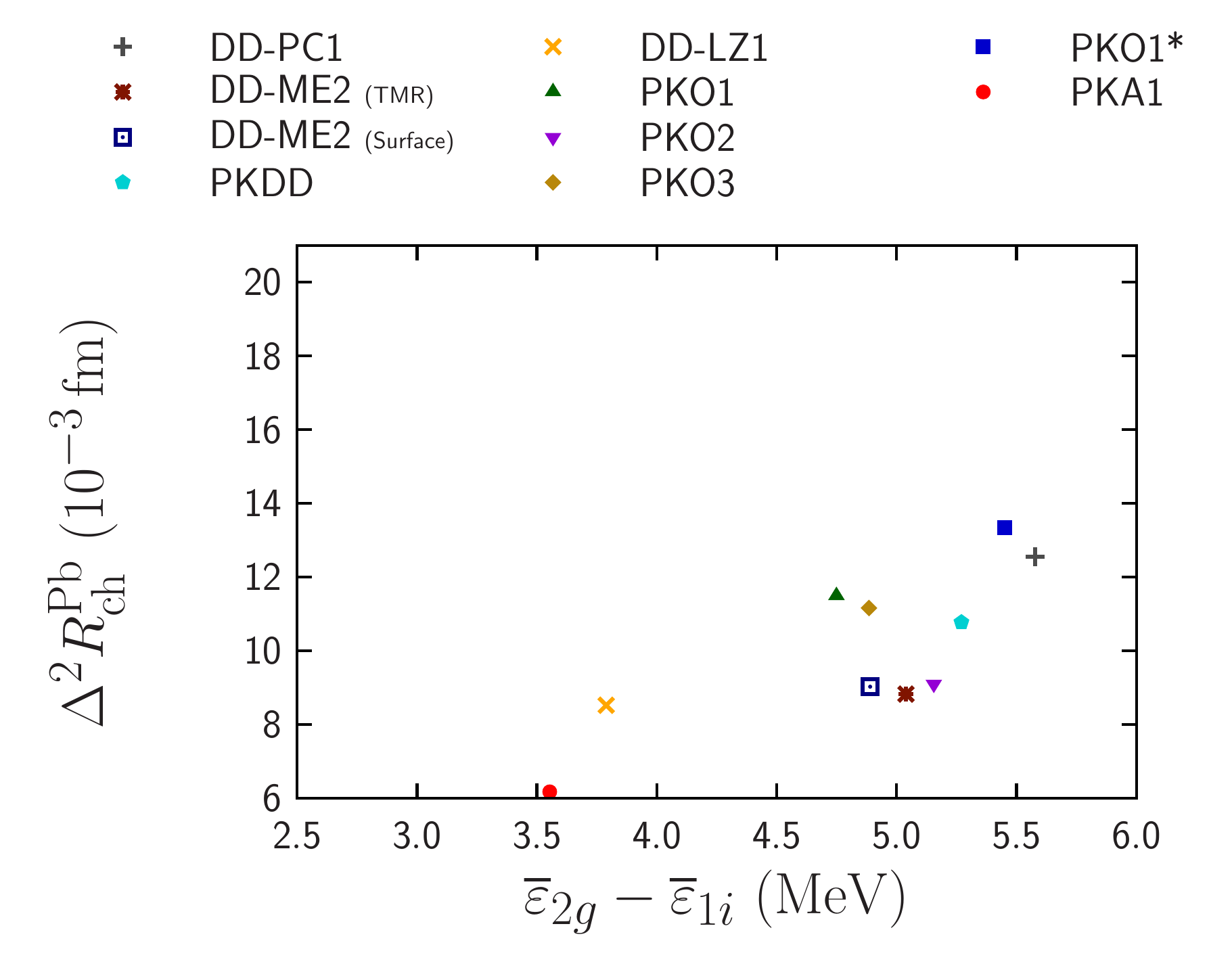}
  \caption{
    Same as Fig.~\ref{fig:average_diff_082_210_nonrel}, but by using relativistic EDFs.}
  \label{fig:average_diff_082_210_rel}
\end{figure}

Figures \ref{fig:average_kink_082_210_nonrel} and \ref{fig:average_kink_082_210_rel} show the correlation between the averaged single-particle energies $ \overline{\epsilon} $ of the $ 1i $ and $ 2g $ orbitals in $ \nuc{Pb}{210}{} $ 
and the kink size $ \Delta^2 R_{\urm{ch}} $.
There is a weak correlation between $ \overline{\epsilon} $ of the $ 1i $ orbital and the kink size.
This can be understood that, as $ \overline{\epsilon} $ of the $ 1i $ orbital is small,
the $ 1i_{11/2} $ orbital becomes lower and accordingly the occupation probability of the $ 1i_{11/2} $ orbital becomes larger.
The occupation probability of the $ 1i_{11/2} $ orbitals is correlated to the kink size as discussed above.
In contrast to the case of $ \nuc{Sn}{134}{} $,
there is also a weak correlation between $ \overline{\epsilon} $ of the $ 2g $ orbital and the kink size;
while the correlation is opposite to that for the $ 1i $ orbital.
In Figs.~\ref{fig:average_diff_082_210_nonrel} and \ref{fig:average_diff_082_210_rel},
the correlation between
the difference between two average energies,
$ \overline{\epsilon}_{2g} - \overline{\epsilon}_{1i} $,
and the kink size is shown.
We found that
there is an obvious correlation between
$ \overline{\epsilon}_{2g} - \overline{\epsilon}_{1i} $
and $ \Delta^2 R_{\urm{ch}}^{\urm{Pb}} $ in both the nonrelativistic and relativistic calculations.
This clearly indicates that the kink size $ \Delta^2 R_{\urm{ch}}^{\urm{Pb}} $ also depends on properties of the central nuclear potential as well.
\begin{table}[tb]
  \centering
  \caption{
    Same as Table~\ref{tab:kink_size_Sn_NR}, but of $ \mathrm{Pb} $
    together with 
    $ \Delta R_{\urm{ch}}^{\urm{Pb}} \left( 208 \right) $
    and
    $ \Delta R_{\urm{ch}}^{\urm{Pb}} \left( 210 \right) $.}
  \label{tab:kink_size_Pb_NR}
  \begin{ruledtabular}
    \begin{tabular}{lddd}
      \multicolumn{1}{l}{EDF} & \multicolumn{1}{c}{$ \Delta R_{\urm{ch}}^{\urm{Pb}} \left( 208 \right) $} & \multicolumn{1}{c}{$ \Delta R_{\urm{ch}}^{\urm{Pb}} \left( 210 \right) $} & \multicolumn{1}{c}{$ \Delta^2 R_{\urm{ch}}^{\urm{Pb}} $} \\
      \hline
      UNEDF0   & +11.158 &  +9.561 & -1.597 \\
      SkM*     &  +9.764 &  +9.491 & -0.273 \\
      UNEDF1   & +11.244 & +11.364 & +0.120 \\
      HFB9     & +10.160 & +10.945 & +0.785 \\
      UNEDF2   & +10.646 & +11.670 & +1.024 \\
      SLy4     &  +9.096 & +11.210 & +2.114 \\
      SLy5     &  +8.920 & +11.507 & +2.587 \\
      SGII     &  +9.396 & +13.625 & +4.229 \\
      SAMi-noT &  +8.683 & +16.457 & +7.774 \\
      SAMi-T   &  +8.595 & +16.829 & +8.234 \\
      SAMi     &  +8.571 & +18.176 & +9.605 \\
      \hline
      Expt.    & +11.0   & +19.6   & +8.6   \\ 
    \end{tabular}
  \end{ruledtabular}
\end{table}
\begin{table}[tb]
  \centering
  \caption{
    Same as Table~\ref{tab:kink_size_Pb_NR}, but by using relativistic EDFs.}
  \label{tab:kink_size_Pb_R}
  \begin{ruledtabular}
    \begin{tabular}{lddd}
      \multicolumn{1}{l}{EDF} & \multicolumn{1}{c}{$ \Delta R_{\urm{ch}}^{\urm{Pb}} \left( 208 \right) $} & \multicolumn{1}{c}{$ \Delta R_{\urm{ch}}^{\urm{Pb}} \left( 210 \right) $} & \multicolumn{1}{c}{$ \Delta^2 R_{\urm{ch}}^{\urm{Pb}} $} \\
      \hline
      PKA1             &  +8.709 & +14.888 &  +6.179 \\
      DD-LZ1           &  +6.935 & +15.462 &  +8.527 \\
      DD-ME2 (TMR)     &  +9.133 & +17.962 &  +8.829 \\
      DD-ME2 (Surface) &  +9.537 & +18.567 &  +9.030 \\
      PKO2             & +10.694 & +19.800 &  +9.106 \\
      PKDD             &  +9.105 & +19.879 & +10.774 \\
      PKO3             & +10.103 & +21.265 & +11.162 \\
      PKO1             &  +9.690 & +21.183 & +11.493 \\
      DD-PC1           &  +7.566 & +20.118 & +12.552 \\
      PKO1*            & +10.553 & +23.893 & +13.340 \\
      \hline
      Expt.            & +11.0   & +19.6   &  +8.6   \\ 
    \end{tabular}
  \end{ruledtabular}
\end{table}
\subsection{Parameters of nuclear equation of state}
\label{sec:calculation_EoS}
\par
\begin{figure}[tb]
  \centering
  \includegraphics[width=1.0\linewidth]{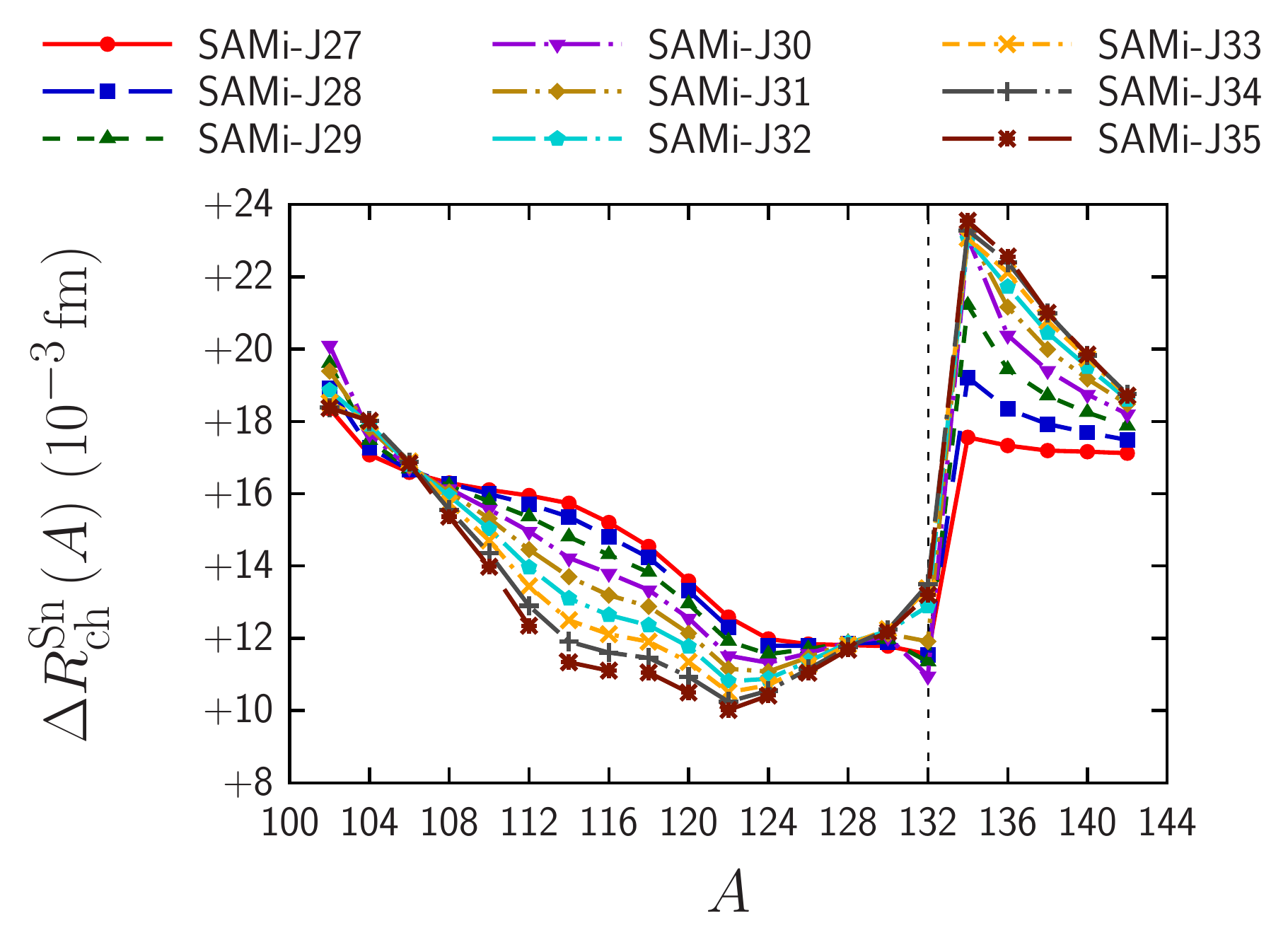}
  \caption{
    Relative change of charge radii $ \Delta R_{\urm{ch}} $ of $ \mathrm{Sn} $ isotopes as a function of $ A $
    calculated by using the SAMi-J family.}
  \label{fig:Rch_Sn_SAMiJ}
\end{figure}
\begin{figure}[tb]
  \centering
  \includegraphics[width=1.0\linewidth]{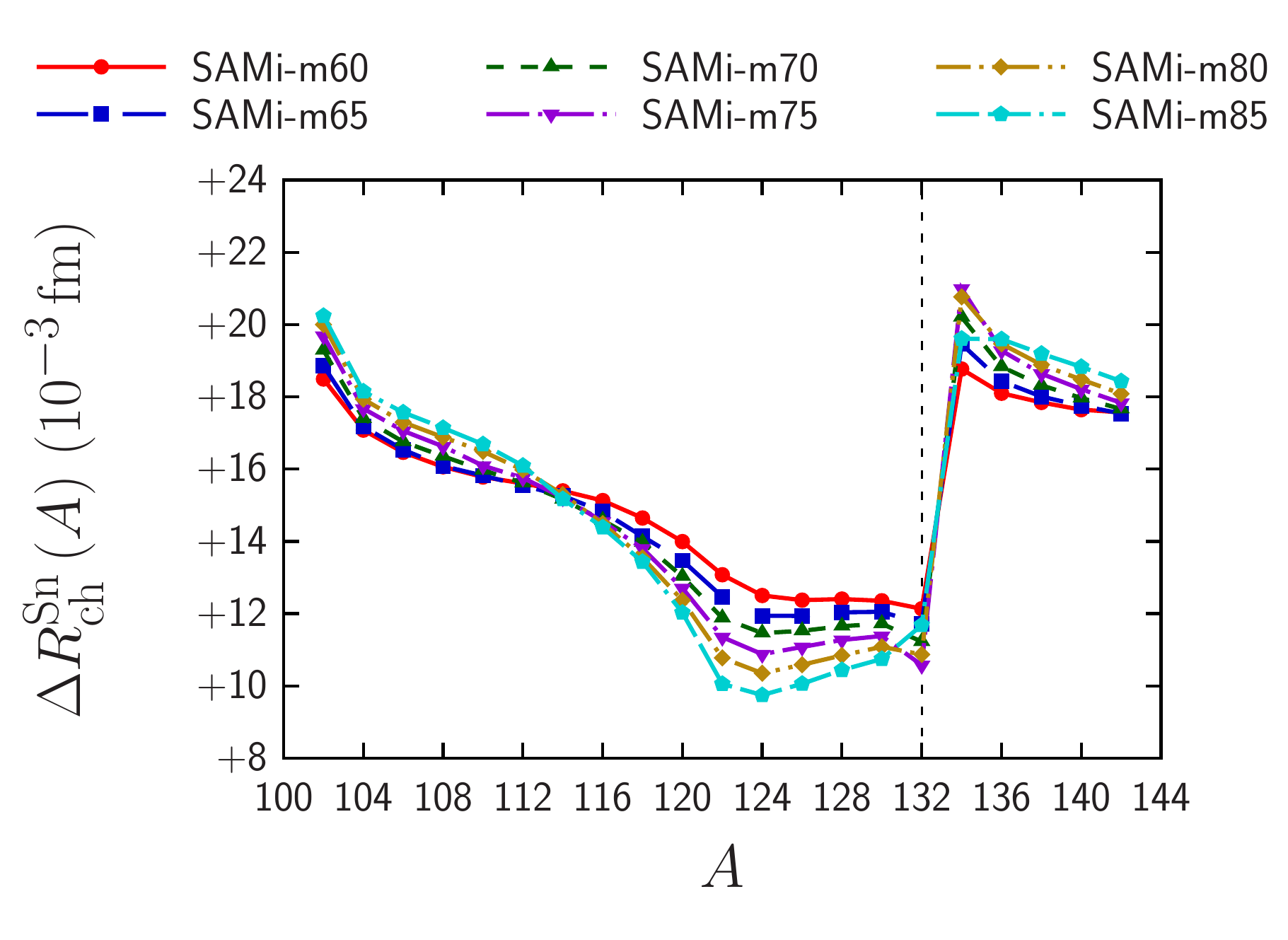}
  \caption{
    Same as Fig.~\ref{fig:Rch_Sn_SAMiJ}, but by using the SAMi-m family.}
  \label{fig:Rch_Sn_SAMim}
\end{figure}
\begin{figure}[tb]
  \centering
  \includegraphics[width=1.0\linewidth]{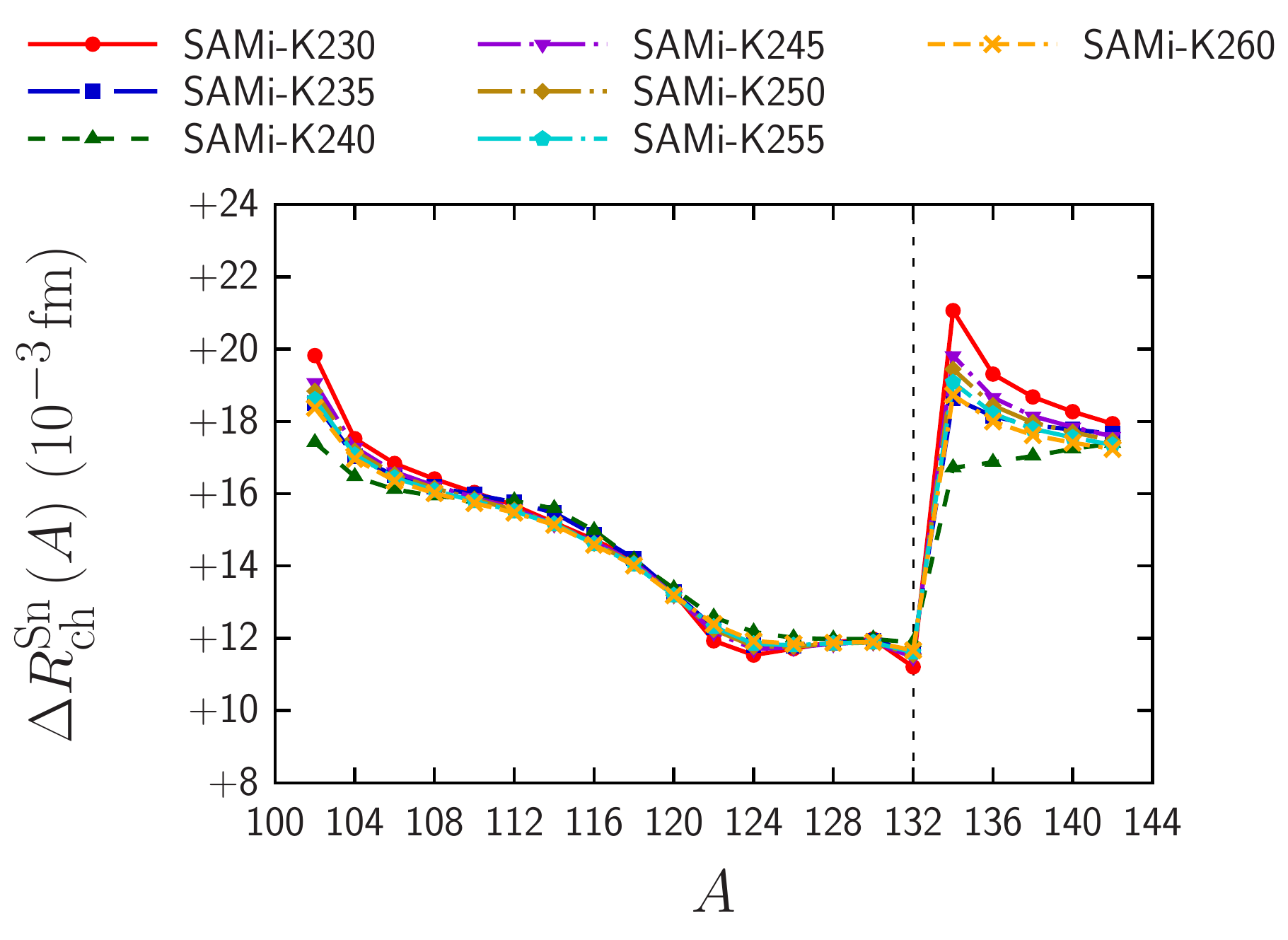}
  \caption{
    Same as Fig.~\ref{fig:Rch_Sn_SAMiJ}, but by using the SAMi-K family.}
  \label{fig:Rch_Sn_SAMiK}
\end{figure}

Next, we will discuss whether parameters of the equation of state (EoS) affect the kink behavior.
The SAMi EDF has three families:
SAMi-J, SAMi-m, and SAMi-K families.
In these families, the fitting protocol of the EDF parametersis basically the same as the original SAMi EDF,
besides one of the EoS parameter is fixed to be a selected value:
each member of the  SAMi-J family assumes a different   symmetry energy coefficient $ J $ at the saturation density,
the SAMi-m member assumes the different effective mass $ m^* $,
and
the SAMi-K member assumes the different incompressibility $ K_{\infty} $.
Figures~\ref{fig:Rch_Sn_SAMiJ}, \ref{fig:Rch_Sn_SAMim}, and \ref{fig:Rch_Sn_SAMiK},
respectively, show the slope $ \Delta R_{\urm{ch}}^{\urm{Sn}} $ as functions of the mass number $ A $
for the SAMi-J, SAMi-m, and SAMi-K families.
If the value of $ \Delta R_{\urm{ch}} $ increases sharply at
$ A = A_{\urm{magic}} $,
the large kink appears at the magic number $ A_{\urm{magic}} $.
The results of the SAMi-K family show an almost complete overlap with each other below the $ N = 82 $ gap,
while they show tiny differences above the gap
although there is no clear tendency of $ K_{\infty} $.
Therefore, it can be concluded that the incompressibility $ K_{\infty} $ scarcely has any visible impact on the value $ R_{\urm{ch}} $.
Different SAMi-m EDF gives similar $ \Delta R_{\urm{ch}} $ above the $ N = 82 $ gap
but slightly different tendency in the region $ 66 \le N \le 82 $.
This may be because the effective mass changes the level distance between single-particle states,
which leads to the different occupancy.
Accordingly, the smaller effective mass gives smaller kink, but its effect is minor.
The size of kink simultaneously depends on the level spacing,
which is not attributable only to the effective mass but also to other parameters~\cite{
  Satula2008Phys.Rev.C78_011302}.
\par
Different SAMi-J EDFs give slightly different behavior of $ \Delta R_{\urm{ch}} $ above the $ N = 82 $ gap.
For simplicity, let us assume that all the SAMi-J EDF gives the same $ R_{\urm{ch}} $ at $ N = 82 $.
Then, naturally, a larger symmetry energy gives a larger neutron radius of $ \nuc{Sn}{132}{} $.
Adding two neutrons to $ \nuc{Sn}{132}{} $, these two neutrons change the proton radius due to the proton-neutron interaction,
which is strongly related to the symmetry energy.
The larger neutron radius leads to the larger proton radius of $ \nuc{Sn}{134}{} $.
In addition, the proton-neutron interaction between the last two neutrons and the protons is important.
As a net effect, the larger $ J $ gives the larger $ R_{\urm{ch}} $ at $ N > 82 $,
as mentioned in Ref.~\cite{
  DayGoodacre2021Phys.Rev.C104_054322}
for mercury isotopes.
\par
\begin{figure}[tb]
  \centering
  \includegraphics[width=1.0\linewidth]{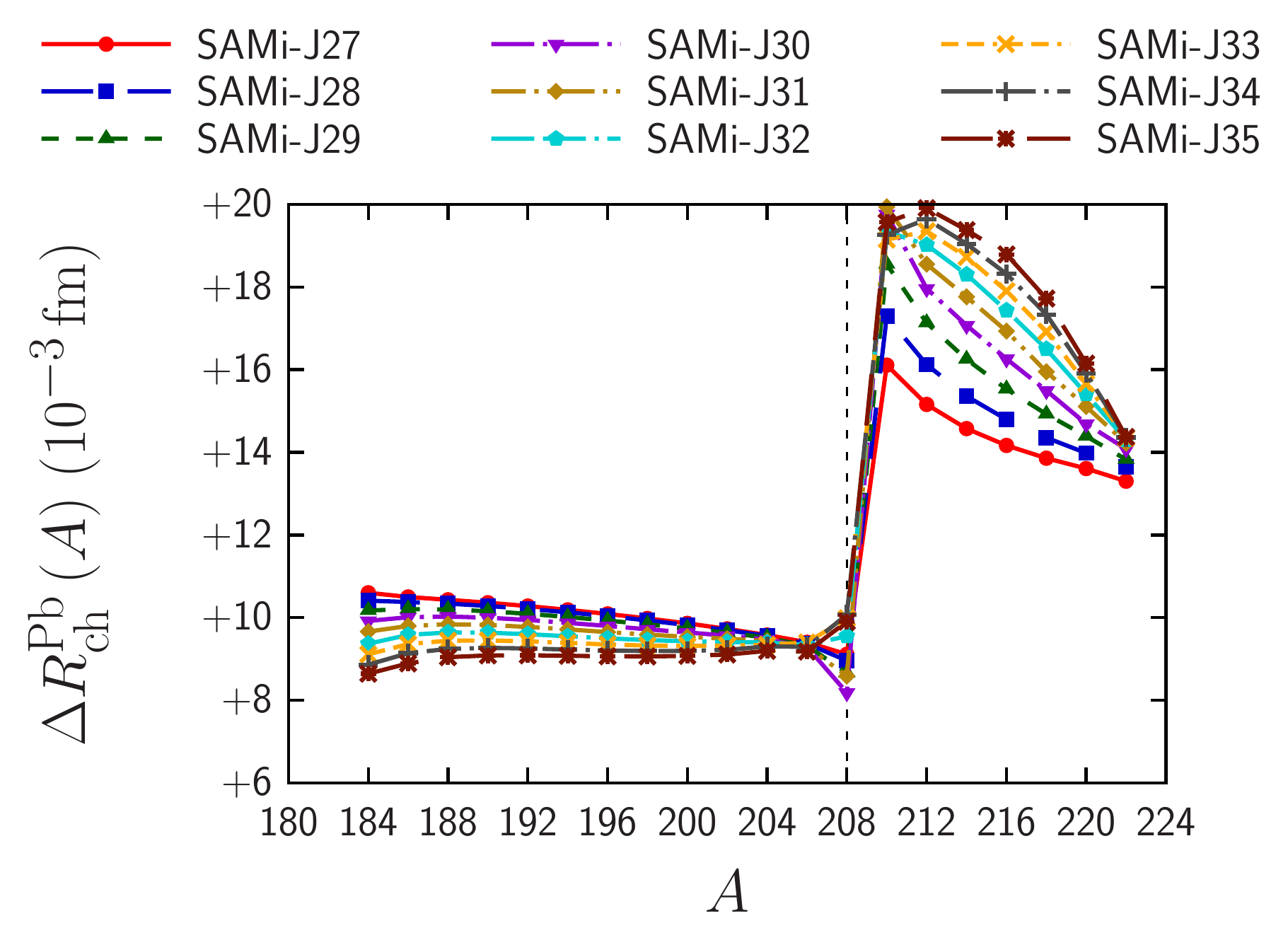}
  \caption{
    Same as Fig.~\ref{fig:Rch_Sn_SAMiJ}, but for $ \mathrm{Pb} $ isotopes.}
  \label{fig:Rch_Pb_SAMiJ}
\end{figure}
\begin{figure}[tb]
  \centering
  \includegraphics[width=1.0\linewidth]{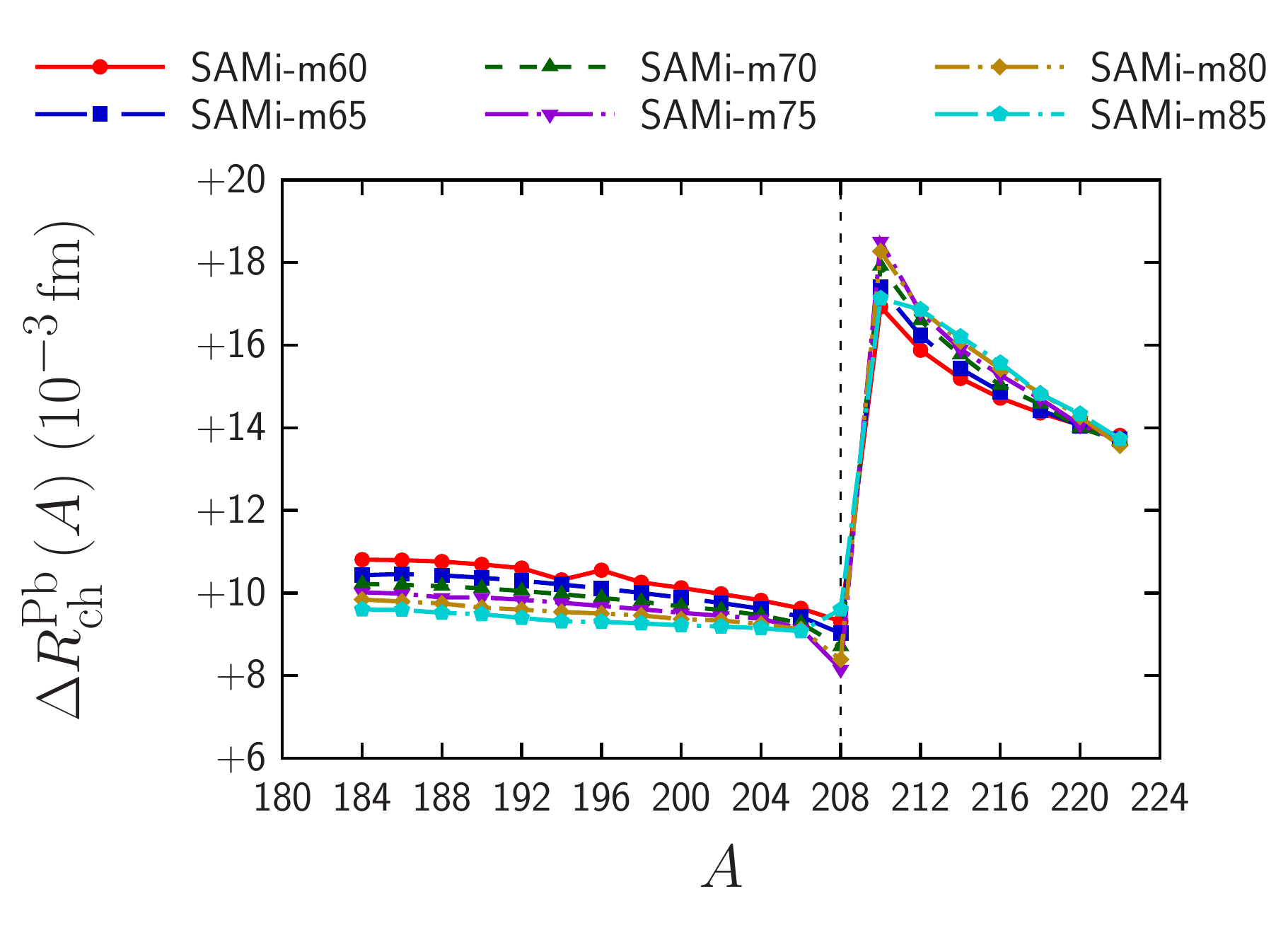}
  \caption{
    Same as Fig.~\ref{fig:Rch_Sn_SAMim}, but for $ \mathrm{Pb} $ isotopes.}
  \label{fig:Rch_Pb_SAMim}
\end{figure}
\begin{figure}[tb]
  \centering
  \includegraphics[width=1.0\linewidth]{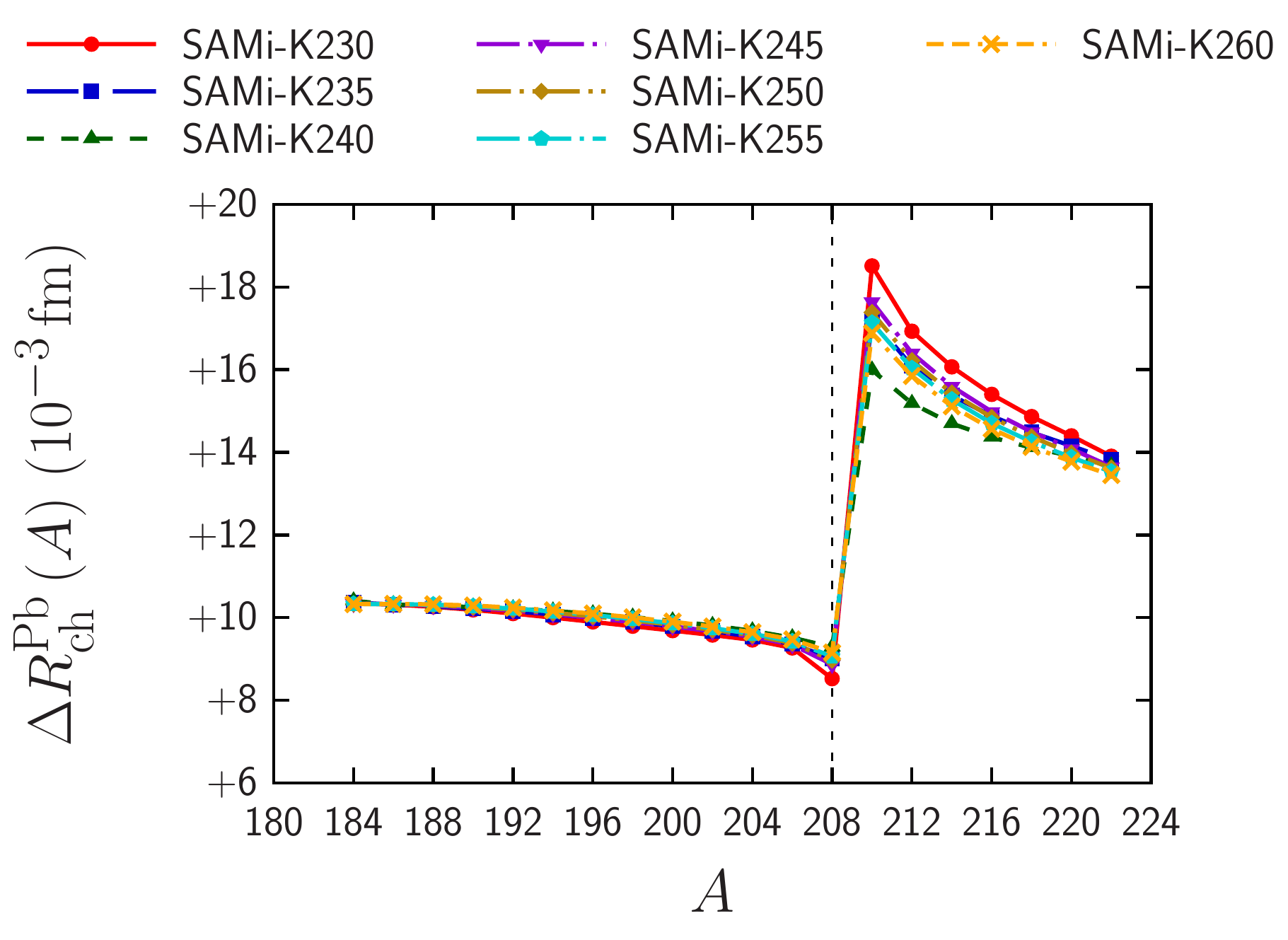}
  \caption{
    Same as Fig.~\ref{fig:Rch_Sn_SAMiK}, but for $ \mathrm{Pb} $ isotopes.}
  \label{fig:Rch_Pb_SAMiK}
\end{figure}

The similar results can be found in $ \mathrm{Pb} $ isotopes,
while all the parameter sets give the similar results below the $ N = 126 $ magic number,
as shown in Figs.~\ref{fig:Rch_Pb_SAMiJ}--\ref{fig:Rch_Pb_SAMiK}.
\subsection{Pairing interaction and its strength}
\label{sec:pairing}
\par
\begin{figure}[tb]
  \centering
  \includegraphics[width=1.0\linewidth]{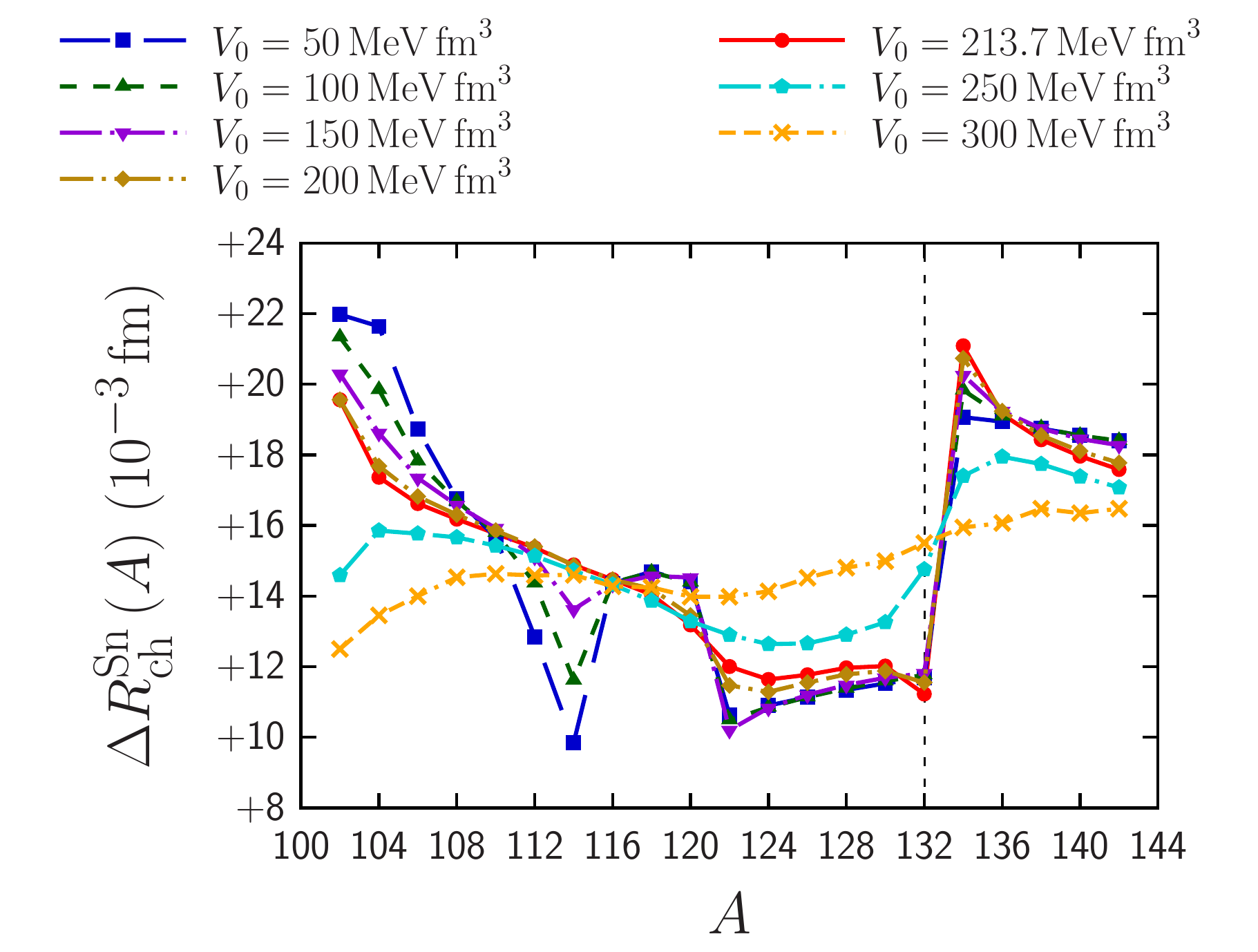}
  \caption{
    Relative change of charge radii $ \Delta R_{\urm{ch}} $ of $ \mathrm{Sn} $ isotopes as a function of $ A $
    for different pairing strengths of the volume-type pairing with the SAMi EDF.}
  \label{fig:Rch_Sn_SAMi_diffStrength}
\end{figure}
\begin{figure}[tb]
  \centering
  \includegraphics[width=1.0\linewidth]{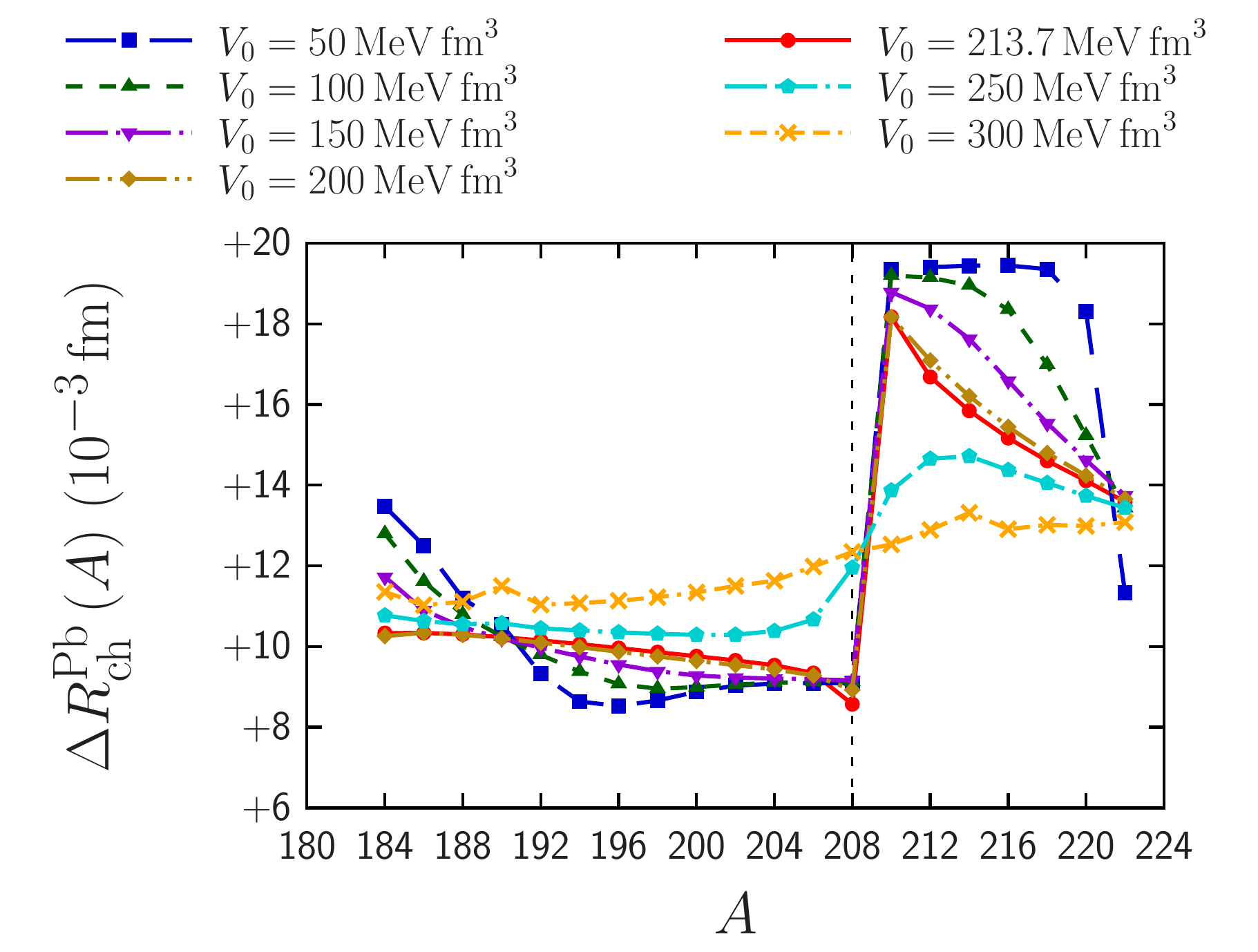}
  \caption{
    Same as Fig.~\ref{fig:Rch_Sn_SAMi_diffStrength}, but for $ \mathrm{Pb} $ isotopes.}
  \label{fig:Rch_Pb_SAMi_diffStrength}
\end{figure}

We will discuss now whether the pairing strength affects the kink behavior.
Figures~\ref{fig:Rch_Sn_SAMi_diffStrength} and \ref{fig:Rch_Pb_SAMi_diffStrength},
respectively,
show the slope $ \Delta R_{\urm{ch}}^{\urm{Sn}} $ and $ \Delta R_{\urm{ch}}^{\urm{Pb}} $
as functions of the mass number $ A $
for different pairing strengths of the volume-type pairing with the SAMi EDF
and
Table~\ref{tab:kink_size_SAMi_strength} shows the kink indicators for both $ \mathrm{Sn} $ and $ \mathrm{Pb} $ isotopes.
Here, $ V_0 = 213.7 \, \mathrm{MeV} \, \mathrm{fm}^3 $ is the adopted strength,
which reproduces the neutron pairing gap of $ \nuc{Sn}{120}{} $ as $ 1.4 \, \mathrm{MeV} $.
Note that the calculation without the pairing ($ V_0 = 0 \, \mathrm{MeV} \, \mathrm{fm}^3 $) does not reach convergence;
thus, the results are not shown here.
Although the behaviors below or above the shell gap are different for different pairing strengths,
results calculated with weaker pairing strengths give similar behavior around the magic numbers.
The kink indicator $ \Delta^2 R_{\urm{ch}} $ is larger if the pairing strength is stronger for $ \mathrm{Sn} $ isotopes,
while the opposite behavior is shown for $ \mathrm{Pb} $ isotopes.
Note that results with $ V_0 = 250 $ and $ 300 \, \mathrm{MeV} \, \mathrm{fm}^3 $ give different behavior.
This is because they give the finite pairing gap even for $ \nuc{Sn}{132}{} $ and $ \nuc{Pb}{208}{} $,
i.e.,
$ \nuc{Sn}{132}{} $ and $ \nuc{Pb}{208}{} $ are no longer magic nuclei.
Therefore, it is concluded that the pairing strength does not affect the kink behavior strongly,
as long as the magicity remains unchanged.
\par
\begin{figure}[tb]
  \centering
  \includegraphics[width=1.0\linewidth]{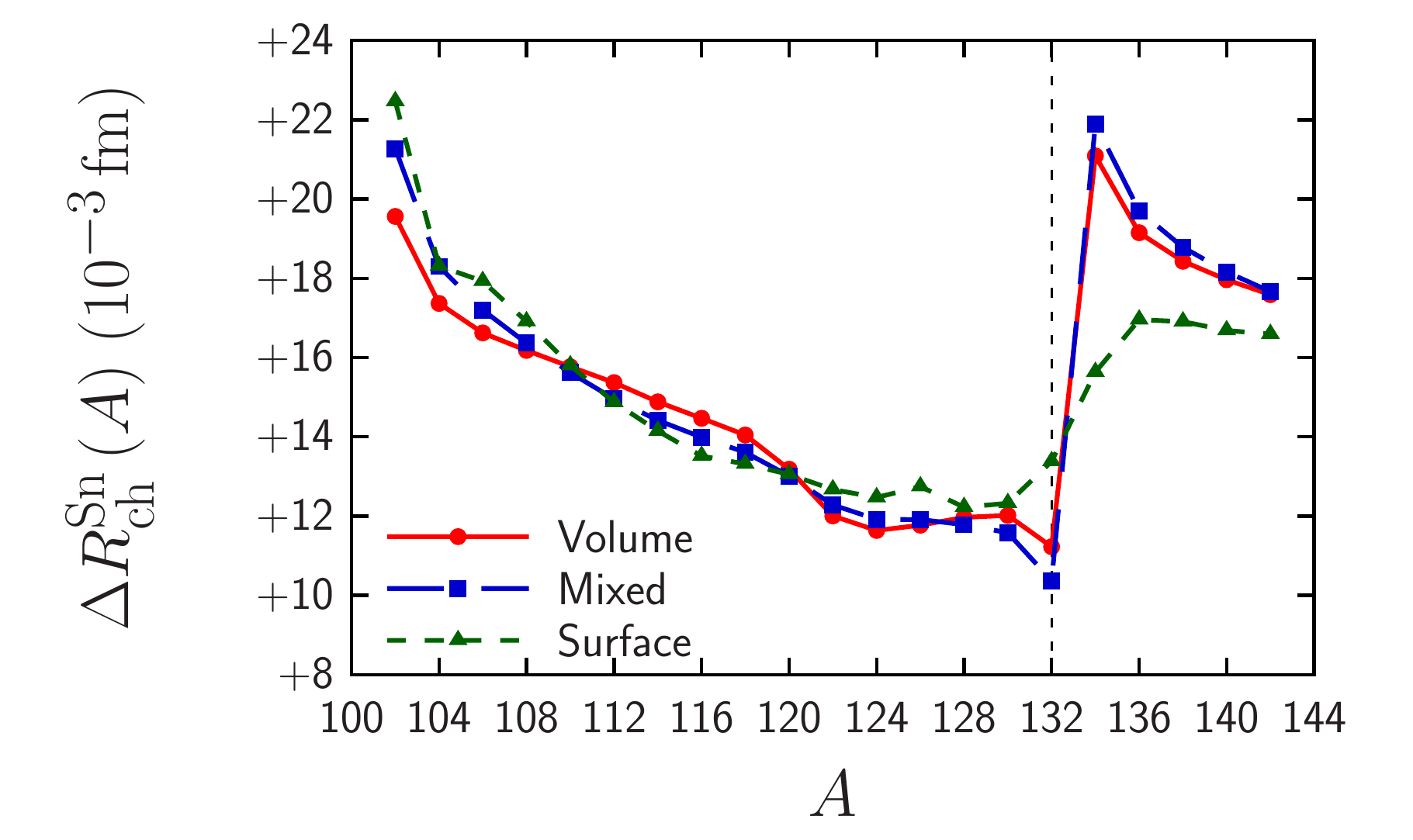}
  \caption{
    Relative change of charge radii $ \Delta R_{\urm{ch}} $ of $ \mathrm{Sn} $ isotopes as a function of $ A $
    for different types of pairing interactions with the SAMi EDF.}
  \label{fig:Rch_Sn_SAMi_diffPairing}
\end{figure}
\begin{figure}[tb]
  \centering
  \includegraphics[width=1.0\linewidth]{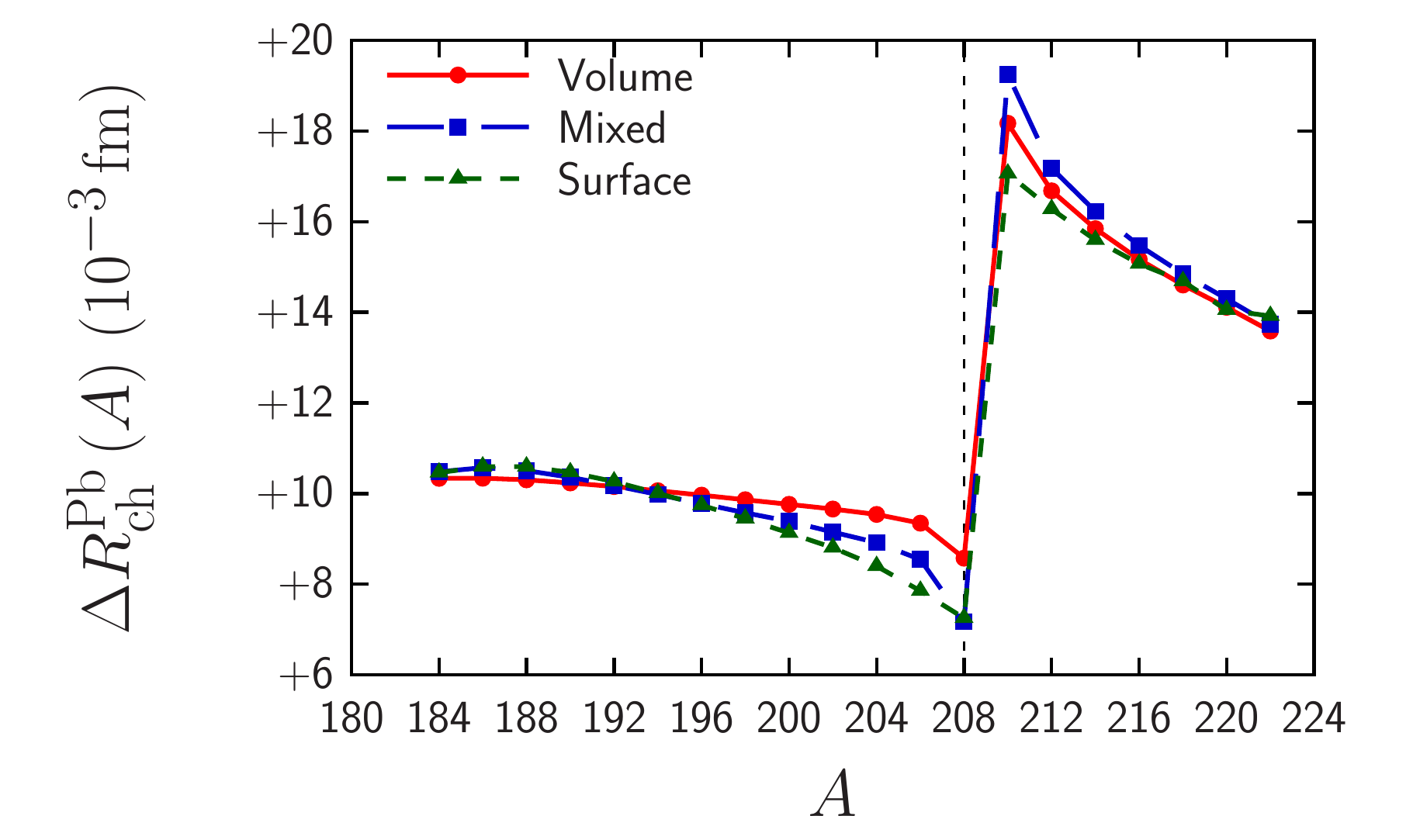}
  \caption{
    Same as Fig.~\ref{fig:Rch_Sn_SAMi_diffPairing}, but for $ \mathrm{Pb} $ isotopes.}
  \label{fig:Rch_Pb_SAMi_diffPairing}
\end{figure}

Figures~\ref{fig:Rch_Sn_SAMi_diffPairing} and \ref{fig:Rch_Pb_SAMi_diffPairing},
respectively,
show the slope $ \Delta R_{\urm{ch}}^{\urm{Sn}} $ and $ \Delta R_{\urm{ch}}^{\urm{Pb}} $
as a function of the mass number $ A $
for different types of pairing interaction with the SAMi EDF.
Table~\ref{tab:kink_size_SAMi_pairing} shows the kink indicates for both $ \mathrm{Sn} $ and $ \mathrm{Pb} $ isotopes.
Here, the pairing interaction used is written as
\begin{equation}
  \label{eq:pairing}
  V_{\urm{pair}} \left( \ve{r} \right)
  =
  -
  V_0
  \left(
    1
    -
    \alpha
    \frac{\rho \left( \ve{r} \right)}{\rho_0}
  \right)
  \delta \left( \ve{r} \right),
\end{equation}
where $ \rho_0 = 0.16 \, \mathrm{fm}^{-3} $ is the saturation density
and
$ \alpha = 0 $, $ 1/2 $, and $ 1 $, respectively, correspond to
the volume-type, mixed-type, and surface-type pairings.
The strengths $ V_0 $ are, respectively,
$ 213.7 $, $ 322.4 $, and $ 558.0 \, \mathrm{MeV} \, \mathrm{fm}^3 $
for volume-type, mixed-type, and surface-type pairings,
which are determined to reproduce the neutron pairing gap of $ \nuc{Sn}{120}{} $ as $ 1.4 \, \mathrm{MeV} $.
It is seen in the figures that, as long as these volume-type, mixed-type, and surface-type pairings are used,
no significant difference can be found,
while the mixed pairing shows the strongest kink indicator.
Note that the pairing gaps of $ \nuc{Sn}{132}{} $ obtained by the mixed-type and surface-type pairings are,
respectively,
$ 0.50 $ and $ 1.47 \, \mathrm{MeV} $
and that of $ \nuc{Pb}{208}{} $ obtained by the surface-type pairing is
$ 0.55 \, \mathrm{MeV} $.
These finite gaps mean that the magicities of these nuclei are effectively broken by the pairing correlations.
\begin{table}[tb]
  \centering
  \caption{
    Kink indicator $ \Delta^2 R_{\urm{ch}}^{\urm{Sn}} $ and $ \Delta^2 R_{\urm{ch}}^{\urm{Pb}} $ for
    $ \mathrm{Sn} $ and $ \mathrm{Pb} $ calculated by the nonrelativistic SAMi EDF
    with various pairing strength of the volume-type pairing 
    are also listed.
    All values are given in units of $ 10^{-3} \, \mathrm{fm} $.}
  \label{tab:kink_size_SAMi_strength}
  \begin{ruledtabular}
    \begin{tabular}{ddd}
      \multicolumn{1}{c}{Strength $ V_0 $} & \multicolumn{1}{c}{$ \Delta^2 R_{\urm{ch}}^{\urm{Sn}} $} & \multicolumn{1}{c}{$ \Delta^2 R_{\urm{ch}}^{\urm{Pb}} $} \\
      \hline
      50    & +7.381 & +10.257 \\
      100   & +8.066 & +10.071 \\
      150   & +8.445 &  +9.626 \\
      200   & +9.185 &  +9.245 \\
      213.7 & +9.863 &  +9.605 \\
      250   & +2.646 &  +1.914 \\
      300   & +0.441 &  +0.183 \\
    \end{tabular}
  \end{ruledtabular}
\end{table}
\begin{table}[tb]
  \centering
  \caption{
    Same as Table \ref{tab:kink_size_SAMi_strength}, but for the different pairing interaction.}
  \label{tab:kink_size_SAMi_pairing}
  \begin{ruledtabular}
    \begin{tabular}{ldd}
      \multicolumn{1}{c}{Type of pairing interaction} & \multicolumn{1}{c}{$ \Delta^2 R_{\urm{ch}}^{\urm{Sn}} $} & \multicolumn{1}{c}{$ \Delta^2 R_{\urm{ch}}^{\urm{Pb}} $} \\
      \hline
      Volume  &  +9.863 &  +9.605 \\
      Mixed   & +11.526 & +12.085 \\ 
      Surface &  +2.251 &  +9.812 \\
    \end{tabular}
  \end{ruledtabular}
\end{table}
\subsection{$ \mathrm{Ca} $ isotopes}
\label{sec:calculation_Ca}
\par
\begin{figure*}[tb]
  \centering
  \includegraphics[width=1.0\linewidth]{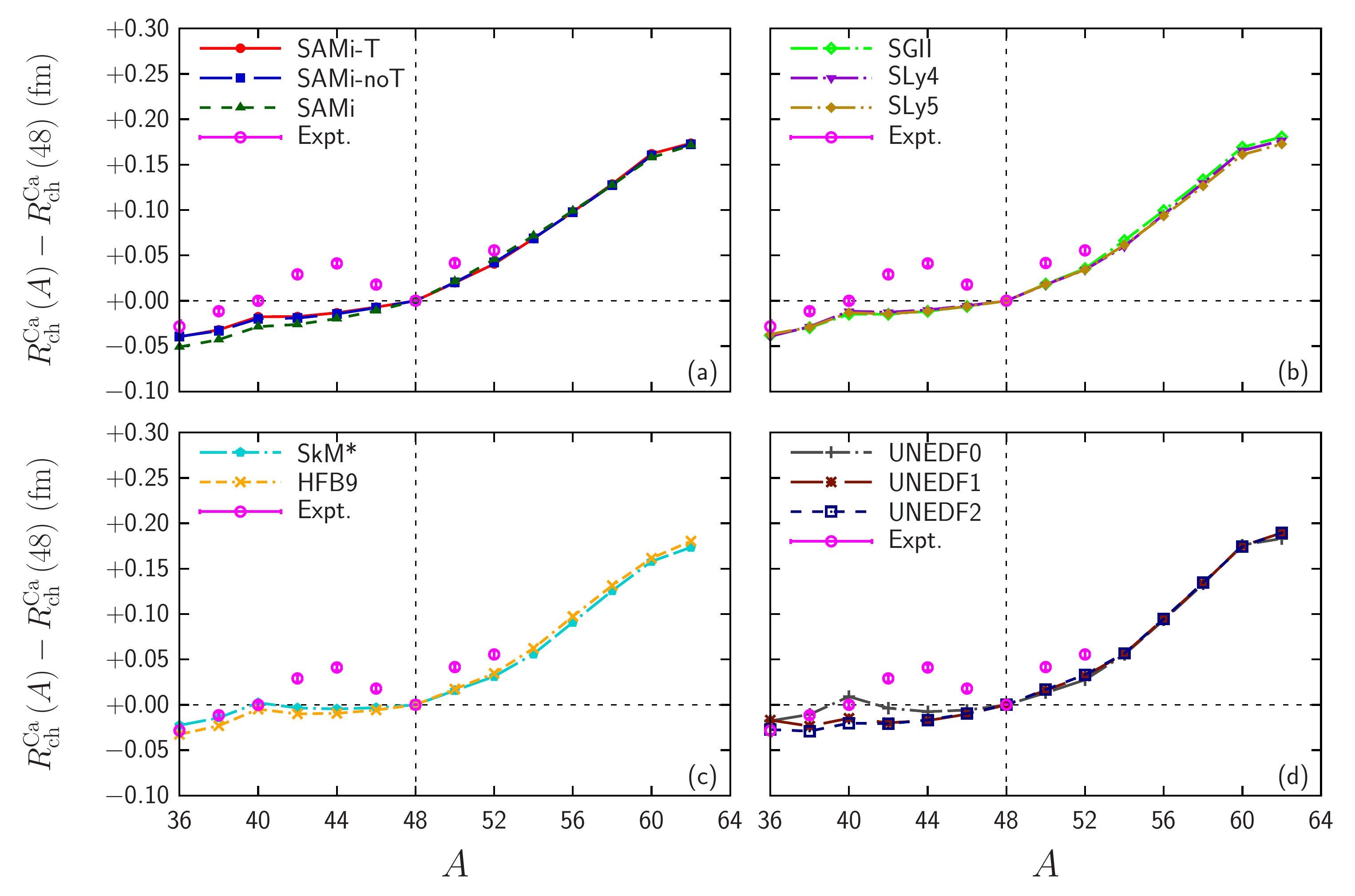}
  \caption{
    Same as Fig.~\ref{fig:Rch_Sn_NR}, but for $ \mathrm{Ca} $ isotopes.
    Experimental data except $ \nuc{Ca}{40}{} $ are taken from Refs.~\cite{
      GarciaRuiz2016Nat.Phys.12_594,
      Miller2019Nat.Phys.15_432},
    instead of Ref.~\cite{
      Angeli2013At.DataNucl.DataTables99_69}.}
  \label{fig:Rch_Ca_NR}
\end{figure*}
\begin{figure*}[tb]
  \centering
  \includegraphics[width=1.0\linewidth]{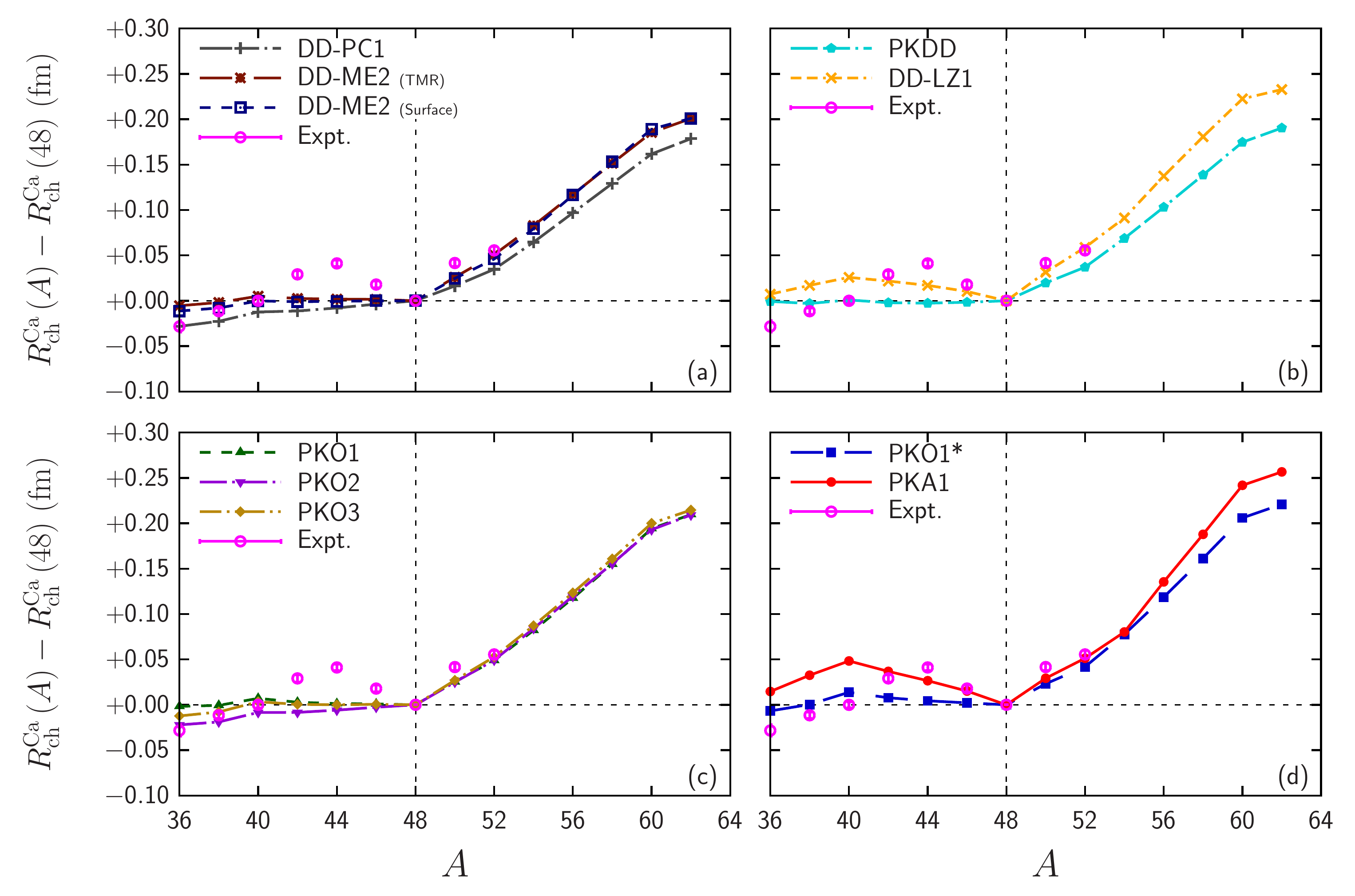}
  \caption{
    Same as Fig.~\ref{fig:Rch_Sn_R}, but for $ \mathrm{Ca} $ isotopes.
    Experimental data except $ \nuc{Ca}{40}{} $ are taken from Refs.~\cite{
      GarciaRuiz2016Nat.Phys.12_594,
      Miller2019Nat.Phys.15_432},
    instead of Ref.~\cite{
      Angeli2013At.DataNucl.DataTables99_69}.}
  \label{fig:Rch_Ca_R}
\end{figure*}

The mass number $ A $ dependence of
$ R_{\urm{ch}}^{\urm{Ca}} \left( A \right) - R_{\urm{ch}}^{\urm{Ca}} \left( 48 \right) $
for $ \mathrm{Ca} $ isotopes 
calculated by nonrelativistic (SHF) and relativistic (RMF and RHF) EDFs are shown in
Figs.~\ref{fig:Rch_Ca_NR} and \ref{fig:Rch_Ca_R}, respectively.
For comparison, experimental data~\cite{
  GarciaRuiz2016Nat.Phys.12_594,
  Miller2019Nat.Phys.15_432}
are also plotted.
The kink sizes for $ \mathrm{Ca} $ isotopes are summarized in Tables~\ref{tab:kink_size_Ca_NR} and \ref{tab:kink_size_Ca_R}.
\par
We can see that the  kink evolution of $ \mathrm{Ca} $ isotopes, especially,
around the $ N = 28 $ kink,
is quite different from those of $ \mathrm{Sn} $ and $ \mathrm{Pb} $ isotopes.
Compared with the calculated results, 
the systematic behavior of experimental $ R_{\urm{ch}} $ values, even $ N < 28 $, is  not well described.
Indeed, it is suggested in several works~\cite{
  Perera2021Phys.Rev.C104_064313,
  Sagawa2022Phys.Lett.B829_137072,
  Yang:2022tjf}
that beyond-mean-field effects are important in $ \mathrm{Ca} $ isotopes.
This is an open and intriguing problem for the microscopic model beyond mean field.
\begin{table}[tb]
  \centering
  \caption{
    Same as Table~\ref{tab:kink_size_Sn_NR}, but of $ \mathrm{Ca} $
    together with 
    $ \Delta R_{\urm{ch}}^{\urm{Ca}} \left( 48 \right) $
    and
    $ \Delta R_{\urm{ch}}^{\urm{Ca}} \left( 50 \right) $.}
  \label{tab:kink_size_Ca_NR}
  \begin{ruledtabular}
    \begin{tabular}{lddd}
      \multicolumn{1}{l}{EDF} & \multicolumn{1}{c}{$ \Delta R_{\urm{ch}}^{\urm{Ca}} \left( 48 \right) $} & \multicolumn{1}{c}{$ \Delta R_{\urm{ch}}^{\urm{Ca}} \left( 50 \right) $} & \multicolumn{1}{c}{$ \Delta^2 R_{\urm{ch}}^{\urm{Ca}} $} \\
      \hline
      UNEDF1   & +10.272 & +15.999 &  +5.727 \\
      UNEDF2   &  +9.611 & +16.829 &  +7.218 \\
      UNEDF0   &  +5.814 & +13.268 &  +7.454 \\
      SAMi     & +11.096 & +21.330 & +10.234 \\
      HFB9     &  +5.729 & +17.407 & +11.678 \\
      SLy5     &  +6.153 & +18.080 & +11.927 \\
      SGII     &  +6.334 & +18.378 & +12.044 \\
      SLy4     &  +5.421 & +17.856 & +12.435 \\
      SAMi-noT &  +7.702 & +20.374 & +12.672 \\
      SKMs     &  +3.182 & +15.861 & +12.679 \\
      SAMi-T   &  +6.976 & +19.841 & +12.865 \\
      \hline
      Expt.    & -17.8   & +41.5   & +59.3   \\
    \end{tabular}
  \end{ruledtabular}
\end{table}
\begin{table}[tb]
  \centering
  \caption{
    Same as Table~\ref{tab:kink_size_Ca_NR}, but by using relativistic EDFs.}
  \label{tab:kink_size_Ca_R}
  \begin{ruledtabular}
    \begin{tabular}{lddd}
      \multicolumn{1}{l}{EDF} & \multicolumn{1}{c}{$ \Delta R_{\urm{ch}}^{\urm{Ca}} \left( 48 \right) $} & \multicolumn{1}{c}{$ \Delta R_{\urm{ch}}^{\urm{Ca}} \left( 50 \right) $} & \multicolumn{1}{c}{$ \Delta^2 R_{\urm{ch}}^{\urm{Ca}} $} \\
     \hline
      DD-PC1           &  +3.671 & +16.293 & +12.622 \\
      PKDD             &  +1.674 & +19.694 & +18.020 \\
      PKO2             &  +2.670 & +24.965 & +22.295 \\
      DD-ME2 (Surface) &  -0.019 & +24.576 & +24.595 \\
      PKO1*            &  -2.174 & +23.201 & +25.375 \\
      PKO1             &  -0.956 & +25.284 & +26.240 \\
      DD-ME2 (TMR)     &  -1.631 & +25.863 & +27.494 \\
      PKO3             &  -0.550 & +27.110 & +27.660 \\
      DD-LZ1           & -10.279 & +31.533 & +41.812 \\
      PKA1             & -15.149 & +29.323 & +44.472 \\
      \hline
      Expt.            & -17.8   & +41.5   & +59.3   \\
    \end{tabular}
  \end{ruledtabular}
\end{table}
%
\section{Conclusion}
\label{sec:conclusion}
\par
In this paper, the sudden change of the mass-number dependence of the charge radius at the neutron shell gap, the so-called kink behavior, is discussed for $ \mathrm{Sn} $ and $ \mathrm{Pb} $ isotopes 
by using the nonrelativistic Skyrme, relativistic mean field (RMF), and the relativistic Hartree-Fock (RHF) calculations.
In general, the RHF calculations give the larger kink with positive kink indicators and the Skyrme calculations give the smaller positive values.  
One exception among the Skyrme EDFs is the SAMi series which gives exceptionally larger kink.  
Abnormal behavior of the kink is induced by some Skyrme EDFs,
for instance, the UNEDF series, which give the opposite behavior, 
a negative value for the kink indicator. 
In order to reproduce the kink behavior for the $ \mathrm{Sn} $ isotopes,
the occupation probability of the $ 1h_{9/2} $ orbital should be large enough.
On the other hand,  
if the occupation probability of the $ 3p_{3/2} $ orbital is large,
the anti kink may appear.
In the case of $ \mathrm{Pb} $ isotopes, a larger occupation probability of the $ 1i_{11/2} $ orbital gives a larger kink indicator at the $ N = 126 $ shell gap.
To make such occupancy, the spin-orbit mean-field must not be too strong.
Note that the averaged value of the single-particle energy for the spin-orbit doublet also affects the occupation probability.
\par
Analyzing the RHF calculation, the tensor interaction, which contributes to the spin-orbit mean-field potential, is concluded as an essential ingredient to produce  such proper occupations of the single-particle states and to reproduce well the kink behavior.
Compared with the tensor effect of RHF,
the effect of the Skyrme tensor interaction in SAMi-T EDF is found to be tiny.
The different strengths between the isoscalar and isovector spin-orbit interactions in the Skyrme EDF,
$ W_0 \ne W'_0 $, 
may not be an absolutely necessary condition to reproduce the kink behavior,
since the kink can be produced even by some Skyrme EDFs with $ W_0 = W'_0 $ spin-orbit interactions.
Hence, the proper determination of the strength of the isoscalar and isovector spin-orbit interactions is demanded.
Considering the reference data for this interest,
experimentally these spin-orbit interactions manifest themselves in the spin-orbit splittings of odd-mass nuclei.
However, these levels are affected by the particle-vibration coupling~\cite{
  Afanasjev2015Phys.Rev.C92_044317},
which is not considered in this paper.
It should be noted that the isovector-scalar $ \delta $ meson can also introduce the spin-orbit interaction with the isospin dependence~\cite{
  Kubis1997Phys.Lett.B399_191,
  Liu2002Phys.Rev.C65_045201,
  Roca-Maza2011Phys.Rev.C84_054309,
  Perera2021Phys.Rev.C104_064313}.
As shown in Ref.~\cite{
  Perera2021Phys.Rev.C104_064313},
the $ \delta $ meson in the RMF calculation is an alternative way to reproduce the kink.
\par
We also investigated whether parameters of nuclear equation of state affect such kink behavior.
It is found that the symmetry energy affects appreciably to change the magnitude of the kink indicator in 
a similar way to induce the neutron-skin thickness.
The effective masses affects the kink behavior slightly;
smaller effective mass gives smaller kink.
In contrast, the nuclear incompressibility scarcely affect the kink behavior.
\par
Mass number dependence of the charge radii of $ \mathrm{Ca} $ isotopes is rather different from those $ \mathrm{Sn} $ or $ \mathrm{Pb} $ isotopes,
which implies the importance of beyond-mean-field effects. 
\par
The form of the pairing interaction is still under debate.
For instance, the derivative dependence such as the Fayans-type~\cite{
  Fayans1998JETPLett.68_169,
  Reinhard2017Phys.Rev.C95_064328},
isoscalar pairing~\cite{
  Bertulani2009Phys.Rev.C80_027303,
  Bertulani2012Phys.Rev.C85_014321,
  Yamagami2012Phys.Rev.C86_034333,
  Sagawa2013Phys.Rev.C87_034310,
  Yoshida2013Prog.Theor.Exp.Phys.2013_113D02,
  Sagawa2016Phys.Scr.91_083011,
  Teeti2021Phys.Rev.C103_034310},
and 
spin-triplet pairing~\cite{
  Oishi2020J.Phys.G47_115106,
  Oishi2021Eur.Phys.J.A57_180,
  Yoshida2021Phys.Rev.C104_014319}
have been discussed recently.
The pairing strength for the deformed nuclei might be different from that for the spherical nuclei~\cite{
  Bertsch2009Phys.Rev.C79_034306,
  Robledo2012Phys.Rev.C86_064313,
  Teeti2021Phys.Rev.C103_034310}.
Effects of them to the kink behavior are left for future perspectives.
The beyond-mean-field effect can also affect the kink behavior~\cite{
  DayGoodacre2021Phys.Rev.Lett.126_032502,
  DayGoodacre2021Phys.Rev.C104_054322,
  Perera2021Phys.Rev.C104_064313},
which is out of the scope of this paper and remains for future perspectives,
since we focused only on the mean-field level in this paper.
%
\begin{acknowledgments}
  The authors acknowledge for the fruitful discussion with
  Andrei Andreyev,
  Nobuo Hinohara,
  Tsunenori Inakura,
  Hitoshi Nakada,
  and
  Qiang Zhao.
  The authors also thank Xavier Roca-Maza for allowing us to use the SAMi-J, SAMi-K, and SAMi-m families.
  T.~N.~acknowledges
  the RIKEN Special Postdoctoral Researcher Program,
  the Science and Technology Hub Collaborative Research Program from RIKEN Cluster for Science, Technology and Innovation Hub (RCSTI),
  the JSPS Grant-in-Aid for Research Activity Start-up under Grant No.~22K20372,
  the JSPS Grant-in-Aid for Transformative Research Areas (A) under Grant No.~23H04526,
  the JSPS Grant-in-Aid for Scientific Research (B) under Grant No.~23H01845,
  and
  the JSPS Grant-in-Aid for Scientific Research (C) under Grant No.~23K03426.
  T.~O.~acknowledges the support of the Yukawa Research Fellow Programme by Yukawa Memorial Foundation.
  H.~S.~acknowledges 
  the JSPS Grant-in-Aid for Scientific Research (C) under Grant No.~19K03858.
  Z. W. acknowledges the support of the Natural Science Foundation of China under Grants Nos.~11905088 and 12275111.
  The numerical calculations were partly performed on cluster computers at the RIKEN iTHEMS program.
\end{acknowledgments}
%
\appendix
\section{Spin-orbit contribution of charge radii}
\label{sec:so_contribution}
\par
The fourth and fifth terms of Eq.~\eqref{eq:Rch} are called the spin-orbit contributions,
originating from the nucleon magnetic moments.
These terms have not been considered in most works~\cite{
  Chabanat1998Nucl.Phys.A635_231},
while it was also discussed that these terms are not negligible if one discusses the tiny contribution or compare with the experimental data precisely~\cite{
  Horowitz2012Phys.Rev.C86_045503,
  Reinhard2021Phys.Rev.C103_054310,
  Naito2021Phys.Rev.C104_024316}.
In this appendix, we will discuss the effect of the spin-orbit contribution to nuclear charge radii and their kink behavior for $ \mathrm{Sn} $ and $ \mathrm{Pb} $ isotopes.
\par
The spin-orbit contribution of charge radii is~\cite{
  Horowitz2012Phys.Rev.C86_045503,
  Naito2021Phys.Rev.C104_024316}
\begin{equation}
  \label{eq:so}
  \avr{r^2}_{\urm{SO} \tau}
  \simeq
  \frac{\kappa_{\tau}}{M_{\tau}^2 N_{\tau}}
  \sum_{a \in \text{occ}}
  \ca{N}_{a \tau}
  \avr{\ve{l} \cdot \ve{\sigma}},
\end{equation}
where 
$ N_{\tau} = Z $ for $ \tau = p $,
$ N_{\tau} = N $ for $ \tau = n $,
$ \kappa_{\tau} $ and $ M_{\tau} $ are, respectively, the magnetic moment and the mass of nucleons $ \tau $,
and
$ \ca{N}_{a \tau} $ is the occupation number of the orbital $ a $.
It is obvious that the $ s $ orbitals do not contribute to $ \avr{r^2}_{\urm{SO} \tau} $;
if the spin-orbit doublets are completely occupied,
contribution of these two orbitals cancel each other out.
Therefore, only the spin-orbit doublets partially occupied contributes to $ \avr{r^2}_{\urm{SO} \tau} $.
\par
The proton spin-orbit contribution to the charge radius, $ \avr{r^2}_{\urm{SO} p} $, is rather simple.
Only the $ 1g_{9/2} $ orbital
($ \avr{\ve{l} \cdot \ve{\sigma}} = +4 $, $ \ca{N}_a = 10 $)
and
only the $ 1h_{11/2} $ orbital
($ \avr{\ve{l} \cdot \ve{\sigma}} = +5 $, $ \ca{N}_a = 12 $),
respectively, contribute to $ \avr{r^2}_{\urm{SO} p} $
in $ \mathrm{Sn} $ and $ \mathrm{Pb} $ isotopes;
hence, the contribution can be, respectively, estimated as
$ \avr{r^2}_{\urm{SO} p} = 0.0634 \, \mathrm{fm}^2 $
and
$ 0.0580 \, \mathrm{fm}^2 $.
These contributions are common for each isotopic chain,
and thus the proton spin-orbit contribution just shifts the graphs of 
Figs.~\ref{fig:Rch_Sn_NR}(a), \ref{fig:Rch_Sn_R}(a), \ref{fig:Rch_Pb_NR}(a), and \ref{fig:Rch_Pb_R}(a) upwards,
and the kink behaviors are not changed.
\par
In contrast, the spin-orbit contributions of the neutrons $ \avr{r^2}_{\urm{SO} n} $ affect the kink behavior.
Note that the magnetic moment of neutrons $ \kappa_n $ is negative.
First, let us focus on $ \mathrm{Sn} $ isotopes.
Below the $ N = 82 $ shell gap, the highest orbital is the $ 1h_{11/2} $ orbital
($ \avr{\ve{l} \cdot \ve{\sigma}} = +5 $),
which gives the negative $ \avr{r^2}_{\urm{SO} n} $ of $ N < 82 $ isotopes,
and thus, the slope below $ N = 82 $ becomes mild
if the spin-orbit contribution is considered. 
In contrast, above the $ N = 82 $ shell gap, either
the $ 1h_{9/2} $ orbital
($ \avr{\ve{l} \cdot \ve{\sigma}} = -6 $),
the $ 2f_{7/2} $ orbital
($ \avr{\ve{l} \cdot \ve{\sigma}} = +3 $),
or
$ 3p_{3/2} $ orbital
($ \avr{\ve{l} \cdot \ve{\sigma}} = +1 $)
are mainly occupied.
If the $ 1h_{9/2} $ orbital is mainly occupied in $ N > 82 $ isotopes,
the spin-orbit contribution leads to the slope above $ N = 82 $ steep, and thus the kink becomes strong.
On the contrary, if the $ 2f_{7/2} $ or $ 3p_{3/2} $ orbital is mainly occupied in $ N > 82 $ isotopes,
the spin-orbit contribution leads to the slope above $ N = 82 $ mild, but the kink becomes slightly strong.
\par
In the case of $ \mathrm{Pb} $ isotopes,
below the $ N = 126 $ shell gap, the highest orbital is the $ 3p_{1/2} $ orbital
($ \avr{\ve{l} \cdot \ve{\sigma}} = -2 $),
which gives the positive $ \avr{r^2}_{\urm{SO} n} $ of $ N < 126 $ isotopes,
and thus, the slope below $ N = 126 $ becomes steep
if the spin-orbit contribution is considered. 
In contrast, above the $ N = 126 $ shell gap, either
the $ 1i_{11/2} $ orbital
($ \avr{\ve{l} \cdot \ve{\sigma}} = -7 $),
the $ 2g_{9/2} $ orbital
($ \avr{\ve{l} \cdot \ve{\sigma}} = +4 $),
and
$ 1j_{15/2} $ orbital
($ \avr{\ve{l} \cdot \ve{\sigma}} = +7 $)
are mainly occupied.
If the $ 1i_{11/2} $ orbital is mainly occupied in $ N > 126 $ isotopes,
the spin-orbit contribution leads to the slope above $ N = 126 $ strong, and thus the kink becomes strong.
On the contrary, if the $ 2g_{9/2} $ or $ 1j_{15/2} $ orbital is mainly occupied in $ N > 126 $ isotopes,
which leads to the slope above $ N = 126 $ mild, the kink becomes weak.
%
%
%
%

\begin{thebibliography}{78}%
\makeatletter
\providecommand \@ifxundefined [1]{%
 \@ifx{#1\undefined}
}%
\providecommand \@ifnum [1]{%
 \ifnum #1\expandafter \@firstoftwo
 \else \expandafter \@secondoftwo
 \fi
}%
\providecommand \@ifx [1]{%
 \ifx #1\expandafter \@firstoftwo
 \else \expandafter \@secondoftwo
 \fi
}%
\providecommand \natexlab [1]{#1}%
\providecommand \enquote  [1]{``#1''}%
\providecommand \bibnamefont  [1]{#1}%
\providecommand \bibfnamefont [1]{#1}%
\providecommand \citenamefont [1]{#1}%
\providecommand \href@noop [0]{\@secondoftwo}%
\providecommand \href [0]{\begingroup \@sanitize@url \@href}%
\providecommand \@href[1]{\@@startlink{#1}\@@href}%
\providecommand \@@href[1]{\endgroup#1\@@endlink}%
\providecommand \@sanitize@url [0]{\catcode `\\12\catcode `\$12\catcode
  `\&12\catcode `\#12\catcode `\^12\catcode `\_12\catcode `\%12\relax}%
\providecommand \@@startlink[1]{}%
\providecommand \@@endlink[0]{}%
\providecommand \url  [0]{\begingroup\@sanitize@url \@url }%
\providecommand \@url [1]{\endgroup\@href {#1}{\urlprefix }}%
\providecommand \urlprefix  [0]{URL }%
\providecommand \Eprint [0]{\href }%
\providecommand \doibase [0]{https://doi.org/}%
\providecommand \selectlanguage [0]{\@gobble}%
\providecommand \bibinfo  [0]{\@secondoftwo}%
\providecommand \bibfield  [0]{\@secondoftwo}%
\providecommand \translation [1]{[#1]}%
\providecommand \BibitemOpen [0]{}%
\providecommand \bibitemStop [0]{}%
\providecommand \bibitemNoStop [0]{.\EOS\space}%
\providecommand \EOS [0]{\spacefactor3000\relax}%
\providecommand \BibitemShut  [1]{\csname bibitem#1\endcsname}%
\let\auto@bib@innerbib\@empty
\bibitem [{\citenamefont {De~Vries}\ \emph {et~al.}(1987)\citenamefont
  {De~Vries}, \citenamefont {De~Jager},\ and\ \citenamefont
  {De~Vries}}]{DeVries1987At.DataNucl.DataTables36_495}%
  \BibitemOpen
  \bibfield  {author} {\bibinfo {author} {\bibfnamefont {H.}~\bibnamefont
  {De~Vries}}, \bibinfo {author} {\bibfnamefont {C.~W.}\ \bibnamefont
  {De~Jager}},\ and\ \bibinfo {author} {\bibfnamefont {C.}~\bibnamefont
  {De~Vries}},\ }\bibfield  {title} {\bibinfo {title} {{Nuclear
  charge-density-distribution parameters from elastic electron scattering}},\
  }\href {https://doi.org/10.1016/0092-640X(87)90013-1} {\bibfield  {journal}
  {\bibinfo  {journal} {At. Data Nucl. Data Tables}\ }\textbf {\bibinfo
  {volume} {36}},\ \bibinfo {pages} {495} (\bibinfo {year} {1987})}\BibitemShut
  {NoStop}%
\bibitem [{\citenamefont {Suda}\ and\ \citenamefont
  {Simon}(2017)}]{Suda2017Prog.Part.Nucl.Phys.96_1}%
  \BibitemOpen
  \bibfield  {author} {\bibinfo {author} {\bibfnamefont {T.}~\bibnamefont
  {Suda}}\ and\ \bibinfo {author} {\bibfnamefont {H.}~\bibnamefont {Simon}},\
  }\bibfield  {title} {\bibinfo {title} {{Prospects for electron scattering on
  unstable, exotic nuclei}},\ }\href
  {https://doi.org/10.1016/j.ppnp.2017.04.002} {\bibfield  {journal} {\bibinfo
  {journal} {Prog. Part. Nucl. Phys.}\ }\textbf {\bibinfo {volume} {96}},\
  \bibinfo {pages} {1} (\bibinfo {year} {2017})}\BibitemShut {NoStop}%
\bibitem [{\citenamefont {Angeli}\ and\ \citenamefont
  {Marinova}(2013)}]{Angeli2013At.DataNucl.DataTables99_69}%
  \BibitemOpen
  \bibfield  {author} {\bibinfo {author} {\bibfnamefont {I.}~\bibnamefont
  {Angeli}}\ and\ \bibinfo {author} {\bibfnamefont {K.~P.}\ \bibnamefont
  {Marinova}},\ }\bibfield  {title} {\bibinfo {title} {{Table of experimental
  nuclear ground state charge radii: An update}},\ }\href
  {https://doi.org/10.1016/j.adt.2011.12.006} {\bibfield  {journal} {\bibinfo
  {journal} {At. Data Nucl. Data Tables}\ }\textbf {\bibinfo {volume} {99}},\
  \bibinfo {pages} {69} (\bibinfo {year} {2013})}\BibitemShut {NoStop}%
\bibitem [{\citenamefont {Campbell}\ \emph {et~al.}(2016)\citenamefont
  {Campbell}, \citenamefont {Moore},\ and\ \citenamefont
  {Pearson}}]{Campbell2016Prog.Part.Nucl.Phys.86_127}%
  \BibitemOpen
  \bibfield  {author} {\bibinfo {author} {\bibfnamefont {P.}~\bibnamefont
  {Campbell}}, \bibinfo {author} {\bibfnamefont {I.~D.}\ \bibnamefont
  {Moore}},\ and\ \bibinfo {author} {\bibfnamefont {M.~R.}\ \bibnamefont
  {Pearson}},\ }\bibfield  {title} {\bibinfo {title} {{Laser spectroscopy for
  nuclear structure physics}},\ }\href
  {https://doi.org/10.1016/j.ppnp.2015.09.003} {\bibfield  {journal} {\bibinfo
  {journal} {Prog. Part. Nucl. Phys.}\ }\textbf {\bibinfo {volume} {86}},\
  \bibinfo {pages} {127} (\bibinfo {year} {2016})}\BibitemShut {NoStop}%
\bibitem [{\citenamefont {Reinhard}\ and\ \citenamefont
  {Flocard}(1995)}]{Reinhard1995Nucl.Phys.A584_467}%
  \BibitemOpen
  \bibfield  {author} {\bibinfo {author} {\bibfnamefont {P.-G.}\ \bibnamefont
  {Reinhard}}\ and\ \bibinfo {author} {\bibfnamefont {H.}~\bibnamefont
  {Flocard}},\ }\bibfield  {title} {\bibinfo {title} {{Nuclear effective forces
  and isotope shifts}},\ }\href {https://doi.org/10.1016/0375-9474(94)00770-N}
  {\bibfield  {journal} {\bibinfo  {journal} {Nucl. Phys. A}\ }\textbf
  {\bibinfo {volume} {584}},\ \bibinfo {pages} {467} (\bibinfo {year}
  {1995})}\BibitemShut {NoStop}%
\bibitem [{\citenamefont {Nakada}(2019)}]{Nakada2019Phys.Rev.C100_044310}%
  \BibitemOpen
  \bibfield  {author} {\bibinfo {author} {\bibfnamefont {H.}~\bibnamefont
  {Nakada}},\ }\bibfield  {title} {\bibinfo {title} {{Irregularities in nuclear
  radii at magic numbers}},\ }\href
  {https://doi.org/10.1103/PhysRevC.100.044310} {\bibfield  {journal} {\bibinfo
   {journal} {Phys. Rev. C}\ }\textbf {\bibinfo {volume} {100}},\ \bibinfo
  {pages} {044310} (\bibinfo {year} {2019})}\BibitemShut {NoStop}%
\bibitem [{\citenamefont {Perera}\ \emph {et~al.}(2021)\citenamefont {Perera},
  \citenamefont {Afanasjev},\ and\ \citenamefont
  {Ring}}]{Perera2021Phys.Rev.C104_064313}%
  \BibitemOpen
  \bibfield  {author} {\bibinfo {author} {\bibfnamefont {U.~C.}\ \bibnamefont
  {Perera}}, \bibinfo {author} {\bibfnamefont {A.~V.}\ \bibnamefont
  {Afanasjev}},\ and\ \bibinfo {author} {\bibfnamefont {P.}~\bibnamefont
  {Ring}},\ }\bibfield  {title} {\bibinfo {title} {{Charge radii in covariant
  density functional theory: A global view}},\ }\href
  {https://doi.org/10.1103/PhysRevC.104.064313} {\bibfield  {journal} {\bibinfo
   {journal} {Phys. Rev. C}\ }\textbf {\bibinfo {volume} {104}},\ \bibinfo
  {pages} {064313} (\bibinfo {year} {2021})}\BibitemShut {NoStop}%
\bibitem [{\citenamefont {Marsh}\ \emph {et~al.}(2018)\citenamefont {Marsh},
  \citenamefont {Day~Goodacre}, \citenamefont {Sels}, \citenamefont {Tsunoda},
  \citenamefont {Andel}, \citenamefont {Andreyev}, \citenamefont {Althubiti},
  \citenamefont {Atanasov}, \citenamefont {Barzakh}, \citenamefont {Billowes},
  \citenamefont {Blaum}, \citenamefont {Cocolios}, \citenamefont {Cubiss},
  \citenamefont {Dobaczewski}, \citenamefont {Farooq-Smith}, \citenamefont
  {Fedorov}, \citenamefont {Fedosseev}, \citenamefont {Flanagan}, \citenamefont
  {Gaffney}, \citenamefont {Ghys}, \citenamefont {Huyse}, \citenamefont
  {Kreim}, \citenamefont {Lunney}, \citenamefont {Lynch}, \citenamefont
  {Manea}, \citenamefont {Martinez~Palenzuela}, \citenamefont {Molkanov},
  \citenamefont {Otsuka}, \citenamefont {Pastore}, \citenamefont {Rosenbusch},
  \citenamefont {Rossel}, \citenamefont {Rothe}, \citenamefont {Schweikhard},
  \citenamefont {Seliverstov}, \citenamefont {Spagnoletti}, \citenamefont
  {Van~Beveren}, \citenamefont {Van~Duppen}, \citenamefont {Veinhard},
  \citenamefont {Verstraelen}, \citenamefont {Welker}, \citenamefont {Wendt},
  \citenamefont {Wienholtz}, \citenamefont {Wolf}, \citenamefont {Zadvornaya},\
  and\ \citenamefont {Zuber}}]{Marsh2018Nat.Phys.14_1163}%
  \BibitemOpen
  \bibfield  {author} {\bibinfo {author} {\bibfnamefont {B.~A.}\ \bibnamefont
  {Marsh}}, \bibinfo {author} {\bibfnamefont {T.}~\bibnamefont {Day~Goodacre}},
  \bibinfo {author} {\bibfnamefont {S.}~\bibnamefont {Sels}}, \bibinfo {author}
  {\bibfnamefont {Y.}~\bibnamefont {Tsunoda}}, \bibinfo {author} {\bibfnamefont
  {B.}~\bibnamefont {Andel}}, \bibinfo {author} {\bibfnamefont {A.~N.}\
  \bibnamefont {Andreyev}}, \bibinfo {author} {\bibfnamefont {N.~A.}\
  \bibnamefont {Althubiti}}, \bibinfo {author} {\bibfnamefont {D.}~\bibnamefont
  {Atanasov}}, \bibinfo {author} {\bibfnamefont {A.~E.}\ \bibnamefont
  {Barzakh}}, \bibinfo {author} {\bibfnamefont {J.}~\bibnamefont {Billowes}},
  \bibinfo {author} {\bibfnamefont {K.}~\bibnamefont {Blaum}}, \bibinfo
  {author} {\bibfnamefont {T.~E.}\ \bibnamefont {Cocolios}}, \bibinfo {author}
  {\bibfnamefont {J.~G.}\ \bibnamefont {Cubiss}}, \bibinfo {author}
  {\bibfnamefont {J.}~\bibnamefont {Dobaczewski}}, \bibinfo {author}
  {\bibfnamefont {G.~J.}\ \bibnamefont {Farooq-Smith}}, \bibinfo {author}
  {\bibfnamefont {D.~V.}\ \bibnamefont {Fedorov}}, \bibinfo {author}
  {\bibfnamefont {V.~N.}\ \bibnamefont {Fedosseev}}, \bibinfo {author}
  {\bibfnamefont {K.~T.}\ \bibnamefont {Flanagan}}, \bibinfo {author}
  {\bibfnamefont {L.~P.}\ \bibnamefont {Gaffney}}, \bibinfo {author}
  {\bibfnamefont {L.}~\bibnamefont {Ghys}}, \bibinfo {author} {\bibfnamefont
  {M.}~\bibnamefont {Huyse}}, \bibinfo {author} {\bibfnamefont
  {S.}~\bibnamefont {Kreim}}, \bibinfo {author} {\bibfnamefont
  {D.}~\bibnamefont {Lunney}}, \bibinfo {author} {\bibfnamefont {K.~M.}\
  \bibnamefont {Lynch}}, \bibinfo {author} {\bibfnamefont {V.}~\bibnamefont
  {Manea}}, \bibinfo {author} {\bibfnamefont {Y.}~\bibnamefont
  {Martinez~Palenzuela}}, \bibinfo {author} {\bibfnamefont {P.~L.}\
  \bibnamefont {Molkanov}}, \bibinfo {author} {\bibfnamefont {T.}~\bibnamefont
  {Otsuka}}, \bibinfo {author} {\bibfnamefont {A.}~\bibnamefont {Pastore}},
  \bibinfo {author} {\bibfnamefont {M.}~\bibnamefont {Rosenbusch}}, \bibinfo
  {author} {\bibfnamefont {R.~E.}\ \bibnamefont {Rossel}}, \bibinfo {author}
  {\bibfnamefont {S.}~\bibnamefont {Rothe}}, \bibinfo {author} {\bibfnamefont
  {L.}~\bibnamefont {Schweikhard}}, \bibinfo {author} {\bibfnamefont {M.~D.}\
  \bibnamefont {Seliverstov}}, \bibinfo {author} {\bibfnamefont
  {P.}~\bibnamefont {Spagnoletti}}, \bibinfo {author} {\bibfnamefont
  {C.}~\bibnamefont {Van~Beveren}}, \bibinfo {author} {\bibfnamefont
  {P.}~\bibnamefont {Van~Duppen}}, \bibinfo {author} {\bibfnamefont
  {M.}~\bibnamefont {Veinhard}}, \bibinfo {author} {\bibfnamefont
  {E.}~\bibnamefont {Verstraelen}}, \bibinfo {author} {\bibfnamefont
  {A.}~\bibnamefont {Welker}}, \bibinfo {author} {\bibfnamefont
  {K.}~\bibnamefont {Wendt}}, \bibinfo {author} {\bibfnamefont
  {F.}~\bibnamefont {Wienholtz}}, \bibinfo {author} {\bibfnamefont {R.~N.}\
  \bibnamefont {Wolf}}, \bibinfo {author} {\bibfnamefont {A.}~\bibnamefont
  {Zadvornaya}},\ and\ \bibinfo {author} {\bibfnamefont {K.}~\bibnamefont
  {Zuber}},\ }\bibfield  {title} {\bibinfo {title} {{Characterization of the
  shape-staggering effect in mercury nuclei}},\ }\href
  {https://doi.org/10.1038/s41567-018-0292-8} {\bibfield  {journal} {\bibinfo
  {journal} {Nat. Phys.}\ }\textbf {\bibinfo {volume} {14}},\ \bibinfo {pages}
  {1163} (\bibinfo {year} {2018})}\BibitemShut {NoStop}%
\bibitem [{\citenamefont {Day~Goodacre}\ \emph
  {et~al.}(2021{\natexlab{a}})\citenamefont {Day~Goodacre}, \citenamefont
  {Afanasjev}, \citenamefont {Barzakh}, \citenamefont {Nies}, \citenamefont
  {Marsh}, \citenamefont {Sels}, \citenamefont {Perera}, \citenamefont {Ring},
  \citenamefont {Wienholtz}, \citenamefont {Andreyev}, \citenamefont
  {Van~Duppen}, \citenamefont {Althubiti}, \citenamefont {Andel}, \citenamefont
  {Atanasov}, \citenamefont {Augusto}, \citenamefont {Billowes}, \citenamefont
  {Blaum}, \citenamefont {Cocolios}, \citenamefont {Cubiss}, \citenamefont
  {Farooq-Smith}, \citenamefont {Fedorov}, \citenamefont {Fedosseev},
  \citenamefont {Flanagan}, \citenamefont {Gaffney}, \citenamefont {Ghys},
  \citenamefont {Gottberg}, \citenamefont {Huyse}, \citenamefont {Kreim},
  \citenamefont {Kunz}, \citenamefont {Lunney}, \citenamefont {Lynch},
  \citenamefont {Manea}, \citenamefont {Palenzuela}, \citenamefont {Medonca},
  \citenamefont {Molkanov}, \citenamefont {Mougeot}, \citenamefont {Ramos},
  \citenamefont {Rosenbusch}, \citenamefont {Rossel}, \citenamefont {Rothe},
  \citenamefont {Schweikhard}, \citenamefont {Seliverstov}, \citenamefont
  {Spagnoletti}, \citenamefont {Van~Beveren}, \citenamefont {Veinhard},
  \citenamefont {Verstraelen}, \citenamefont {Welker}, \citenamefont {Wendt},
  \citenamefont {Wolf}, \citenamefont {Zadvornaya},\ and\ \citenamefont
  {Zuber}}]{DayGoodacre2021Phys.Rev.C104_054322}%
  \BibitemOpen
  \bibfield  {author} {\bibinfo {author} {\bibfnamefont {T.}~\bibnamefont
  {Day~Goodacre}}, \bibinfo {author} {\bibfnamefont {A.~V.}\ \bibnamefont
  {Afanasjev}}, \bibinfo {author} {\bibfnamefont {A.~E.}\ \bibnamefont
  {Barzakh}}, \bibinfo {author} {\bibfnamefont {L.}~\bibnamefont {Nies}},
  \bibinfo {author} {\bibfnamefont {B.~A.}\ \bibnamefont {Marsh}}, \bibinfo
  {author} {\bibfnamefont {S.}~\bibnamefont {Sels}}, \bibinfo {author}
  {\bibfnamefont {U.~C.}\ \bibnamefont {Perera}}, \bibinfo {author}
  {\bibfnamefont {P.}~\bibnamefont {Ring}}, \bibinfo {author} {\bibfnamefont
  {F.}~\bibnamefont {Wienholtz}}, \bibinfo {author} {\bibfnamefont {A.~N.}\
  \bibnamefont {Andreyev}}, \bibinfo {author} {\bibfnamefont {P.}~\bibnamefont
  {Van~Duppen}}, \bibinfo {author} {\bibfnamefont {N.~A.}\ \bibnamefont
  {Althubiti}}, \bibinfo {author} {\bibfnamefont {B.}~\bibnamefont {Andel}},
  \bibinfo {author} {\bibfnamefont {D.}~\bibnamefont {Atanasov}}, \bibinfo
  {author} {\bibfnamefont {R.~S.}\ \bibnamefont {Augusto}}, \bibinfo {author}
  {\bibfnamefont {J.}~\bibnamefont {Billowes}}, \bibinfo {author}
  {\bibfnamefont {K.}~\bibnamefont {Blaum}}, \bibinfo {author} {\bibfnamefont
  {T.~E.}\ \bibnamefont {Cocolios}}, \bibinfo {author} {\bibfnamefont {J.~G.}\
  \bibnamefont {Cubiss}}, \bibinfo {author} {\bibfnamefont {G.~J.}\
  \bibnamefont {Farooq-Smith}}, \bibinfo {author} {\bibfnamefont {D.~V.}\
  \bibnamefont {Fedorov}}, \bibinfo {author} {\bibfnamefont {V.~N.}\
  \bibnamefont {Fedosseev}}, \bibinfo {author} {\bibfnamefont {K.~T.}\
  \bibnamefont {Flanagan}}, \bibinfo {author} {\bibfnamefont {L.~P.}\
  \bibnamefont {Gaffney}}, \bibinfo {author} {\bibfnamefont {L.}~\bibnamefont
  {Ghys}}, \bibinfo {author} {\bibfnamefont {A.}~\bibnamefont {Gottberg}},
  \bibinfo {author} {\bibfnamefont {M.}~\bibnamefont {Huyse}}, \bibinfo
  {author} {\bibfnamefont {S.}~\bibnamefont {Kreim}}, \bibinfo {author}
  {\bibfnamefont {P.}~\bibnamefont {Kunz}}, \bibinfo {author} {\bibfnamefont
  {D.}~\bibnamefont {Lunney}}, \bibinfo {author} {\bibfnamefont {K.~M.}\
  \bibnamefont {Lynch}}, \bibinfo {author} {\bibfnamefont {V.}~\bibnamefont
  {Manea}}, \bibinfo {author} {\bibfnamefont {Y.~M.}\ \bibnamefont
  {Palenzuela}}, \bibinfo {author} {\bibfnamefont {T.~M.}\ \bibnamefont
  {Medonca}}, \bibinfo {author} {\bibfnamefont {P.~L.}\ \bibnamefont
  {Molkanov}}, \bibinfo {author} {\bibfnamefont {M.}~\bibnamefont {Mougeot}},
  \bibinfo {author} {\bibfnamefont {J.~P.}\ \bibnamefont {Ramos}}, \bibinfo
  {author} {\bibfnamefont {M.}~\bibnamefont {Rosenbusch}}, \bibinfo {author}
  {\bibfnamefont {R.~E.}\ \bibnamefont {Rossel}}, \bibinfo {author}
  {\bibfnamefont {S.}~\bibnamefont {Rothe}}, \bibinfo {author} {\bibfnamefont
  {L.}~\bibnamefont {Schweikhard}}, \bibinfo {author} {\bibfnamefont {M.~D.}\
  \bibnamefont {Seliverstov}}, \bibinfo {author} {\bibfnamefont
  {P.}~\bibnamefont {Spagnoletti}}, \bibinfo {author} {\bibfnamefont
  {C.}~\bibnamefont {Van~Beveren}}, \bibinfo {author} {\bibfnamefont
  {M.}~\bibnamefont {Veinhard}}, \bibinfo {author} {\bibfnamefont
  {E.}~\bibnamefont {Verstraelen}}, \bibinfo {author} {\bibfnamefont
  {A.}~\bibnamefont {Welker}}, \bibinfo {author} {\bibfnamefont
  {K.}~\bibnamefont {Wendt}}, \bibinfo {author} {\bibfnamefont {R.~N.}\
  \bibnamefont {Wolf}}, \bibinfo {author} {\bibfnamefont {A.}~\bibnamefont
  {Zadvornaya}},\ and\ \bibinfo {author} {\bibfnamefont {K.}~\bibnamefont
  {Zuber}},\ }\bibfield  {title} {\bibinfo {title} {{Charge radii, moments, and
  masses of mercury isotopes across the $N=126$ shell closure}},\ }\href
  {https://doi.org/10.1103/PhysRevC.104.054322} {\bibfield  {journal} {\bibinfo
   {journal} {Phys. Rev. C}\ }\textbf {\bibinfo {volume} {104}},\ \bibinfo
  {pages} {054322} (\bibinfo {year} {2021}{\natexlab{a}})}\BibitemShut
  {NoStop}%
\bibitem [{\citenamefont {Day~Goodacre}\ \emph
  {et~al.}(2021{\natexlab{b}})\citenamefont {Day~Goodacre}, \citenamefont
  {Afanasjev}, \citenamefont {Barzakh}, \citenamefont {Marsh}, \citenamefont
  {Sels}, \citenamefont {Ring}, \citenamefont {Nakada}, \citenamefont
  {Andreyev}, \citenamefont {Van~Duppen}, \citenamefont {Althubiti},
  \citenamefont {Andel}, \citenamefont {Atanasov}, \citenamefont {Billowes},
  \citenamefont {Blaum}, \citenamefont {Cocolios}, \citenamefont {Cubiss},
  \citenamefont {Farooq-Smith}, \citenamefont {Fedorov}, \citenamefont
  {Fedosseev}, \citenamefont {Flanagan}, \citenamefont {Gaffney}, \citenamefont
  {Ghys}, \citenamefont {Huyse}, \citenamefont {Kreim}, \citenamefont {Lunney},
  \citenamefont {Lynch}, \citenamefont {Manea}, \citenamefont
  {Martinez~Palenzuela}, \citenamefont {Molkanov}, \citenamefont {Rosenbusch},
  \citenamefont {Rossel}, \citenamefont {Rothe}, \citenamefont {Schweikhard},
  \citenamefont {Seliverstov}, \citenamefont {Spagnoletti}, \citenamefont
  {Van~Beveren}, \citenamefont {Veinhard}, \citenamefont {Verstraelen},
  \citenamefont {Welker}, \citenamefont {Wendt}, \citenamefont {Wienholtz},
  \citenamefont {Wolf}, \citenamefont {Zadvornaya},\ and\ \citenamefont
  {Zuber}}]{DayGoodacre2021Phys.Rev.Lett.126_032502}%
  \BibitemOpen
  \bibfield  {author} {\bibinfo {author} {\bibfnamefont {T.}~\bibnamefont
  {Day~Goodacre}}, \bibinfo {author} {\bibfnamefont {A.~V.}\ \bibnamefont
  {Afanasjev}}, \bibinfo {author} {\bibfnamefont {A.~E.}\ \bibnamefont
  {Barzakh}}, \bibinfo {author} {\bibfnamefont {B.~A.}\ \bibnamefont {Marsh}},
  \bibinfo {author} {\bibfnamefont {S.}~\bibnamefont {Sels}}, \bibinfo {author}
  {\bibfnamefont {P.}~\bibnamefont {Ring}}, \bibinfo {author} {\bibfnamefont
  {H.}~\bibnamefont {Nakada}}, \bibinfo {author} {\bibfnamefont {A.~N.}\
  \bibnamefont {Andreyev}}, \bibinfo {author} {\bibfnamefont {P.}~\bibnamefont
  {Van~Duppen}}, \bibinfo {author} {\bibfnamefont {N.~A.}\ \bibnamefont
  {Althubiti}}, \bibinfo {author} {\bibfnamefont {B.}~\bibnamefont {Andel}},
  \bibinfo {author} {\bibfnamefont {D.}~\bibnamefont {Atanasov}}, \bibinfo
  {author} {\bibfnamefont {J.}~\bibnamefont {Billowes}}, \bibinfo {author}
  {\bibfnamefont {K.}~\bibnamefont {Blaum}}, \bibinfo {author} {\bibfnamefont
  {T.~E.}\ \bibnamefont {Cocolios}}, \bibinfo {author} {\bibfnamefont {J.~G.}\
  \bibnamefont {Cubiss}}, \bibinfo {author} {\bibfnamefont {G.~J.}\
  \bibnamefont {Farooq-Smith}}, \bibinfo {author} {\bibfnamefont {D.~V.}\
  \bibnamefont {Fedorov}}, \bibinfo {author} {\bibfnamefont {V.~N.}\
  \bibnamefont {Fedosseev}}, \bibinfo {author} {\bibfnamefont {K.~T.}\
  \bibnamefont {Flanagan}}, \bibinfo {author} {\bibfnamefont {L.~P.}\
  \bibnamefont {Gaffney}}, \bibinfo {author} {\bibfnamefont {L.}~\bibnamefont
  {Ghys}}, \bibinfo {author} {\bibfnamefont {M.}~\bibnamefont {Huyse}},
  \bibinfo {author} {\bibfnamefont {S.}~\bibnamefont {Kreim}}, \bibinfo
  {author} {\bibfnamefont {D.}~\bibnamefont {Lunney}}, \bibinfo {author}
  {\bibfnamefont {K.~M.}\ \bibnamefont {Lynch}}, \bibinfo {author}
  {\bibfnamefont {V.}~\bibnamefont {Manea}}, \bibinfo {author} {\bibfnamefont
  {Y.}~\bibnamefont {Martinez~Palenzuela}}, \bibinfo {author} {\bibfnamefont
  {P.~L.}\ \bibnamefont {Molkanov}}, \bibinfo {author} {\bibfnamefont
  {M.}~\bibnamefont {Rosenbusch}}, \bibinfo {author} {\bibfnamefont {R.~E.}\
  \bibnamefont {Rossel}}, \bibinfo {author} {\bibfnamefont {S.}~\bibnamefont
  {Rothe}}, \bibinfo {author} {\bibfnamefont {L.}~\bibnamefont {Schweikhard}},
  \bibinfo {author} {\bibfnamefont {M.~D.}\ \bibnamefont {Seliverstov}},
  \bibinfo {author} {\bibfnamefont {P.}~\bibnamefont {Spagnoletti}}, \bibinfo
  {author} {\bibfnamefont {C.}~\bibnamefont {Van~Beveren}}, \bibinfo {author}
  {\bibfnamefont {M.}~\bibnamefont {Veinhard}}, \bibinfo {author}
  {\bibfnamefont {E.}~\bibnamefont {Verstraelen}}, \bibinfo {author}
  {\bibfnamefont {A.}~\bibnamefont {Welker}}, \bibinfo {author} {\bibfnamefont
  {K.}~\bibnamefont {Wendt}}, \bibinfo {author} {\bibfnamefont
  {F.}~\bibnamefont {Wienholtz}}, \bibinfo {author} {\bibfnamefont {R.~N.}\
  \bibnamefont {Wolf}}, \bibinfo {author} {\bibfnamefont {A.}~\bibnamefont
  {Zadvornaya}},\ and\ \bibinfo {author} {\bibfnamefont {K.}~\bibnamefont
  {Zuber}},\ }\bibfield  {title} {\bibinfo {title} {{Laser Spectroscopy of
  Neutron-Rich $^{207,208}\mathrm{Hg}$ Isotopes: Illuminating the Kink and
  Odd-Even Staggering in Charge Radii across the $N=126$ Shell Closure}},\
  }\href {https://doi.org/10.1103/PhysRevLett.126.032502} {\bibfield  {journal}
  {\bibinfo  {journal} {Phys. Rev. Lett.}\ }\textbf {\bibinfo {volume} {126}},\
  \bibinfo {pages} {032502} (\bibinfo {year} {2021}{\natexlab{b}})}\BibitemShut
  {NoStop}%
\bibitem [{\citenamefont {Vautherin}\ and\ \citenamefont
  {Brink}(1972)}]{Vautherin1972Phys.Rev.C5_626}%
  \BibitemOpen
  \bibfield  {author} {\bibinfo {author} {\bibfnamefont {D.}~\bibnamefont
  {Vautherin}}\ and\ \bibinfo {author} {\bibfnamefont {D.~M.}\ \bibnamefont
  {Brink}},\ }\bibfield  {title} {\bibinfo {title} {{Hartree-Fock Calculations
  with Skyrme's Interaction. I. Spherical Nuclei}},\ }\href
  {https://doi.org/10.1103/PhysRevC.5.626} {\bibfield  {journal} {\bibinfo
  {journal} {Phys. Rev. C}\ }\textbf {\bibinfo {volume} {5}},\ \bibinfo {pages}
  {626} (\bibinfo {year} {1972})}\BibitemShut {NoStop}%
\bibitem [{\citenamefont {Miller}\ and\ \citenamefont
  {Green}(1972)}]{Miller1972Phys.Rev.C5_241}%
  \BibitemOpen
  \bibfield  {author} {\bibinfo {author} {\bibfnamefont {L.~D.}\ \bibnamefont
  {Miller}}\ and\ \bibinfo {author} {\bibfnamefont {A.~E.~S.}\ \bibnamefont
  {Green}},\ }\bibfield  {title} {\bibinfo {title} {{Relativistic
  Self-Consistent Meson Field Theory of Spherical Nuclei}},\ }\href
  {https://doi.org/10.1103/PhysRevC.5.241} {\bibfield  {journal} {\bibinfo
  {journal} {Phys. Rev. C}\ }\textbf {\bibinfo {volume} {5}},\ \bibinfo {pages}
  {241} (\bibinfo {year} {1972})}\BibitemShut {NoStop}%
\bibitem [{\citenamefont {Walecka}(1974)}]{Walecka1974Ann.Phys.83_491}%
  \BibitemOpen
  \bibfield  {author} {\bibinfo {author} {\bibfnamefont {J.~D.}\ \bibnamefont
  {Walecka}},\ }\bibfield  {title} {\bibinfo {title} {{A theory of highly
  condensed matter}},\ }\href {https://doi.org/10.1016/0003-4916(74)90208-5}
  {\bibfield  {journal} {\bibinfo  {journal} {Ann. Phys.}\ }\textbf {\bibinfo
  {volume} {83}},\ \bibinfo {pages} {491} (\bibinfo {year} {1974})}\BibitemShut
  {NoStop}%
\bibitem [{\citenamefont {Long}\ \emph {et~al.}(2006)\citenamefont {Long},
  \citenamefont {Van~Giai},\ and\ \citenamefont
  {Meng}}]{Long2006Phys.Lett.B640_150}%
  \BibitemOpen
  \bibfield  {author} {\bibinfo {author} {\bibfnamefont {W.-H.}\ \bibnamefont
  {Long}}, \bibinfo {author} {\bibfnamefont {N.}~\bibnamefont {Van~Giai}},\
  and\ \bibinfo {author} {\bibfnamefont {J.}~\bibnamefont {Meng}},\ }\bibfield
  {title} {\bibinfo {title} {{Density-dependent relativistic Hartree-Fock
  approach}},\ }\href {https://doi.org/10.1016/j.physletb.2006.07.064}
  {\bibfield  {journal} {\bibinfo  {journal} {Phys. Lett. B}\ }\textbf
  {\bibinfo {volume} {640}},\ \bibinfo {pages} {150} (\bibinfo {year}
  {2006})}\BibitemShut {NoStop}%
\bibitem [{\citenamefont {Long}\ \emph {et~al.}(2010)\citenamefont {Long},
  \citenamefont {Ring}, \citenamefont {Giai},\ and\ \citenamefont
  {Meng}}]{Long2010Phys.Rev.C81_024308}%
  \BibitemOpen
  \bibfield  {author} {\bibinfo {author} {\bibfnamefont {W.~H.}\ \bibnamefont
  {Long}}, \bibinfo {author} {\bibfnamefont {P.}~\bibnamefont {Ring}}, \bibinfo
  {author} {\bibfnamefont {N.~V.}\ \bibnamefont {Giai}},\ and\ \bibinfo
  {author} {\bibfnamefont {J.}~\bibnamefont {Meng}},\ }\bibfield  {title}
  {\bibinfo {title} {{Relativistic Hartree-Fock-Bogoliubov theory with density
  dependent meson-nucleon couplings}},\ }\href
  {https://doi.org/10.1103/PhysRevC.81.024308} {\bibfield  {journal} {\bibinfo
  {journal} {Phys. Rev. C}\ }\textbf {\bibinfo {volume} {81}},\ \bibinfo
  {pages} {024308} (\bibinfo {year} {2010})}\BibitemShut {NoStop}%
\bibitem [{\citenamefont {Roca-Maza}\ \emph {et~al.}(2012)\citenamefont
  {Roca-Maza}, \citenamefont {Col\`o},\ and\ \citenamefont
  {Sagawa}}]{Roca-Maza2012Phys.Rev.C86_031306}%
  \BibitemOpen
  \bibfield  {author} {\bibinfo {author} {\bibfnamefont {X.}~\bibnamefont
  {Roca-Maza}}, \bibinfo {author} {\bibfnamefont {G.}~\bibnamefont {Col\`o}},\
  and\ \bibinfo {author} {\bibfnamefont {H.}~\bibnamefont {Sagawa}},\
  }\bibfield  {title} {\bibinfo {title} {{New Skyrme interaction with improved
  spin-isospin properties}},\ }\href
  {https://doi.org/10.1103/PhysRevC.86.031306} {\bibfield  {journal} {\bibinfo
  {journal} {Phys. Rev. C}\ }\textbf {\bibinfo {volume} {86}},\ \bibinfo
  {pages} {031306} (\bibinfo {year} {2012})}\BibitemShut {NoStop}%
\bibitem [{\citenamefont {Van~Giai}\ and\ \citenamefont
  {Sagawa}(1981)}]{VanGiai1981Phys.Lett.B106_379}%
  \BibitemOpen
  \bibfield  {author} {\bibinfo {author} {\bibfnamefont {N.}~\bibnamefont
  {Van~Giai}}\ and\ \bibinfo {author} {\bibfnamefont {H.}~\bibnamefont
  {Sagawa}},\ }\bibfield  {title} {\bibinfo {title} {{Spin-isospin and pairing
  properties of modified Skyrme interactions}},\ }\href
  {https://doi.org/10.1016/0370-2693(81)90646-8} {\bibfield  {journal}
  {\bibinfo  {journal} {Phys. Lett. B}\ }\textbf {\bibinfo {volume} {106}},\
  \bibinfo {pages} {379} (\bibinfo {year} {1981})}\BibitemShut {NoStop}%
\bibitem [{\citenamefont {Chabanat}\ \emph {et~al.}(1998)\citenamefont
  {Chabanat}, \citenamefont {Bonche}, \citenamefont {Haensel}, \citenamefont
  {Meyer},\ and\ \citenamefont {Schaeffer}}]{Chabanat1998Nucl.Phys.A635_231}%
  \BibitemOpen
  \bibfield  {author} {\bibinfo {author} {\bibfnamefont {E.}~\bibnamefont
  {Chabanat}}, \bibinfo {author} {\bibfnamefont {P.}~\bibnamefont {Bonche}},
  \bibinfo {author} {\bibfnamefont {P.}~\bibnamefont {Haensel}}, \bibinfo
  {author} {\bibfnamefont {J.}~\bibnamefont {Meyer}},\ and\ \bibinfo {author}
  {\bibfnamefont {R.}~\bibnamefont {Schaeffer}},\ }\bibfield  {title} {\bibinfo
  {title} {{A Skyrme parametrization from subnuclear to neutron star densities
  Part II. Nuclei far from stabilities}},\ }\href
  {https://doi.org/10.1016/S0375-9474(98)00180-8} {\bibfield  {journal}
  {\bibinfo  {journal} {Nucl. Phys. A}\ }\textbf {\bibinfo {volume} {635}},\
  \bibinfo {pages} {231} (\bibinfo {year} {1998})}\BibitemShut {NoStop}%
\bibitem [{\citenamefont {Bartel}\ \emph {et~al.}(1982)\citenamefont {Bartel},
  \citenamefont {Quentin}, \citenamefont {Brack}, \citenamefont {Guet},\ and\
  \citenamefont {H{\aa}kansson}}]{Bartel1982Nucl.Phys.A386_79}%
  \BibitemOpen
  \bibfield  {author} {\bibinfo {author} {\bibfnamefont {J.}~\bibnamefont
  {Bartel}}, \bibinfo {author} {\bibfnamefont {P.}~\bibnamefont {Quentin}},
  \bibinfo {author} {\bibfnamefont {M.}~\bibnamefont {Brack}}, \bibinfo
  {author} {\bibfnamefont {C.}~\bibnamefont {Guet}},\ and\ \bibinfo {author}
  {\bibfnamefont {H.-B.}\ \bibnamefont {H{\aa}kansson}},\ }\bibfield  {title}
  {\bibinfo {title} {{Towards a better parametrisation of Skyrme-like effective
  forces: A critical study of the SkM force}},\ }\href
  {https://doi.org/10.1016/0375-9474(82)90403-1} {\bibfield  {journal}
  {\bibinfo  {journal} {Nucl. Phys. A}\ }\textbf {\bibinfo {volume} {386}},\
  \bibinfo {pages} {79} (\bibinfo {year} {1982})}\BibitemShut {NoStop}%
\bibitem [{\citenamefont {Goriely}\ \emph {et~al.}(2005)\citenamefont
  {Goriely}, \citenamefont {Samyn}, \citenamefont {Pearson},\ and\
  \citenamefont {Onsi}}]{Goriely2005Nucl.Phys.A750_425}%
  \BibitemOpen
  \bibfield  {author} {\bibinfo {author} {\bibfnamefont {S.}~\bibnamefont
  {Goriely}}, \bibinfo {author} {\bibfnamefont {M.}~\bibnamefont {Samyn}},
  \bibinfo {author} {\bibfnamefont {J.~M.}\ \bibnamefont {Pearson}},\ and\
  \bibinfo {author} {\bibfnamefont {M.}~\bibnamefont {Onsi}},\ }\bibfield
  {title} {\bibinfo {title} {{Further explorations of
  Skyrme-Hartree-Fock-Bogoliubov mass formulas. IV: Neutron-matter
  constraint}},\ }\href {https://doi.org/10.1016/j.nuclphysa.2005.01.009}
  {\bibfield  {journal} {\bibinfo  {journal} {Nucl. Phys. A}\ }\textbf
  {\bibinfo {volume} {750}},\ \bibinfo {pages} {425} (\bibinfo {year}
  {2005})}\BibitemShut {NoStop}%
\bibitem [{\citenamefont {Kortelainen}\ \emph {et~al.}(2010)\citenamefont
  {Kortelainen}, \citenamefont {Lesinski}, \citenamefont {Mor\'e},
  \citenamefont {Nazarewicz}, \citenamefont {Sarich}, \citenamefont {Schunck},
  \citenamefont {Stoitsov},\ and\ \citenamefont
  {Wild}}]{Kortelainen2010Phys.Rev.C82_024313}%
  \BibitemOpen
  \bibfield  {author} {\bibinfo {author} {\bibfnamefont {M.}~\bibnamefont
  {Kortelainen}}, \bibinfo {author} {\bibfnamefont {T.}~\bibnamefont
  {Lesinski}}, \bibinfo {author} {\bibfnamefont {J.}~\bibnamefont {Mor\'e}},
  \bibinfo {author} {\bibfnamefont {W.}~\bibnamefont {Nazarewicz}}, \bibinfo
  {author} {\bibfnamefont {J.}~\bibnamefont {Sarich}}, \bibinfo {author}
  {\bibfnamefont {N.}~\bibnamefont {Schunck}}, \bibinfo {author} {\bibfnamefont
  {M.~V.}\ \bibnamefont {Stoitsov}},\ and\ \bibinfo {author} {\bibfnamefont
  {S.}~\bibnamefont {Wild}},\ }\bibfield  {title} {\bibinfo {title} {{Nuclear
  energy density optimization}},\ }\href
  {https://doi.org/10.1103/PhysRevC.82.024313} {\bibfield  {journal} {\bibinfo
  {journal} {Phys. Rev. C}\ }\textbf {\bibinfo {volume} {82}},\ \bibinfo
  {pages} {024313} (\bibinfo {year} {2010})}\BibitemShut {NoStop}%
\bibitem [{\citenamefont {Kortelainen}\ \emph {et~al.}(2012)\citenamefont
  {Kortelainen}, \citenamefont {McDonnell}, \citenamefont {Nazarewicz},
  \citenamefont {Reinhard}, \citenamefont {Sarich}, \citenamefont {Schunck},
  \citenamefont {Stoitsov},\ and\ \citenamefont
  {Wild}}]{Kortelainen2012Phys.Rev.C85_024304}%
  \BibitemOpen
  \bibfield  {author} {\bibinfo {author} {\bibfnamefont {M.}~\bibnamefont
  {Kortelainen}}, \bibinfo {author} {\bibfnamefont {J.}~\bibnamefont
  {McDonnell}}, \bibinfo {author} {\bibfnamefont {W.}~\bibnamefont
  {Nazarewicz}}, \bibinfo {author} {\bibfnamefont {P.-G.}\ \bibnamefont
  {Reinhard}}, \bibinfo {author} {\bibfnamefont {J.}~\bibnamefont {Sarich}},
  \bibinfo {author} {\bibfnamefont {N.}~\bibnamefont {Schunck}}, \bibinfo
  {author} {\bibfnamefont {M.~V.}\ \bibnamefont {Stoitsov}},\ and\ \bibinfo
  {author} {\bibfnamefont {S.~M.}\ \bibnamefont {Wild}},\ }\bibfield  {title}
  {\bibinfo {title} {{Nuclear energy density optimization: Large
  deformations}},\ }\href {https://doi.org/10.1103/PhysRevC.85.024304}
  {\bibfield  {journal} {\bibinfo  {journal} {Phys. Rev. C}\ }\textbf {\bibinfo
  {volume} {85}},\ \bibinfo {pages} {024304} (\bibinfo {year}
  {2012})}\BibitemShut {NoStop}%
\bibitem [{\citenamefont {Kortelainen}\ \emph {et~al.}(2014)\citenamefont
  {Kortelainen}, \citenamefont {McDonnell}, \citenamefont {Nazarewicz},
  \citenamefont {Olsen}, \citenamefont {Reinhard}, \citenamefont {Sarich},
  \citenamefont {Schunck}, \citenamefont {Wild}, \citenamefont {Davesne},
  \citenamefont {Erler},\ and\ \citenamefont
  {Pastore}}]{Kortelainen2014Phys.Rev.C89_054314}%
  \BibitemOpen
  \bibfield  {author} {\bibinfo {author} {\bibfnamefont {M.}~\bibnamefont
  {Kortelainen}}, \bibinfo {author} {\bibfnamefont {J.}~\bibnamefont
  {McDonnell}}, \bibinfo {author} {\bibfnamefont {W.}~\bibnamefont
  {Nazarewicz}}, \bibinfo {author} {\bibfnamefont {E.}~\bibnamefont {Olsen}},
  \bibinfo {author} {\bibfnamefont {P.-G.}\ \bibnamefont {Reinhard}}, \bibinfo
  {author} {\bibfnamefont {J.}~\bibnamefont {Sarich}}, \bibinfo {author}
  {\bibfnamefont {N.}~\bibnamefont {Schunck}}, \bibinfo {author} {\bibfnamefont
  {S.~M.}\ \bibnamefont {Wild}}, \bibinfo {author} {\bibfnamefont
  {D.}~\bibnamefont {Davesne}}, \bibinfo {author} {\bibfnamefont
  {J.}~\bibnamefont {Erler}},\ and\ \bibinfo {author} {\bibfnamefont
  {A.}~\bibnamefont {Pastore}},\ }\bibfield  {title} {\bibinfo {title}
  {{Nuclear energy density optimization: Shell structure}},\ }\href
  {https://doi.org/10.1103/PhysRevC.89.054314} {\bibfield  {journal} {\bibinfo
  {journal} {Phys. Rev. C}\ }\textbf {\bibinfo {volume} {89}},\ \bibinfo
  {pages} {054314} (\bibinfo {year} {2014})}\BibitemShut {NoStop}%
\bibitem [{\citenamefont {Shen}\ \emph {et~al.}(2019)\citenamefont {Shen},
  \citenamefont {Col\`o},\ and\ \citenamefont
  {Roca-Maza}}]{Shen2019Phys.Rev.C99_034322}%
  \BibitemOpen
  \bibfield  {author} {\bibinfo {author} {\bibfnamefont {S.}~\bibnamefont
  {Shen}}, \bibinfo {author} {\bibfnamefont {G.}~\bibnamefont {Col\`o}},\ and\
  \bibinfo {author} {\bibfnamefont {X.}~\bibnamefont {Roca-Maza}},\ }\bibfield
  {title} {\bibinfo {title} {{Skyrme functional with tensor terms from
  \textit{ab initio} calculations of neutron-proton drops}},\ }\href
  {https://doi.org/10.1103/PhysRevC.99.034322} {\bibfield  {journal} {\bibinfo
  {journal} {Phys. Rev. C}\ }\textbf {\bibinfo {volume} {99}},\ \bibinfo
  {pages} {034322} (\bibinfo {year} {2019})}\BibitemShut {NoStop}%
\bibitem [{\citenamefont {Roca-Maza}\ \emph {et~al.}(2013)\citenamefont
  {Roca-Maza}, \citenamefont {Brenna}, \citenamefont {Agrawal}, \citenamefont
  {Bortignon}, \citenamefont {Col\`{o}}, \citenamefont {Cao}, \citenamefont
  {Paar},\ and\ \citenamefont {Vretenar}}]{Roca-Maza2013Phys.Rev.C87_034301}%
  \BibitemOpen
  \bibfield  {author} {\bibinfo {author} {\bibfnamefont {X.}~\bibnamefont
  {Roca-Maza}}, \bibinfo {author} {\bibfnamefont {M.}~\bibnamefont {Brenna}},
  \bibinfo {author} {\bibfnamefont {B.~K.}\ \bibnamefont {Agrawal}}, \bibinfo
  {author} {\bibfnamefont {P.~F.}\ \bibnamefont {Bortignon}}, \bibinfo {author}
  {\bibfnamefont {G.}~\bibnamefont {Col\`{o}}}, \bibinfo {author}
  {\bibfnamefont {L.-G.}\ \bibnamefont {Cao}}, \bibinfo {author} {\bibfnamefont
  {N.}~\bibnamefont {Paar}},\ and\ \bibinfo {author} {\bibfnamefont
  {D.}~\bibnamefont {Vretenar}},\ }\bibfield  {title} {\bibinfo {title} {{Giant
  quadrupole resonances in ${}^{208} \mathrm{Pb} $, the nuclear symmetry
  energy, and the neutron skin thickness}},\ }\href
  {https://doi.org/10.1103/PhysRevC.87.034301} {\bibfield  {journal} {\bibinfo
  {journal} {Phys. Rev. C}\ }\textbf {\bibinfo {volume} {87}},\ \bibinfo
  {pages} {034301} (\bibinfo {year} {2013})}\BibitemShut {NoStop}%
\bibitem [{\citenamefont {Roca-Maza}()}]{Roca-Maza_}%
  \BibitemOpen
  \bibfield  {author} {\bibinfo {author} {\bibfnamefont {X.}~\bibnamefont
  {Roca-Maza}},\ }\href@noop {} {}\bibinfo {note} {Private
  communication}\BibitemShut {NoStop}%
\bibitem [{\citenamefont {Dobaczewski}\ \emph {et~al.}(1984)\citenamefont
  {Dobaczewski}, \citenamefont {Flocard},\ and\ \citenamefont
  {Treiner}}]{Dobaczewski1984Nucl.Phys.A422_103}%
  \BibitemOpen
  \bibfield  {author} {\bibinfo {author} {\bibfnamefont {J.}~\bibnamefont
  {Dobaczewski}}, \bibinfo {author} {\bibfnamefont {H.}~\bibnamefont
  {Flocard}},\ and\ \bibinfo {author} {\bibfnamefont {J.}~\bibnamefont
  {Treiner}},\ }\bibfield  {title} {\bibinfo {title} {{Hartree-Fock-Bogolyubov
  description of nuclei near the neutron-drip line}},\ }\href
  {https://doi.org/10.1016/0375-9474(84)90433-0} {\bibfield  {journal}
  {\bibinfo  {journal} {Nucl. Phys. A}\ }\textbf {\bibinfo {volume} {422}},\
  \bibinfo {pages} {103} (\bibinfo {year} {1984})}\BibitemShut {NoStop}%
\bibitem [{\citenamefont {Navarro~Perez}\ \emph {et~al.}(2017)\citenamefont
  {Navarro~Perez}, \citenamefont {Schunck}, \citenamefont {Lasseri},
  \citenamefont {Zhang},\ and\ \citenamefont
  {Sarich}}]{NavarroPerez2017Comput.Phys.Commun.220_363}%
  \BibitemOpen
  \bibfield  {author} {\bibinfo {author} {\bibfnamefont {R.}~\bibnamefont
  {Navarro~Perez}}, \bibinfo {author} {\bibfnamefont {N.}~\bibnamefont
  {Schunck}}, \bibinfo {author} {\bibfnamefont {R.-D.}\ \bibnamefont
  {Lasseri}}, \bibinfo {author} {\bibfnamefont {C.}~\bibnamefont {Zhang}},\
  and\ \bibinfo {author} {\bibfnamefont {J.}~\bibnamefont {Sarich}},\
  }\bibfield  {title} {\bibinfo {title} {{Axially deformed solution of the
  Skyrme-Hartree-Fock-Bogolyubov equations using the transformed harmonic
  oscillator basis (III) \textsc{hfbtho} (v3.00): A new version of the
  program}},\ }\href {https://doi.org/10.1016/j.cpc.2017.06.022} {\bibfield
  {journal} {\bibinfo  {journal} {Comput. Phys. Commun.}\ }\textbf {\bibinfo
  {volume} {220}},\ \bibinfo {pages} {363} (\bibinfo {year}
  {2017})}\BibitemShut {NoStop}%
\bibitem [{\citenamefont {Nik\v{s}i\'{c}}\ \emph {et~al.}(2008)\citenamefont
  {Nik\v{s}i\'{c}}, \citenamefont {Vretenar},\ and\ \citenamefont
  {Ring}}]{Niksic2008Phys.Rev.C78_034318}%
  \BibitemOpen
  \bibfield  {author} {\bibinfo {author} {\bibfnamefont {T.}~\bibnamefont
  {Nik\v{s}i\'{c}}}, \bibinfo {author} {\bibfnamefont {D.}~\bibnamefont
  {Vretenar}},\ and\ \bibinfo {author} {\bibfnamefont {P.}~\bibnamefont
  {Ring}},\ }\bibfield  {title} {\bibinfo {title} {{Relativistic nuclear energy
  density functionals: Adjusting parameters to binding energies}},\ }\href
  {https://doi.org/10.1103/PhysRevC.78.034318} {\bibfield  {journal} {\bibinfo
  {journal} {Phys. Rev. C}\ }\textbf {\bibinfo {volume} {78}},\ \bibinfo
  {pages} {034318} (\bibinfo {year} {2008})}\BibitemShut {NoStop}%
\bibitem [{\citenamefont {Lalazissis}\ \emph {et~al.}(2005)\citenamefont
  {Lalazissis}, \citenamefont {Nik\v{s}i\'{c}}, \citenamefont {Vretenar},\ and\
  \citenamefont {Ring}}]{Lalazissis2005Phys.Rev.C71_024312}%
  \BibitemOpen
  \bibfield  {author} {\bibinfo {author} {\bibfnamefont {G.~A.}\ \bibnamefont
  {Lalazissis}}, \bibinfo {author} {\bibfnamefont {T.}~\bibnamefont
  {Nik\v{s}i\'{c}}}, \bibinfo {author} {\bibfnamefont {D.}~\bibnamefont
  {Vretenar}},\ and\ \bibinfo {author} {\bibfnamefont {P.}~\bibnamefont
  {Ring}},\ }\bibfield  {title} {\bibinfo {title} {{New relativistic mean-field
  interaction with density-dependent meson-nucleon couplings}},\ }\href
  {https://doi.org/10.1103/PhysRevC.71.024312} {\bibfield  {journal} {\bibinfo
  {journal} {Phys. Rev. C}\ }\textbf {\bibinfo {volume} {71}},\ \bibinfo
  {pages} {024312} (\bibinfo {year} {2005})}\BibitemShut {NoStop}%
\bibitem [{\citenamefont {Long}\ \emph {et~al.}(2004)\citenamefont {Long},
  \citenamefont {Meng}, \citenamefont {Van~Giai},\ and\ \citenamefont
  {Zhou}}]{Long2004Phys.Rev.C69_034319}%
  \BibitemOpen
  \bibfield  {author} {\bibinfo {author} {\bibfnamefont {W.}~\bibnamefont
  {Long}}, \bibinfo {author} {\bibfnamefont {J.}~\bibnamefont {Meng}}, \bibinfo
  {author} {\bibfnamefont {N.}~\bibnamefont {Van~Giai}},\ and\ \bibinfo
  {author} {\bibfnamefont {S.-G.}\ \bibnamefont {Zhou}},\ }\bibfield  {title}
  {\bibinfo {title} {{New effective interactions in relativistic mean field
  theory with nonlinear terms and density-dependent meson-nucleon coupling}},\
  }\href {https://doi.org/10.1103/PhysRevC.69.034319} {\bibfield  {journal}
  {\bibinfo  {journal} {Phys. Rev. C}\ }\textbf {\bibinfo {volume} {69}},\
  \bibinfo {pages} {034319} (\bibinfo {year} {2004})}\BibitemShut {NoStop}%
\bibitem [{\citenamefont {Wei}\ \emph {et~al.}(2020)\citenamefont {Wei},
  \citenamefont {Zhao}, \citenamefont {Wang}, \citenamefont {Geng},
  \citenamefont {Sun}, \citenamefont {Niu},\ and\ \citenamefont
  {Long}}]{Wei2020Chin.Phys.C44_074107}%
  \BibitemOpen
  \bibfield  {author} {\bibinfo {author} {\bibfnamefont {B.}~\bibnamefont
  {Wei}}, \bibinfo {author} {\bibfnamefont {Q.}~\bibnamefont {Zhao}}, \bibinfo
  {author} {\bibfnamefont {Z.-H.}\ \bibnamefont {Wang}}, \bibinfo {author}
  {\bibfnamefont {J.}~\bibnamefont {Geng}}, \bibinfo {author} {\bibfnamefont
  {B.-Y.}\ \bibnamefont {Sun}}, \bibinfo {author} {\bibfnamefont {Y.-F.}\
  \bibnamefont {Niu}},\ and\ \bibinfo {author} {\bibfnamefont {W.-H.}\
  \bibnamefont {Long}},\ }\bibfield  {title} {\bibinfo {title} {{Novel
  relativistic mean field Lagrangian guided by pseudo-spin symmetry
  restoration}},\ }\href {https://doi.org/10.1088/1674-1137/44/7/074107}
  {\bibfield  {journal} {\bibinfo  {journal} {Chin. Phys. C}\ }\textbf
  {\bibinfo {volume} {44}},\ \bibinfo {pages} {074107} (\bibinfo {year}
  {2020})}\BibitemShut {NoStop}%
\bibitem [{\citenamefont {Long}\ \emph {et~al.}(2008)\citenamefont {Long},
  \citenamefont {Sagawa}, \citenamefont {Meng},\ and\ \citenamefont
  {Van~Giai}}]{Long2008Europhys.Lett.82_12001}%
  \BibitemOpen
  \bibfield  {author} {\bibinfo {author} {\bibfnamefont {W.~H.}\ \bibnamefont
  {Long}}, \bibinfo {author} {\bibfnamefont {H.}~\bibnamefont {Sagawa}},
  \bibinfo {author} {\bibfnamefont {J.}~\bibnamefont {Meng}},\ and\ \bibinfo
  {author} {\bibfnamefont {N.}~\bibnamefont {Van~Giai}},\ }\bibfield  {title}
  {\bibinfo {title} {{Evolution of nuclear shell structure due to the pion
  exchange potential}},\ }\href {https://doi.org/10.1209/0295-5075/82/12001}
  {\bibfield  {journal} {\bibinfo  {journal} {Europhys. Lett.}\ }\textbf
  {\bibinfo {volume} {82}},\ \bibinfo {pages} {12001} (\bibinfo {year}
  {2008})}\BibitemShut {NoStop}%
\bibitem [{\citenamefont {Long}\ \emph {et~al.}(2007)\citenamefont {Long},
  \citenamefont {Sagawa}, \citenamefont {Van~Giai},\ and\ \citenamefont
  {Meng}}]{Long2007Phys.Rev.C76_034314}%
  \BibitemOpen
  \bibfield  {author} {\bibinfo {author} {\bibfnamefont {W.}~\bibnamefont
  {Long}}, \bibinfo {author} {\bibfnamefont {H.}~\bibnamefont {Sagawa}},
  \bibinfo {author} {\bibfnamefont {N.}~\bibnamefont {Van~Giai}},\ and\
  \bibinfo {author} {\bibfnamefont {J.}~\bibnamefont {Meng}},\ }\bibfield
  {title} {\bibinfo {title} {{Shell structure and $ \rho $-tensor correlations
  in density dependent relativistic Hartree-Fock theory}},\ }\href
  {https://doi.org/10.1103/PhysRevC.76.034314} {\bibfield  {journal} {\bibinfo
  {journal} {Phys. Rev. C}\ }\textbf {\bibinfo {volume} {76}},\ \bibinfo
  {pages} {034314} (\bibinfo {year} {2007})}\BibitemShut {NoStop}%
\bibitem [{\citenamefont {Wang}\ \emph {et~al.}(2021)\citenamefont {Wang},
  \citenamefont {Naito},\ and\ \citenamefont
  {Liang}}]{Wang2021Phys.Rev.C103_064326}%
  \BibitemOpen
  \bibfield  {author} {\bibinfo {author} {\bibfnamefont {Z.}~\bibnamefont
  {Wang}}, \bibinfo {author} {\bibfnamefont {T.}~\bibnamefont {Naito}},\ and\
  \bibinfo {author} {\bibfnamefont {H.}~\bibnamefont {Liang}},\ }\bibfield
  {title} {\bibinfo {title} {{Tensor-force effects on shell-structure evolution
  in $N=82$ isotones and $Z=50$ isotopes in the relativistic Hartree-Fock
  theory}},\ }\href {https://doi.org/10.1103/PhysRevC.103.064326} {\bibfield
  {journal} {\bibinfo  {journal} {Phys. Rev. C}\ }\textbf {\bibinfo {volume}
  {103}},\ \bibinfo {pages} {064326} (\bibinfo {year} {2021})}\BibitemShut
  {NoStop}%
\bibitem [{\citenamefont {Tian}\ \emph {et~al.}(2009)\citenamefont {Tian},
  \citenamefont {Ma},\ and\ \citenamefont {Ring}}]{Tian2009Phys.Lett.B676_44}%
  \BibitemOpen
  \bibfield  {author} {\bibinfo {author} {\bibfnamefont {Y.}~\bibnamefont
  {Tian}}, \bibinfo {author} {\bibfnamefont {Z.~Y.}\ \bibnamefont {Ma}},\ and\
  \bibinfo {author} {\bibfnamefont {P.}~\bibnamefont {Ring}},\ }\bibfield
  {title} {\bibinfo {title} {{A finite range pairing force for density
  functional theory in superfluid nuclei}},\ }\href
  {https://doi.org/10.1016/j.physletb.2009.04.067} {\bibfield  {journal}
  {\bibinfo  {journal} {Phys. Lett. B}\ }\textbf {\bibinfo {volume} {676}},\
  \bibinfo {pages} {44} (\bibinfo {year} {2009})}\BibitemShut {NoStop}%
\bibitem [{\citenamefont {Dobaczewski}\ \emph {et~al.}(1996)\citenamefont
  {Dobaczewski}, \citenamefont {Nazarewicz}, \citenamefont {Werner},
  \citenamefont {Berger}, \citenamefont {Chinn},\ and\ \citenamefont
  {Decharg\'e}}]{Dobaczewski1996Phys.Rev.C53_2809}%
  \BibitemOpen
  \bibfield  {author} {\bibinfo {author} {\bibfnamefont {J.}~\bibnamefont
  {Dobaczewski}}, \bibinfo {author} {\bibfnamefont {W.}~\bibnamefont
  {Nazarewicz}}, \bibinfo {author} {\bibfnamefont {T.~R.}\ \bibnamefont
  {Werner}}, \bibinfo {author} {\bibfnamefont {J.~F.}\ \bibnamefont {Berger}},
  \bibinfo {author} {\bibfnamefont {C.~R.}\ \bibnamefont {Chinn}},\ and\
  \bibinfo {author} {\bibfnamefont {J.}~\bibnamefont {Decharg\'e}},\ }\bibfield
   {title} {\bibinfo {title} {{Mean-field description of ground-state
  properties of drip-line nuclei: Pairing and continuum effects}},\ }\href
  {https://doi.org/10.1103/PhysRevC.53.2809} {\bibfield  {journal} {\bibinfo
  {journal} {Phys. Rev. C}\ }\textbf {\bibinfo {volume} {53}},\ \bibinfo
  {pages} {2809} (\bibinfo {year} {1996})}\BibitemShut {NoStop}%
\bibitem [{\citenamefont {Nakada}\ and\ \citenamefont
  {Yamagami}(2011)}]{Nakada2011Phys.Rev.C83_031302}%
  \BibitemOpen
  \bibfield  {author} {\bibinfo {author} {\bibfnamefont {H.}~\bibnamefont
  {Nakada}}\ and\ \bibinfo {author} {\bibfnamefont {M.}~\bibnamefont
  {Yamagami}},\ }\bibfield  {title} {\bibinfo {title} {{Coulombic effect and
  renormalization in nuclear pairing}},\ }\href
  {https://doi.org/10.1103/PhysRevC.83.031302} {\bibfield  {journal} {\bibinfo
  {journal} {Phys. Rev. C}\ }\textbf {\bibinfo {volume} {83}},\ \bibinfo
  {pages} {031302} (\bibinfo {year} {2011})}\BibitemShut {NoStop}%
\bibitem [{\citenamefont {Zyla}\ \emph {et~al.}(2020)\citenamefont {Zyla},
  \citenamefont {Barnett}, \citenamefont {Beringer}, \citenamefont {Dahl},
  \citenamefont {Dwyer}, \citenamefont {Groom}, \citenamefont {Lin},
  \citenamefont {Lugovsky}, \citenamefont {Pianori}, \citenamefont {Robinson},
  \citenamefont {Wohl}, \citenamefont {Yao}, \citenamefont {Agashe},
  \citenamefont {Aielli}, \citenamefont {Allanach}, \citenamefont {Amsler},
  \citenamefont {Antonelli}, \citenamefont {Aschenauer}, \citenamefont {Asner},
  \citenamefont {Baer}, \citenamefont {Banerjee}, \citenamefont {Baudis},
  \citenamefont {Bauer}, \citenamefont {Beatty}, \citenamefont {Belousov},
  \citenamefont {Bethke}, \citenamefont {Bettini}, \citenamefont {Biebel},
  \citenamefont {Black}, \citenamefont {Blucher}, \citenamefont {Buchmuller},
  \citenamefont {Burkert}, \citenamefont {Bychkov}, \citenamefont {Cahn},
  \citenamefont {Carena}, \citenamefont {Ceccucci}, \citenamefont {Cerri},
  \citenamefont {Chakraborty}, \citenamefont {Chivukula}, \citenamefont
  {Cowan}, \citenamefont {D'Ambrosio}, \citenamefont {Damour}, \citenamefont
  {de~Florian}, \citenamefont {de~Gouv\^{e}a}, \citenamefont {DeGrand},
  \citenamefont {de~Jong}, \citenamefont {Dissertori}, \citenamefont
  {Dobrescu}, \citenamefont {D'Onofrio}, \citenamefont {Doser}, \citenamefont
  {Drees}, \citenamefont {Dreiner}, \citenamefont {Eerola}, \citenamefont
  {Egede}, \citenamefont {Eidelman}, \citenamefont {Ellis}, \citenamefont
  {Erler}, \citenamefont {Ezhela}, \citenamefont {Fetscher}, \citenamefont
  {Fields}, \citenamefont {Foster}, \citenamefont {Freitas}, \citenamefont
  {Gallagher}, \citenamefont {Garren}, \citenamefont {Gerber}, \citenamefont
  {Gerbier}, \citenamefont {Gershon}, \citenamefont {Gershtein}, \citenamefont
  {Gherghetta}, \citenamefont {Godizov}, \citenamefont {Gonzalez-Garcia},
  \citenamefont {Goodman}, \citenamefont {Grab}, \citenamefont {Gritsan},
  \citenamefont {Grojean}, \citenamefont {Gr\"{u}newald}, \citenamefont
  {Gurtu}, \citenamefont {Gutsche}, \citenamefont {Haber}, \citenamefont
  {Hanhart}, \citenamefont {Hashimoto}, \citenamefont {Hayato}, \citenamefont
  {Hebecker}, \citenamefont {Heinemeyer}, \citenamefont {Heltsley},
  \citenamefont {Hern\'{a}ndez-Rey}, \citenamefont {Hikasa}, \citenamefont
  {Hisano}, \citenamefont {H\"{o}cker}, \citenamefont {Holder}, \citenamefont
  {Holtkamp}, \citenamefont {Huston}, \citenamefont {Hyodo}, \citenamefont
  {Johnson}, \citenamefont {Kado}, \citenamefont {Karliner}, \citenamefont
  {Katz}, \citenamefont {Kenzie}, \citenamefont {Khoze}, \citenamefont {Klein},
  \citenamefont {Klempt}, \citenamefont {Kowalewski}, \citenamefont {Krauss},
  \citenamefont {Kreps}, \citenamefont {Krusche}, \citenamefont {Kwon},
  \citenamefont {Lahav}, \citenamefont {Laiho}, \citenamefont {Lellouch},
  \citenamefont {Lesgourgues}, \citenamefont {Liddle}, \citenamefont {Ligeti},
  \citenamefont {Lippmann}, \citenamefont {Liss}, \citenamefont {Littenberg},
  \citenamefont {Lourengo}, \citenamefont {Lugovsky}, \citenamefont {Lusiani},
  \citenamefont {Makida}, \citenamefont {Maltoni}, \citenamefont {Mannel},
  \citenamefont {Manohar}, \citenamefont {Marciano}, \citenamefont {Masoni},
  \citenamefont {Matthews}, \citenamefont {Mei{\ss}ner}, \citenamefont
  {Mikhasenko}, \citenamefont {Miller}, \citenamefont {Milstead}, \citenamefont
  {Mitchell}, \citenamefont {M\"{o}nig}, \citenamefont {Molaro}, \citenamefont
  {Moortgat}, \citenamefont {Moskovic}, \citenamefont {Nakamura}, \citenamefont
  {Narain}, \citenamefont {Nason}, \citenamefont {Navas}, \citenamefont
  {Neubert}, \citenamefont {Nevski}, \citenamefont {Nir}, \citenamefont
  {Olive}, \citenamefont {Patrignani}, \citenamefont {Peacock}, \citenamefont
  {Petcov}, \citenamefont {Petrov}, \citenamefont {Pich}, \citenamefont
  {Piepke}, \citenamefont {Pomarol}, \citenamefont {Profumo}, \citenamefont
  {Quadt}, \citenamefont {Rabbertz}, \citenamefont {Rademacker}, \citenamefont
  {Raffelt}, \citenamefont {Ramani}, \citenamefont {Ramsey-Musolf},
  \citenamefont {Ratcliff}, \citenamefont {Richardson}, \citenamefont
  {Ringwald}, \citenamefont {Roesler}, \citenamefont {Rolli}, \citenamefont
  {Romaniouk}, \citenamefont {Rosenberg}, \citenamefont {Rosner}, \citenamefont
  {Rybka}, \citenamefont {Ryskin}, \citenamefont {Ryutin}, \citenamefont
  {Sakai}, \citenamefont {Salam}, \citenamefont {Sarkar}, \citenamefont
  {Sauli}, \citenamefont {Schneider}, \citenamefont {Scholberg}, \citenamefont
  {Schwartz}, \citenamefont {Schwiening}, \citenamefont {Scott}, \citenamefont
  {Sharma}, \citenamefont {Sharpe}, \citenamefont {Shutt}, \citenamefont
  {Silari}, \citenamefont {Sj\"{o}strand}, \citenamefont {Skands},
  \citenamefont {Skwarnicki}, \citenamefont {Smoot}, \citenamefont {Soffer},
  \citenamefont {Sozzi}, \citenamefont {Spanier}, \citenamefont {Spiering},
  \citenamefont {Stahl}, \citenamefont {Stone}, \citenamefont {Sumino},
  \citenamefont {Sumiyoshi}, \citenamefont {Syphers}, \citenamefont
  {Takahashi}, \citenamefont {Tanabashi}, \citenamefont {Tanaka}, \citenamefont
  {Ta\v{s}evsk\'{y}}, \citenamefont {Terashi}, \citenamefont {Terning},
  \citenamefont {Thoma}, \citenamefont {Thorne}, \citenamefont {Tiator},
  \citenamefont {Titov}, \citenamefont {Tkachenko}, \citenamefont {Tovey},
  \citenamefont {Trabelsi}, \citenamefont {Urquijo}, \citenamefont {Valencia},
  \citenamefont {Van~de Water}, \citenamefont {Varelas}, \citenamefont
  {Venanzoni}, \citenamefont {Verde}, \citenamefont {Vincter}, \citenamefont
  {Vogel}, \citenamefont {Vogelsang}, \citenamefont {Vogt}, \citenamefont
  {Vorobyev}, \citenamefont {Wakely}, \citenamefont {Walkowiak}, \citenamefont
  {Walter}, \citenamefont {Wands}, \citenamefont {Wascko}, \citenamefont
  {Weinberg}, \citenamefont {Weinberg}, \citenamefont {White}, \citenamefont
  {Wiencke}, \citenamefont {Willocq}, \citenamefont {Woody}, \citenamefont
  {Workman}, \citenamefont {Yokoyama}, \citenamefont {Yoshida}, \citenamefont
  {Zanderighi}, \citenamefont {Zeller}, \citenamefont {Zenin}, \citenamefont
  {Zhu}, \citenamefont {Zhu}, \citenamefont {Zimmermann}, \citenamefont
  {Anderson}, \citenamefont {Basaglia}, \citenamefont {Lugovsky}, \citenamefont
  {Schaffner},\ and\ \citenamefont
  {Zheng}}]{Zyla2020Prog.Theor.Exp.Phys.2020_083C01}%
  \BibitemOpen
  \bibfield  {author} {\bibinfo {author} {\bibfnamefont {P.~A.}\ \bibnamefont
  {Zyla}}, \bibinfo {author} {\bibfnamefont {R.~M.}\ \bibnamefont {Barnett}},
  \bibinfo {author} {\bibfnamefont {J.}~\bibnamefont {Beringer}}, \bibinfo
  {author} {\bibfnamefont {O.}~\bibnamefont {Dahl}}, \bibinfo {author}
  {\bibfnamefont {D.~A.}\ \bibnamefont {Dwyer}}, \bibinfo {author}
  {\bibfnamefont {D.~E.}\ \bibnamefont {Groom}}, \bibinfo {author}
  {\bibfnamefont {C.-J.}\ \bibnamefont {Lin}}, \bibinfo {author} {\bibfnamefont
  {K.~S.}\ \bibnamefont {Lugovsky}}, \bibinfo {author} {\bibfnamefont
  {E.}~\bibnamefont {Pianori}}, \bibinfo {author} {\bibfnamefont {D.~J.}\
  \bibnamefont {Robinson}}, \bibinfo {author} {\bibfnamefont {C.~G.}\
  \bibnamefont {Wohl}}, \bibinfo {author} {\bibfnamefont {W.-M.}\ \bibnamefont
  {Yao}}, \bibinfo {author} {\bibfnamefont {K.}~\bibnamefont {Agashe}},
  \bibinfo {author} {\bibfnamefont {G.}~\bibnamefont {Aielli}}, \bibinfo
  {author} {\bibfnamefont {B.~C.}\ \bibnamefont {Allanach}}, \bibinfo {author}
  {\bibfnamefont {C.}~\bibnamefont {Amsler}}, \bibinfo {author} {\bibfnamefont
  {M.}~\bibnamefont {Antonelli}}, \bibinfo {author} {\bibfnamefont {E.~C.}\
  \bibnamefont {Aschenauer}}, \bibinfo {author} {\bibfnamefont {D.~M.}\
  \bibnamefont {Asner}}, \bibinfo {author} {\bibfnamefont {H.}~\bibnamefont
  {Baer}}, \bibinfo {author} {\bibfnamefont {S.}~\bibnamefont {Banerjee}},
  \bibinfo {author} {\bibfnamefont {L.}~\bibnamefont {Baudis}}, \bibinfo
  {author} {\bibfnamefont {C.~W.}\ \bibnamefont {Bauer}}, \bibinfo {author}
  {\bibfnamefont {J.~J.}\ \bibnamefont {Beatty}}, \bibinfo {author}
  {\bibfnamefont {V.~I.}\ \bibnamefont {Belousov}}, \bibinfo {author}
  {\bibfnamefont {S.}~\bibnamefont {Bethke}}, \bibinfo {author} {\bibfnamefont
  {A.}~\bibnamefont {Bettini}}, \bibinfo {author} {\bibfnamefont
  {O.}~\bibnamefont {Biebel}}, \bibinfo {author} {\bibfnamefont {K.~M.}\
  \bibnamefont {Black}}, \bibinfo {author} {\bibfnamefont {E.}~\bibnamefont
  {Blucher}}, \bibinfo {author} {\bibfnamefont {O.}~\bibnamefont {Buchmuller}},
  \bibinfo {author} {\bibfnamefont {V.}~\bibnamefont {Burkert}}, \bibinfo
  {author} {\bibfnamefont {M.~A.}\ \bibnamefont {Bychkov}}, \bibinfo {author}
  {\bibfnamefont {R.~N.}\ \bibnamefont {Cahn}}, \bibinfo {author}
  {\bibfnamefont {M.}~\bibnamefont {Carena}}, \bibinfo {author} {\bibfnamefont
  {A.}~\bibnamefont {Ceccucci}}, \bibinfo {author} {\bibfnamefont
  {A.}~\bibnamefont {Cerri}}, \bibinfo {author} {\bibfnamefont
  {D.}~\bibnamefont {Chakraborty}}, \bibinfo {author} {\bibfnamefont {R.~S.}\
  \bibnamefont {Chivukula}}, \bibinfo {author} {\bibfnamefont {G.}~\bibnamefont
  {Cowan}}, \bibinfo {author} {\bibfnamefont {G.}~\bibnamefont {D'Ambrosio}},
  \bibinfo {author} {\bibfnamefont {T.}~\bibnamefont {Damour}}, \bibinfo
  {author} {\bibfnamefont {D.}~\bibnamefont {de~Florian}}, \bibinfo {author}
  {\bibfnamefont {A.}~\bibnamefont {de~Gouv\^{e}a}}, \bibinfo {author}
  {\bibfnamefont {T.}~\bibnamefont {DeGrand}}, \bibinfo {author} {\bibfnamefont
  {P.}~\bibnamefont {de~Jong}}, \bibinfo {author} {\bibfnamefont
  {G.}~\bibnamefont {Dissertori}}, \bibinfo {author} {\bibfnamefont {B.~A.}\
  \bibnamefont {Dobrescu}}, \bibinfo {author} {\bibfnamefont {M.}~\bibnamefont
  {D'Onofrio}}, \bibinfo {author} {\bibfnamefont {M.}~\bibnamefont {Doser}},
  \bibinfo {author} {\bibfnamefont {M.}~\bibnamefont {Drees}}, \bibinfo
  {author} {\bibfnamefont {H.~K.}\ \bibnamefont {Dreiner}}, \bibinfo {author}
  {\bibfnamefont {P.}~\bibnamefont {Eerola}}, \bibinfo {author} {\bibfnamefont
  {U.}~\bibnamefont {Egede}}, \bibinfo {author} {\bibfnamefont
  {S.}~\bibnamefont {Eidelman}}, \bibinfo {author} {\bibfnamefont
  {J.}~\bibnamefont {Ellis}}, \bibinfo {author} {\bibfnamefont
  {J.}~\bibnamefont {Erler}}, \bibinfo {author} {\bibfnamefont {V.~V.}\
  \bibnamefont {Ezhela}}, \bibinfo {author} {\bibfnamefont {W.}~\bibnamefont
  {Fetscher}}, \bibinfo {author} {\bibfnamefont {B.~D.}\ \bibnamefont
  {Fields}}, \bibinfo {author} {\bibfnamefont {B.}~\bibnamefont {Foster}},
  \bibinfo {author} {\bibfnamefont {A.}~\bibnamefont {Freitas}}, \bibinfo
  {author} {\bibfnamefont {H.}~\bibnamefont {Gallagher}}, \bibinfo {author}
  {\bibfnamefont {L.}~\bibnamefont {Garren}}, \bibinfo {author} {\bibfnamefont
  {H.-J.}\ \bibnamefont {Gerber}}, \bibinfo {author} {\bibfnamefont
  {G.}~\bibnamefont {Gerbier}}, \bibinfo {author} {\bibfnamefont
  {T.}~\bibnamefont {Gershon}}, \bibinfo {author} {\bibfnamefont
  {Y.}~\bibnamefont {Gershtein}}, \bibinfo {author} {\bibfnamefont
  {T.}~\bibnamefont {Gherghetta}}, \bibinfo {author} {\bibfnamefont {A.~A.}\
  \bibnamefont {Godizov}}, \bibinfo {author} {\bibfnamefont {M.~C.}\
  \bibnamefont {Gonzalez-Garcia}}, \bibinfo {author} {\bibfnamefont
  {M.}~\bibnamefont {Goodman}}, \bibinfo {author} {\bibfnamefont
  {C.}~\bibnamefont {Grab}}, \bibinfo {author} {\bibfnamefont {A.~V.}\
  \bibnamefont {Gritsan}}, \bibinfo {author} {\bibfnamefont {C.}~\bibnamefont
  {Grojean}}, \bibinfo {author} {\bibfnamefont {M.}~\bibnamefont
  {Gr\"{u}newald}}, \bibinfo {author} {\bibfnamefont {A.}~\bibnamefont
  {Gurtu}}, \bibinfo {author} {\bibfnamefont {T.}~\bibnamefont {Gutsche}},
  \bibinfo {author} {\bibfnamefont {H.~E.}\ \bibnamefont {Haber}}, \bibinfo
  {author} {\bibfnamefont {C.}~\bibnamefont {Hanhart}}, \bibinfo {author}
  {\bibfnamefont {S.}~\bibnamefont {Hashimoto}}, \bibinfo {author}
  {\bibfnamefont {Y.}~\bibnamefont {Hayato}}, \bibinfo {author} {\bibfnamefont
  {A.}~\bibnamefont {Hebecker}}, \bibinfo {author} {\bibfnamefont
  {S.}~\bibnamefont {Heinemeyer}}, \bibinfo {author} {\bibfnamefont
  {B.}~\bibnamefont {Heltsley}}, \bibinfo {author} {\bibfnamefont {J.~J.}\
  \bibnamefont {Hern\'{a}ndez-Rey}}, \bibinfo {author} {\bibfnamefont
  {K.}~\bibnamefont {Hikasa}}, \bibinfo {author} {\bibfnamefont
  {J.}~\bibnamefont {Hisano}}, \bibinfo {author} {\bibfnamefont
  {A.}~\bibnamefont {H\"{o}cker}}, \bibinfo {author} {\bibfnamefont
  {J.}~\bibnamefont {Holder}}, \bibinfo {author} {\bibfnamefont
  {A.}~\bibnamefont {Holtkamp}}, \bibinfo {author} {\bibfnamefont
  {J.}~\bibnamefont {Huston}}, \bibinfo {author} {\bibfnamefont
  {T.}~\bibnamefont {Hyodo}}, \bibinfo {author} {\bibfnamefont {K.~F.}\
  \bibnamefont {Johnson}}, \bibinfo {author} {\bibfnamefont {M.}~\bibnamefont
  {Kado}}, \bibinfo {author} {\bibfnamefont {M.}~\bibnamefont {Karliner}},
  \bibinfo {author} {\bibfnamefont {U.~F.}\ \bibnamefont {Katz}}, \bibinfo
  {author} {\bibfnamefont {M.}~\bibnamefont {Kenzie}}, \bibinfo {author}
  {\bibfnamefont {V.~A.}\ \bibnamefont {Khoze}}, \bibinfo {author}
  {\bibfnamefont {S.~R.}\ \bibnamefont {Klein}}, \bibinfo {author}
  {\bibfnamefont {E.}~\bibnamefont {Klempt}}, \bibinfo {author} {\bibfnamefont
  {R.~V.}\ \bibnamefont {Kowalewski}}, \bibinfo {author} {\bibfnamefont
  {F.}~\bibnamefont {Krauss}}, \bibinfo {author} {\bibfnamefont
  {M.}~\bibnamefont {Kreps}}, \bibinfo {author} {\bibfnamefont
  {B.}~\bibnamefont {Krusche}}, \bibinfo {author} {\bibfnamefont
  {Y.}~\bibnamefont {Kwon}}, \bibinfo {author} {\bibfnamefont {O.}~\bibnamefont
  {Lahav}}, \bibinfo {author} {\bibfnamefont {J.}~\bibnamefont {Laiho}},
  \bibinfo {author} {\bibfnamefont {L.~P.}\ \bibnamefont {Lellouch}}, \bibinfo
  {author} {\bibfnamefont {J.}~\bibnamefont {Lesgourgues}}, \bibinfo {author}
  {\bibfnamefont {A.~R.}\ \bibnamefont {Liddle}}, \bibinfo {author}
  {\bibfnamefont {Z.}~\bibnamefont {Ligeti}}, \bibinfo {author} {\bibfnamefont
  {C.}~\bibnamefont {Lippmann}}, \bibinfo {author} {\bibfnamefont {T.~M.}\
  \bibnamefont {Liss}}, \bibinfo {author} {\bibfnamefont {L.}~\bibnamefont
  {Littenberg}}, \bibinfo {author} {\bibfnamefont {C.}~\bibnamefont
  {Lourengo}}, \bibinfo {author} {\bibfnamefont {S.~B.}\ \bibnamefont
  {Lugovsky}}, \bibinfo {author} {\bibfnamefont {A.}~\bibnamefont {Lusiani}},
  \bibinfo {author} {\bibfnamefont {Y.}~\bibnamefont {Makida}}, \bibinfo
  {author} {\bibfnamefont {F.}~\bibnamefont {Maltoni}}, \bibinfo {author}
  {\bibfnamefont {T.}~\bibnamefont {Mannel}}, \bibinfo {author} {\bibfnamefont
  {A.~V.}\ \bibnamefont {Manohar}}, \bibinfo {author} {\bibfnamefont {W.~J.}\
  \bibnamefont {Marciano}}, \bibinfo {author} {\bibfnamefont {A.}~\bibnamefont
  {Masoni}}, \bibinfo {author} {\bibfnamefont {J.}~\bibnamefont {Matthews}},
  \bibinfo {author} {\bibfnamefont {U.-G.}\ \bibnamefont {Mei{\ss}ner}},
  \bibinfo {author} {\bibfnamefont {M.}~\bibnamefont {Mikhasenko}}, \bibinfo
  {author} {\bibfnamefont {D.~J.}\ \bibnamefont {Miller}}, \bibinfo {author}
  {\bibfnamefont {D.}~\bibnamefont {Milstead}}, \bibinfo {author}
  {\bibfnamefont {R.~E.}\ \bibnamefont {Mitchell}}, \bibinfo {author}
  {\bibfnamefont {K.}~\bibnamefont {M\"{o}nig}}, \bibinfo {author}
  {\bibfnamefont {P.}~\bibnamefont {Molaro}}, \bibinfo {author} {\bibfnamefont
  {F.}~\bibnamefont {Moortgat}}, \bibinfo {author} {\bibfnamefont
  {M.}~\bibnamefont {Moskovic}}, \bibinfo {author} {\bibfnamefont
  {K.}~\bibnamefont {Nakamura}}, \bibinfo {author} {\bibfnamefont
  {M.}~\bibnamefont {Narain}}, \bibinfo {author} {\bibfnamefont
  {P.}~\bibnamefont {Nason}}, \bibinfo {author} {\bibfnamefont
  {S.}~\bibnamefont {Navas}}, \bibinfo {author} {\bibfnamefont
  {M.}~\bibnamefont {Neubert}}, \bibinfo {author} {\bibfnamefont
  {P.}~\bibnamefont {Nevski}}, \bibinfo {author} {\bibfnamefont
  {Y.}~\bibnamefont {Nir}}, \bibinfo {author} {\bibfnamefont {K.~A.}\
  \bibnamefont {Olive}}, \bibinfo {author} {\bibfnamefont {C.}~\bibnamefont
  {Patrignani}}, \bibinfo {author} {\bibfnamefont {J.~A.}\ \bibnamefont
  {Peacock}}, \bibinfo {author} {\bibfnamefont {S.~T.}\ \bibnamefont {Petcov}},
  \bibinfo {author} {\bibfnamefont {V.~A.}\ \bibnamefont {Petrov}}, \bibinfo
  {author} {\bibfnamefont {A.}~\bibnamefont {Pich}}, \bibinfo {author}
  {\bibfnamefont {A.}~\bibnamefont {Piepke}}, \bibinfo {author} {\bibfnamefont
  {A.}~\bibnamefont {Pomarol}}, \bibinfo {author} {\bibfnamefont
  {S.}~\bibnamefont {Profumo}}, \bibinfo {author} {\bibfnamefont
  {A.}~\bibnamefont {Quadt}}, \bibinfo {author} {\bibfnamefont
  {K.}~\bibnamefont {Rabbertz}}, \bibinfo {author} {\bibfnamefont
  {J.}~\bibnamefont {Rademacker}}, \bibinfo {author} {\bibfnamefont
  {G.}~\bibnamefont {Raffelt}}, \bibinfo {author} {\bibfnamefont
  {H.}~\bibnamefont {Ramani}}, \bibinfo {author} {\bibfnamefont
  {M.}~\bibnamefont {Ramsey-Musolf}}, \bibinfo {author} {\bibfnamefont {B.~N.}\
  \bibnamefont {Ratcliff}}, \bibinfo {author} {\bibfnamefont {P.}~\bibnamefont
  {Richardson}}, \bibinfo {author} {\bibfnamefont {A.}~\bibnamefont
  {Ringwald}}, \bibinfo {author} {\bibfnamefont {S.}~\bibnamefont {Roesler}},
  \bibinfo {author} {\bibfnamefont {S.}~\bibnamefont {Rolli}}, \bibinfo
  {author} {\bibfnamefont {A.}~\bibnamefont {Romaniouk}}, \bibinfo {author}
  {\bibfnamefont {L.~J.}\ \bibnamefont {Rosenberg}}, \bibinfo {author}
  {\bibfnamefont {J.~L.}\ \bibnamefont {Rosner}}, \bibinfo {author}
  {\bibfnamefont {G.}~\bibnamefont {Rybka}}, \bibinfo {author} {\bibfnamefont
  {M.}~\bibnamefont {Ryskin}}, \bibinfo {author} {\bibfnamefont {R.~A.}\
  \bibnamefont {Ryutin}}, \bibinfo {author} {\bibfnamefont {Y.}~\bibnamefont
  {Sakai}}, \bibinfo {author} {\bibfnamefont {G.~P.}\ \bibnamefont {Salam}},
  \bibinfo {author} {\bibfnamefont {S.}~\bibnamefont {Sarkar}}, \bibinfo
  {author} {\bibfnamefont {F.}~\bibnamefont {Sauli}}, \bibinfo {author}
  {\bibfnamefont {O.}~\bibnamefont {Schneider}}, \bibinfo {author}
  {\bibfnamefont {K.}~\bibnamefont {Scholberg}}, \bibinfo {author}
  {\bibfnamefont {A.~J.}\ \bibnamefont {Schwartz}}, \bibinfo {author}
  {\bibfnamefont {J.}~\bibnamefont {Schwiening}}, \bibinfo {author}
  {\bibfnamefont {D.}~\bibnamefont {Scott}}, \bibinfo {author} {\bibfnamefont
  {V.}~\bibnamefont {Sharma}}, \bibinfo {author} {\bibfnamefont {S.~R.}\
  \bibnamefont {Sharpe}}, \bibinfo {author} {\bibfnamefont {T.}~\bibnamefont
  {Shutt}}, \bibinfo {author} {\bibfnamefont {M.}~\bibnamefont {Silari}},
  \bibinfo {author} {\bibfnamefont {T.}~\bibnamefont {Sj\"{o}strand}}, \bibinfo
  {author} {\bibfnamefont {P.}~\bibnamefont {Skands}}, \bibinfo {author}
  {\bibfnamefont {T.}~\bibnamefont {Skwarnicki}}, \bibinfo {author}
  {\bibfnamefont {G.~F.}\ \bibnamefont {Smoot}}, \bibinfo {author}
  {\bibfnamefont {A.}~\bibnamefont {Soffer}}, \bibinfo {author} {\bibfnamefont
  {M.~S.}\ \bibnamefont {Sozzi}}, \bibinfo {author} {\bibfnamefont
  {S.}~\bibnamefont {Spanier}}, \bibinfo {author} {\bibfnamefont
  {C.}~\bibnamefont {Spiering}}, \bibinfo {author} {\bibfnamefont
  {A.}~\bibnamefont {Stahl}}, \bibinfo {author} {\bibfnamefont {S.~L.}\
  \bibnamefont {Stone}}, \bibinfo {author} {\bibfnamefont {Y.}~\bibnamefont
  {Sumino}}, \bibinfo {author} {\bibfnamefont {T.}~\bibnamefont {Sumiyoshi}},
  \bibinfo {author} {\bibfnamefont {M.~J.}\ \bibnamefont {Syphers}}, \bibinfo
  {author} {\bibfnamefont {F.}~\bibnamefont {Takahashi}}, \bibinfo {author}
  {\bibfnamefont {M.}~\bibnamefont {Tanabashi}}, \bibinfo {author}
  {\bibfnamefont {J.}~\bibnamefont {Tanaka}}, \bibinfo {author} {\bibfnamefont
  {M.}~\bibnamefont {Ta\v{s}evsk\'{y}}}, \bibinfo {author} {\bibfnamefont
  {K.}~\bibnamefont {Terashi}}, \bibinfo {author} {\bibfnamefont
  {J.}~\bibnamefont {Terning}}, \bibinfo {author} {\bibfnamefont
  {U.}~\bibnamefont {Thoma}}, \bibinfo {author} {\bibfnamefont {R.~S.}\
  \bibnamefont {Thorne}}, \bibinfo {author} {\bibfnamefont {L.}~\bibnamefont
  {Tiator}}, \bibinfo {author} {\bibfnamefont {M.}~\bibnamefont {Titov}},
  \bibinfo {author} {\bibfnamefont {N.~P.}\ \bibnamefont {Tkachenko}}, \bibinfo
  {author} {\bibfnamefont {D.~R.}\ \bibnamefont {Tovey}}, \bibinfo {author}
  {\bibfnamefont {K.}~\bibnamefont {Trabelsi}}, \bibinfo {author}
  {\bibfnamefont {P.}~\bibnamefont {Urquijo}}, \bibinfo {author} {\bibfnamefont
  {G.}~\bibnamefont {Valencia}}, \bibinfo {author} {\bibfnamefont
  {R.}~\bibnamefont {Van~de Water}}, \bibinfo {author} {\bibfnamefont
  {N.}~\bibnamefont {Varelas}}, \bibinfo {author} {\bibfnamefont
  {G.}~\bibnamefont {Venanzoni}}, \bibinfo {author} {\bibfnamefont
  {L.}~\bibnamefont {Verde}}, \bibinfo {author} {\bibfnamefont {M.~G.}\
  \bibnamefont {Vincter}}, \bibinfo {author} {\bibfnamefont {P.}~\bibnamefont
  {Vogel}}, \bibinfo {author} {\bibfnamefont {W.}~\bibnamefont {Vogelsang}},
  \bibinfo {author} {\bibfnamefont {A.}~\bibnamefont {Vogt}}, \bibinfo {author}
  {\bibfnamefont {V.}~\bibnamefont {Vorobyev}}, \bibinfo {author}
  {\bibfnamefont {S.~P.}\ \bibnamefont {Wakely}}, \bibinfo {author}
  {\bibfnamefont {W.}~\bibnamefont {Walkowiak}}, \bibinfo {author}
  {\bibfnamefont {C.~W.}\ \bibnamefont {Walter}}, \bibinfo {author}
  {\bibfnamefont {D.}~\bibnamefont {Wands}}, \bibinfo {author} {\bibfnamefont
  {M.~O.}\ \bibnamefont {Wascko}}, \bibinfo {author} {\bibfnamefont {D.~H.}\
  \bibnamefont {Weinberg}}, \bibinfo {author} {\bibfnamefont {E.~J.}\
  \bibnamefont {Weinberg}}, \bibinfo {author} {\bibfnamefont {M.}~\bibnamefont
  {White}}, \bibinfo {author} {\bibfnamefont {L.~R.}\ \bibnamefont {Wiencke}},
  \bibinfo {author} {\bibfnamefont {S.}~\bibnamefont {Willocq}}, \bibinfo
  {author} {\bibfnamefont {C.~L.}\ \bibnamefont {Woody}}, \bibinfo {author}
  {\bibfnamefont {R.~L.}\ \bibnamefont {Workman}}, \bibinfo {author}
  {\bibfnamefont {M.}~\bibnamefont {Yokoyama}}, \bibinfo {author}
  {\bibfnamefont {R.}~\bibnamefont {Yoshida}}, \bibinfo {author} {\bibfnamefont
  {G.}~\bibnamefont {Zanderighi}}, \bibinfo {author} {\bibfnamefont {G.~P.}\
  \bibnamefont {Zeller}}, \bibinfo {author} {\bibfnamefont {O.~V.}\
  \bibnamefont {Zenin}}, \bibinfo {author} {\bibfnamefont {R.-Y.}\ \bibnamefont
  {Zhu}}, \bibinfo {author} {\bibfnamefont {S.-L.}\ \bibnamefont {Zhu}},
  \bibinfo {author} {\bibfnamefont {F.}~\bibnamefont {Zimmermann}}, \bibinfo
  {author} {\bibfnamefont {J.}~\bibnamefont {Anderson}}, \bibinfo {author}
  {\bibfnamefont {T.}~\bibnamefont {Basaglia}}, \bibinfo {author}
  {\bibfnamefont {V.~S.}\ \bibnamefont {Lugovsky}}, \bibinfo {author}
  {\bibfnamefont {P.}~\bibnamefont {Schaffner}},\ and\ \bibinfo {author}
  {\bibfnamefont {W.}~\bibnamefont {Zheng}} (\bibinfo {collaboration} {Particle
  Data Group}),\ }\bibfield  {title} {\bibinfo {title} {{Review of Particle
  Physics}},\ }\href {https://doi.org/10.1093/ptep/ptaa104} {\bibfield
  {journal} {\bibinfo  {journal} {Prog. Theor. Exp. Phys.}\ }\textbf {\bibinfo
  {volume} {2020}},\ \bibinfo {pages} {083C01} (\bibinfo {year}
  {2020})}\BibitemShut {NoStop}%
\bibitem [{\citenamefont {Naito}\ \emph {et~al.}(2021)\citenamefont {Naito},
  \citenamefont {Col\`o}, \citenamefont {Liang},\ and\ \citenamefont
  {Roca-Maza}}]{Naito2021Phys.Rev.C104_024316}%
  \BibitemOpen
  \bibfield  {author} {\bibinfo {author} {\bibfnamefont {T.}~\bibnamefont
  {Naito}}, \bibinfo {author} {\bibfnamefont {G.}~\bibnamefont {Col\`o}},
  \bibinfo {author} {\bibfnamefont {H.}~\bibnamefont {Liang}},\ and\ \bibinfo
  {author} {\bibfnamefont {X.}~\bibnamefont {Roca-Maza}},\ }\bibfield  {title}
  {\bibinfo {title} {{Second and fourth moments of the charge density and
  neutron-skin thickness of atomic nuclei}},\ }\href
  {https://doi.org/10.1103/PhysRevC.104.024316} {\bibfield  {journal} {\bibinfo
   {journal} {Phys. Rev. C}\ }\textbf {\bibinfo {volume} {104}},\ \bibinfo
  {pages} {024316} (\bibinfo {year} {2021})}\BibitemShut {NoStop}%
\bibitem [{\citenamefont {Gorges}\ \emph {et~al.}(2019)\citenamefont {Gorges},
  \citenamefont {Rodr\'{\i}guez}, \citenamefont {Balabanski}, \citenamefont
  {Bissell}, \citenamefont {Blaum}, \citenamefont {Cheal}, \citenamefont
  {Garcia~Ruiz}, \citenamefont {Georgiev}, \citenamefont {Gins}, \citenamefont
  {Heylen}, \citenamefont {Kanellakopoulos}, \citenamefont {Kaufmann},
  \citenamefont {Kowalska}, \citenamefont {Lagaki}, \citenamefont {Lechner},
  \citenamefont {Maa\ss{}}, \citenamefont {Malbrunot-Ettenauer}, \citenamefont
  {Nazarewicz}, \citenamefont {Neugart}, \citenamefont {Neyens}, \citenamefont
  {N\"ortersh\"auser}, \citenamefont {Reinhard}, \citenamefont {Sailer},
  \citenamefont {S\'anchez}, \citenamefont {Schmidt}, \citenamefont {Wehner},
  \citenamefont {Wraith}, \citenamefont {Xie}, \citenamefont {Xu},
  \citenamefont {Yang},\ and\ \citenamefont
  {Yordanov}}]{Gorges2019Phys.Rev.Lett.122_192502}%
  \BibitemOpen
  \bibfield  {author} {\bibinfo {author} {\bibfnamefont {C.}~\bibnamefont
  {Gorges}}, \bibinfo {author} {\bibfnamefont {L.~V.}\ \bibnamefont
  {Rodr\'{\i}guez}}, \bibinfo {author} {\bibfnamefont {D.~L.}\ \bibnamefont
  {Balabanski}}, \bibinfo {author} {\bibfnamefont {M.~L.}\ \bibnamefont
  {Bissell}}, \bibinfo {author} {\bibfnamefont {K.}~\bibnamefont {Blaum}},
  \bibinfo {author} {\bibfnamefont {B.}~\bibnamefont {Cheal}}, \bibinfo
  {author} {\bibfnamefont {R.~F.}\ \bibnamefont {Garcia~Ruiz}}, \bibinfo
  {author} {\bibfnamefont {G.}~\bibnamefont {Georgiev}}, \bibinfo {author}
  {\bibfnamefont {W.}~\bibnamefont {Gins}}, \bibinfo {author} {\bibfnamefont
  {H.}~\bibnamefont {Heylen}}, \bibinfo {author} {\bibfnamefont
  {A.}~\bibnamefont {Kanellakopoulos}}, \bibinfo {author} {\bibfnamefont
  {S.}~\bibnamefont {Kaufmann}}, \bibinfo {author} {\bibfnamefont
  {M.}~\bibnamefont {Kowalska}}, \bibinfo {author} {\bibfnamefont
  {V.}~\bibnamefont {Lagaki}}, \bibinfo {author} {\bibfnamefont
  {S.}~\bibnamefont {Lechner}}, \bibinfo {author} {\bibfnamefont
  {B.}~\bibnamefont {Maa\ss{}}}, \bibinfo {author} {\bibfnamefont
  {S.}~\bibnamefont {Malbrunot-Ettenauer}}, \bibinfo {author} {\bibfnamefont
  {W.}~\bibnamefont {Nazarewicz}}, \bibinfo {author} {\bibfnamefont
  {R.}~\bibnamefont {Neugart}}, \bibinfo {author} {\bibfnamefont
  {G.}~\bibnamefont {Neyens}}, \bibinfo {author} {\bibfnamefont
  {W.}~\bibnamefont {N\"ortersh\"auser}}, \bibinfo {author} {\bibfnamefont
  {P.-G.}\ \bibnamefont {Reinhard}}, \bibinfo {author} {\bibfnamefont
  {S.}~\bibnamefont {Sailer}}, \bibinfo {author} {\bibfnamefont
  {R.}~\bibnamefont {S\'anchez}}, \bibinfo {author} {\bibfnamefont
  {S.}~\bibnamefont {Schmidt}}, \bibinfo {author} {\bibfnamefont
  {L.}~\bibnamefont {Wehner}}, \bibinfo {author} {\bibfnamefont
  {C.}~\bibnamefont {Wraith}}, \bibinfo {author} {\bibfnamefont
  {L.}~\bibnamefont {Xie}}, \bibinfo {author} {\bibfnamefont {Z.~Y.}\
  \bibnamefont {Xu}}, \bibinfo {author} {\bibfnamefont {X.~F.}\ \bibnamefont
  {Yang}},\ and\ \bibinfo {author} {\bibfnamefont {D.~T.}\ \bibnamefont
  {Yordanov}},\ }\bibfield  {title} {\bibinfo {title} {{Laser Spectroscopy of
  Neutron-Rich Tin Isotopes: A Discontinuity in Charge Radii across the $N=82$
  Shell Closure}},\ }\href {https://doi.org/10.1103/PhysRevLett.122.192502}
  {\bibfield  {journal} {\bibinfo  {journal} {Phys. Rev. Lett.}\ }\textbf
  {\bibinfo {volume} {122}},\ \bibinfo {pages} {192502} (\bibinfo {year}
  {2019})}\BibitemShut {NoStop}%
\bibitem [{\citenamefont {Pearson}\ and\ \citenamefont
  {Farine}(1994)}]{Pearson1994Phys.Rev.C50_185}%
  \BibitemOpen
  \bibfield  {author} {\bibinfo {author} {\bibfnamefont {J.~M.}\ \bibnamefont
  {Pearson}}\ and\ \bibinfo {author} {\bibfnamefont {M.}~\bibnamefont
  {Farine}},\ }\bibfield  {title} {\bibinfo {title} {{Relativistic mean-field
  theory and a density-dependent spin-orbit Skyrme force}},\ }\href
  {https://doi.org/10.1103/PhysRevC.50.185} {\bibfield  {journal} {\bibinfo
  {journal} {Phys. Rev. C}\ }\textbf {\bibinfo {volume} {50}},\ \bibinfo
  {pages} {185} (\bibinfo {year} {1994})}\BibitemShut {NoStop}%
\bibitem [{\citenamefont {Pudliner}\ \emph {et~al.}(1996)\citenamefont
  {Pudliner}, \citenamefont {Smerzi}, \citenamefont {Carlson}, \citenamefont
  {Pandharipande}, \citenamefont {Pieper},\ and\ \citenamefont
  {Ravenhall}}]{Pudliner1996Phys.Rev.Lett.76_2416}%
  \BibitemOpen
  \bibfield  {author} {\bibinfo {author} {\bibfnamefont {B.~S.}\ \bibnamefont
  {Pudliner}}, \bibinfo {author} {\bibfnamefont {A.}~\bibnamefont {Smerzi}},
  \bibinfo {author} {\bibfnamefont {J.}~\bibnamefont {Carlson}}, \bibinfo
  {author} {\bibfnamefont {V.~R.}\ \bibnamefont {Pandharipande}}, \bibinfo
  {author} {\bibfnamefont {S.~C.}\ \bibnamefont {Pieper}},\ and\ \bibinfo
  {author} {\bibfnamefont {D.~G.}\ \bibnamefont {Ravenhall}},\ }\bibfield
  {title} {\bibinfo {title} {{Neutron Drops and Skyrme Energy-Density
  Functionals}},\ }\href {https://doi.org/10.1103/PhysRevLett.76.2416}
  {\bibfield  {journal} {\bibinfo  {journal} {Phys. Rev. Lett.}\ }\textbf
  {\bibinfo {volume} {76}},\ \bibinfo {pages} {2416} (\bibinfo {year}
  {1996})}\BibitemShut {NoStop}%
\bibitem [{\citenamefont {Bender}\ \emph {et~al.}(2003)\citenamefont {Bender},
  \citenamefont {Heenen},\ and\ \citenamefont
  {Reinhard}}]{Bender2003Rev.Mod.Phys.75_121}%
  \BibitemOpen
  \bibfield  {author} {\bibinfo {author} {\bibfnamefont {M.}~\bibnamefont
  {Bender}}, \bibinfo {author} {\bibfnamefont {P.-H.}\ \bibnamefont {Heenen}},\
  and\ \bibinfo {author} {\bibfnamefont {P.-G.}\ \bibnamefont {Reinhard}},\
  }\bibfield  {title} {\bibinfo {title} {{Self-consistent mean-field models for
  nuclear structure}},\ }\href {https://doi.org/10.1103/RevModPhys.75.121}
  {\bibfield  {journal} {\bibinfo  {journal} {Rev. Mod. Phys.}\ }\textbf
  {\bibinfo {volume} {75}},\ \bibinfo {pages} {121} (\bibinfo {year}
  {2003})}\BibitemShut {NoStop}%
\bibitem [{\citenamefont {Kanada-En'yo}(2022)}]{Kanada-Enyo:2022dhj}%
  \BibitemOpen
  \bibfield  {author} {\bibinfo {author} {\bibfnamefont {Y.}~\bibnamefont
  {Kanada-En'yo}},\ }\bibfield  {title} {\bibinfo {title} {{Effects of
  density-dependent spin-orbit interactions in Skyrme-Hartree-Fock-Bogoliubov
  calculations of the charge radii and densities of Pb isotopes}},\ }\Eprint
  {https://arxiv.org/abs/2209.11411} {arXiv:2209.11411 [nucl-th]}  (\bibinfo
  {year} {2022})\BibitemShut {NoStop}%
\bibitem [{\citenamefont {Sharma}\ \emph {et~al.}(1995)\citenamefont {Sharma},
  \citenamefont {Lalazissis}, \citenamefont {K\"onig},\ and\ \citenamefont
  {Ring}}]{Sharma1995Phys.Rev.Lett.74_3744}%
  \BibitemOpen
  \bibfield  {author} {\bibinfo {author} {\bibfnamefont {M.~M.}\ \bibnamefont
  {Sharma}}, \bibinfo {author} {\bibfnamefont {G.}~\bibnamefont {Lalazissis}},
  \bibinfo {author} {\bibfnamefont {J.}~\bibnamefont {K\"onig}},\ and\ \bibinfo
  {author} {\bibfnamefont {P.}~\bibnamefont {Ring}},\ }\bibfield  {title}
  {\bibinfo {title} {{Isospin Dependence of the Spin-Orbit Force and Effective
  Nuclear Potentials}},\ }\href {https://doi.org/10.1103/PhysRevLett.74.3744}
  {\bibfield  {journal} {\bibinfo  {journal} {Phys. Rev. Lett.}\ }\textbf
  {\bibinfo {volume} {74}},\ \bibinfo {pages} {3744} (\bibinfo {year}
  {1995})}\BibitemShut {NoStop}%
\bibitem [{\citenamefont {Reinhard}(1989)}]{Reinhard1989Rep.Prog.Phys.52_439}%
  \BibitemOpen
  \bibfield  {author} {\bibinfo {author} {\bibfnamefont {P.-G.}\ \bibnamefont
  {Reinhard}},\ }\bibfield  {title} {\bibinfo {title} {{The relativistic
  mean-field description of nuclei and nuclear dynamics}},\ }\href
  {https://doi.org/10.1088/0034-4885/52/4/002} {\bibfield  {journal} {\bibinfo
  {journal} {Rep. Prog. Phys.}\ }\textbf {\bibinfo {volume} {52}},\ \bibinfo
  {pages} {439} (\bibinfo {year} {1989})}\BibitemShut {NoStop}%
\bibitem [{\citenamefont {Onsi}\ \emph {et~al.}(1997)\citenamefont {Onsi},
  \citenamefont {Nayak}, \citenamefont {Pearson}, \citenamefont {Freyer},\ and\
  \citenamefont {Stocker}}]{Onsi1997Phys.Rev.C55_3166}%
  \BibitemOpen
  \bibfield  {author} {\bibinfo {author} {\bibfnamefont {M.}~\bibnamefont
  {Onsi}}, \bibinfo {author} {\bibfnamefont {R.~C.}\ \bibnamefont {Nayak}},
  \bibinfo {author} {\bibfnamefont {J.~M.}\ \bibnamefont {Pearson}}, \bibinfo
  {author} {\bibfnamefont {H.}~\bibnamefont {Freyer}},\ and\ \bibinfo {author}
  {\bibfnamefont {W.}~\bibnamefont {Stocker}},\ }\bibfield  {title} {\bibinfo
  {title} {{Skyrme representation of a relativistic spin-orbit field}},\ }\href
  {https://doi.org/10.1103/PhysRevC.55.3166} {\bibfield  {journal} {\bibinfo
  {journal} {Phys. Rev. C}\ }\textbf {\bibinfo {volume} {55}},\ \bibinfo
  {pages} {3166} (\bibinfo {year} {1997})}\BibitemShut {NoStop}%
\bibitem [{\citenamefont {Nayak}\ and\ \citenamefont
  {Pearson}(1998)}]{Nayak1998Phys.Rev.C58_878}%
  \BibitemOpen
  \bibfield  {author} {\bibinfo {author} {\bibfnamefont {R.~C.}\ \bibnamefont
  {Nayak}}\ and\ \bibinfo {author} {\bibfnamefont {J.~M.}\ \bibnamefont
  {Pearson}},\ }\bibfield  {title} {\bibinfo {title} {{Spin-orbit field and
  extrapolated properties of exotic nuclei}},\ }\href
  {https://doi.org/10.1103/PhysRevC.58.878} {\bibfield  {journal} {\bibinfo
  {journal} {Phys. Rev. C}\ }\textbf {\bibinfo {volume} {58}},\ \bibinfo
  {pages} {878} (\bibinfo {year} {1998})}\BibitemShut {NoStop}%
\bibitem [{\citenamefont {Pearson}(2001)}]{Pearson2001Phys.Lett.B513_319}%
  \BibitemOpen
  \bibfield  {author} {\bibinfo {author} {\bibfnamefont {J.~M.}\ \bibnamefont
  {Pearson}},\ }\bibfield  {title} {\bibinfo {title} {{Skyrme Hartree--Fock
  method and the spin--orbit term of the relativistic mean field}},\ }\href
  {https://doi.org/10.1016/S0370-2693(01)00375-6} {\bibfield  {journal}
  {\bibinfo  {journal} {Phys. Lett. B}\ }\textbf {\bibinfo {volume} {513}},\
  \bibinfo {pages} {319} (\bibinfo {year} {2001})}\BibitemShut {NoStop}%
\bibitem [{\citenamefont {Geng}\ \emph {et~al.}(2019)\citenamefont {Geng},
  \citenamefont {Li}, \citenamefont {Long}, \citenamefont {Niu},\ and\
  \citenamefont {Chang}}]{Geng2019Phys.Rev.C100_051301}%
  \BibitemOpen
  \bibfield  {author} {\bibinfo {author} {\bibfnamefont {J.}~\bibnamefont
  {Geng}}, \bibinfo {author} {\bibfnamefont {J.~J.}\ \bibnamefont {Li}},
  \bibinfo {author} {\bibfnamefont {W.~H.}\ \bibnamefont {Long}}, \bibinfo
  {author} {\bibfnamefont {Y.~F.}\ \bibnamefont {Niu}},\ and\ \bibinfo {author}
  {\bibfnamefont {S.~Y.}\ \bibnamefont {Chang}},\ }\bibfield  {title} {\bibinfo
  {title} {{Pseudospin symmetry restoration and the in-medium balance between
  nuclear attractive and repulsive interactions}},\ }\href
  {https://doi.org/10.1103/PhysRevC.100.051301} {\bibfield  {journal} {\bibinfo
   {journal} {Phys. Rev. C}\ }\textbf {\bibinfo {volume} {100}},\ \bibinfo
  {pages} {051301} (\bibinfo {year} {2019})}\BibitemShut {NoStop}%
\bibitem [{\citenamefont {Liang}\ \emph {et~al.}(2015)\citenamefont {Liang},
  \citenamefont {Meng},\ and\ \citenamefont {Zhou}}]{Liang2015Phys.Rep.570_1}%
  \BibitemOpen
  \bibfield  {author} {\bibinfo {author} {\bibfnamefont {H.}~\bibnamefont
  {Liang}}, \bibinfo {author} {\bibfnamefont {J.}~\bibnamefont {Meng}},\ and\
  \bibinfo {author} {\bibfnamefont {S.-G.}\ \bibnamefont {Zhou}},\ }\bibfield
  {title} {\bibinfo {title} {{Hidden pseudospin and spin symmetries and their
  origins in atomic nuclei}},\ }\href
  {https://doi.org/10.1016/j.physrep.2014.12.005} {\bibfield  {journal}
  {\bibinfo  {journal} {Phys. Rep.}\ }\textbf {\bibinfo {volume} {570}},\
  \bibinfo {pages} {1} (\bibinfo {year} {2015})}\BibitemShut {NoStop}%
\bibitem [{\citenamefont {Afanasjev}(2008)}]{Afanasjev2008Phys.Rev.C78_054303}%
  \BibitemOpen
  \bibfield  {author} {\bibinfo {author} {\bibfnamefont {A.~V.}\ \bibnamefont
  {Afanasjev}},\ }\bibfield  {title} {\bibinfo {title} {{Band terminations in
  density functional theory}},\ }\href
  {https://doi.org/10.1103/PhysRevC.78.054303} {\bibfield  {journal} {\bibinfo
  {journal} {Phys. Rev. C}\ }\textbf {\bibinfo {volume} {78}},\ \bibinfo
  {pages} {054303} (\bibinfo {year} {2008})}\BibitemShut {NoStop}%
\bibitem [{\citenamefont {Satu\l{}a}\ \emph {et~al.}(2008)\citenamefont
  {Satu\l{}a}, \citenamefont {Wyss},\ and\ \citenamefont
  {Zalewski}}]{Satula2008Phys.Rev.C78_011302}%
  \BibitemOpen
  \bibfield  {author} {\bibinfo {author} {\bibfnamefont {W.}~\bibnamefont
  {Satu\l{}a}}, \bibinfo {author} {\bibfnamefont {R.~A.}\ \bibnamefont
  {Wyss}},\ and\ \bibinfo {author} {\bibfnamefont {M.}~\bibnamefont
  {Zalewski}},\ }\bibfield  {title} {\bibinfo {title} {{Contradicting effective
  mass scalings in the single-particle spectra calculated using the Skyrme
  energy density functional method}},\ }\href
  {https://doi.org/10.1103/PhysRevC.78.011302} {\bibfield  {journal} {\bibinfo
  {journal} {Phys. Rev. C}\ }\textbf {\bibinfo {volume} {78}},\ \bibinfo
  {pages} {011302} (\bibinfo {year} {2008})}\BibitemShut {NoStop}%
\bibitem [{\citenamefont {Garcia~Ruiz}\ \emph {et~al.}(2016)\citenamefont
  {Garcia~Ruiz}, \citenamefont {Bissell}, \citenamefont {Blaum}, \citenamefont
  {Ekstr\"{o}m}, \citenamefont {Fr\"{o}mmgen}, \citenamefont {Hagen},
  \citenamefont {Hammen}, \citenamefont {Hebeler}, \citenamefont {Holt},
  \citenamefont {Jansen}, \citenamefont {Kowalska}, \citenamefont {Kreim},
  \citenamefont {Nazarewicz}, \citenamefont {Neugart}, \citenamefont {Neyens},
  \citenamefont {N\"{o}rtersh\"{a}user}, \citenamefont {Papenbrock},
  \citenamefont {Papuga}, \citenamefont {Schwenk}, \citenamefont {Simonis},
  \citenamefont {Wendt},\ and\ \citenamefont
  {Yordanov}}]{GarciaRuiz2016Nat.Phys.12_594}%
  \BibitemOpen
  \bibfield  {author} {\bibinfo {author} {\bibfnamefont {R.~F.}\ \bibnamefont
  {Garcia~Ruiz}}, \bibinfo {author} {\bibfnamefont {M.~L.}\ \bibnamefont
  {Bissell}}, \bibinfo {author} {\bibfnamefont {K.}~\bibnamefont {Blaum}},
  \bibinfo {author} {\bibfnamefont {A.}~\bibnamefont {Ekstr\"{o}m}}, \bibinfo
  {author} {\bibfnamefont {N.}~\bibnamefont {Fr\"{o}mmgen}}, \bibinfo {author}
  {\bibfnamefont {G.}~\bibnamefont {Hagen}}, \bibinfo {author} {\bibfnamefont
  {M.}~\bibnamefont {Hammen}}, \bibinfo {author} {\bibfnamefont
  {K.}~\bibnamefont {Hebeler}}, \bibinfo {author} {\bibfnamefont {J.~D.}\
  \bibnamefont {Holt}}, \bibinfo {author} {\bibfnamefont {G.~R.}\ \bibnamefont
  {Jansen}}, \bibinfo {author} {\bibfnamefont {M.}~\bibnamefont {Kowalska}},
  \bibinfo {author} {\bibfnamefont {K.}~\bibnamefont {Kreim}}, \bibinfo
  {author} {\bibfnamefont {W.}~\bibnamefont {Nazarewicz}}, \bibinfo {author}
  {\bibfnamefont {R.}~\bibnamefont {Neugart}}, \bibinfo {author} {\bibfnamefont
  {G.}~\bibnamefont {Neyens}}, \bibinfo {author} {\bibfnamefont
  {W.}~\bibnamefont {N\"{o}rtersh\"{a}user}}, \bibinfo {author} {\bibfnamefont
  {T.}~\bibnamefont {Papenbrock}}, \bibinfo {author} {\bibfnamefont
  {J.}~\bibnamefont {Papuga}}, \bibinfo {author} {\bibfnamefont
  {A.}~\bibnamefont {Schwenk}}, \bibinfo {author} {\bibfnamefont
  {J.}~\bibnamefont {Simonis}}, \bibinfo {author} {\bibfnamefont {K.~A.}\
  \bibnamefont {Wendt}},\ and\ \bibinfo {author} {\bibfnamefont {D.~T.}\
  \bibnamefont {Yordanov}},\ }\bibfield  {title} {\bibinfo {title}
  {{Unexpectedly large charge radii of neutron-rich calcium isotopes}},\ }\href
  {https://doi.org/10.1038/NPHYS3645} {\bibfield  {journal} {\bibinfo
  {journal} {Nat. Phys.}\ }\textbf {\bibinfo {volume} {12}},\ \bibinfo {pages}
  {594} (\bibinfo {year} {2016})}\BibitemShut {NoStop}%
\bibitem [{\citenamefont {Miller}\ \emph {et~al.}(2019)\citenamefont {Miller},
  \citenamefont {Minamisono}, \citenamefont {Klose}, \citenamefont {Garand},
  \citenamefont {Kujawa}, \citenamefont {Lantis}, \citenamefont {Liu},
  \citenamefont {Maa{\ss}}, \citenamefont {Mantica}, \citenamefont
  {Nazarewicz}, \citenamefont {N\"{o}rtersh\"{a}user}, \citenamefont {Pineda},
  \citenamefont {Reinhard}, \citenamefont {Rossi}, \citenamefont {Sommer},
  \citenamefont {Sumithrarachchi}, \citenamefont {Teigelh\"{o}fer},\ and\
  \citenamefont {Watkins}}]{Miller2019Nat.Phys.15_432}%
  \BibitemOpen
  \bibfield  {author} {\bibinfo {author} {\bibfnamefont {A.~J.}\ \bibnamefont
  {Miller}}, \bibinfo {author} {\bibfnamefont {K.}~\bibnamefont {Minamisono}},
  \bibinfo {author} {\bibfnamefont {A.}~\bibnamefont {Klose}}, \bibinfo
  {author} {\bibfnamefont {D.}~\bibnamefont {Garand}}, \bibinfo {author}
  {\bibfnamefont {C.}~\bibnamefont {Kujawa}}, \bibinfo {author} {\bibfnamefont
  {J.~D.}\ \bibnamefont {Lantis}}, \bibinfo {author} {\bibfnamefont
  {Y.}~\bibnamefont {Liu}}, \bibinfo {author} {\bibfnamefont {B.}~\bibnamefont
  {Maa{\ss}}}, \bibinfo {author} {\bibfnamefont {P.~F.}\ \bibnamefont
  {Mantica}}, \bibinfo {author} {\bibfnamefont {W.}~\bibnamefont {Nazarewicz}},
  \bibinfo {author} {\bibfnamefont {W.}~\bibnamefont {N\"{o}rtersh\"{a}user}},
  \bibinfo {author} {\bibfnamefont {S.~V.}\ \bibnamefont {Pineda}}, \bibinfo
  {author} {\bibfnamefont {P.-G.}\ \bibnamefont {Reinhard}}, \bibinfo {author}
  {\bibfnamefont {D.~M.}\ \bibnamefont {Rossi}}, \bibinfo {author}
  {\bibfnamefont {F.}~\bibnamefont {Sommer}}, \bibinfo {author} {\bibfnamefont
  {C.}~\bibnamefont {Sumithrarachchi}}, \bibinfo {author} {\bibfnamefont
  {A.}~\bibnamefont {Teigelh\"{o}fer}},\ and\ \bibinfo {author} {\bibfnamefont
  {J.}~\bibnamefont {Watkins}},\ }\bibfield  {title} {\bibinfo {title} {{Proton
  superfluidity and charge radii in proton-rich calcium isotopes}},\ }\href
  {https://doi.org/10.1038/s41567-019-0416-9} {\bibfield  {journal} {\bibinfo
  {journal} {Nat. Phys.}\ }\textbf {\bibinfo {volume} {15}},\ \bibinfo {pages}
  {432} (\bibinfo {year} {2019})}\BibitemShut {NoStop}%
\bibitem [{\citenamefont {Sagawa}\ \emph {et~al.}(2022)\citenamefont {Sagawa},
  \citenamefont {Yoshida}, \citenamefont {Naito}, \citenamefont {Uesaka},
  \citenamefont {Zenihiro}, \citenamefont {Tanaka},\ and\ \citenamefont
  {Suzuki}}]{Sagawa2022Phys.Lett.B829_137072}%
  \BibitemOpen
  \bibfield  {author} {\bibinfo {author} {\bibfnamefont {H.}~\bibnamefont
  {Sagawa}}, \bibinfo {author} {\bibfnamefont {S.}~\bibnamefont {Yoshida}},
  \bibinfo {author} {\bibfnamefont {T.}~\bibnamefont {Naito}}, \bibinfo
  {author} {\bibfnamefont {T.}~\bibnamefont {Uesaka}}, \bibinfo {author}
  {\bibfnamefont {J.}~\bibnamefont {Zenihiro}}, \bibinfo {author}
  {\bibfnamefont {J.}~\bibnamefont {Tanaka}},\ and\ \bibinfo {author}
  {\bibfnamefont {T.}~\bibnamefont {Suzuki}},\ }\bibfield  {title} {\bibinfo
  {title} {{Isovector density and isospin impurity in $ {}^{40} \mathrm{Ca}
  $}},\ }\href {https://doi.org/10.1016/j.physletb.2022.137072} {\bibfield
  {journal} {\bibinfo  {journal} {Phys. Lett. B}\ }\textbf {\bibinfo {volume}
  {829}},\ \bibinfo {pages} {137072} (\bibinfo {year} {2022})}\BibitemShut
  {NoStop}%
\bibitem [{\citenamefont {Yang}\ \emph {et~al.}(2022)\citenamefont {Yang},
  \citenamefont {Fan}, \citenamefont {Naito}, \citenamefont {Niu},
  \citenamefont {Li},\ and\ \citenamefont {Liang}}]{Yang:2022tjf}%
  \BibitemOpen
  \bibfield  {author} {\bibinfo {author} {\bibfnamefont {Z.-X.}\ \bibnamefont
  {Yang}}, \bibinfo {author} {\bibfnamefont {X.-H.}\ \bibnamefont {Fan}},
  \bibinfo {author} {\bibfnamefont {T.}~\bibnamefont {Naito}}, \bibinfo
  {author} {\bibfnamefont {Z.-M.}\ \bibnamefont {Niu}}, \bibinfo {author}
  {\bibfnamefont {Z.-P.}\ \bibnamefont {Li}},\ and\ \bibinfo {author}
  {\bibfnamefont {H.}~\bibnamefont {Liang}},\ }\bibfield  {title} {\bibinfo
  {title} {{Calibration of nuclear charge density distribution by
  back-propagation neural networks}},\ }\Eprint
  {https://arxiv.org/abs/2205.15649} {arXiv:2205.15649 [nucl-th]}  (\bibinfo
  {year} {2022})\BibitemShut {NoStop}%
\bibitem [{\citenamefont {Afanasjev}\ and\ \citenamefont
  {Litvinova}(2015)}]{Afanasjev2015Phys.Rev.C92_044317}%
  \BibitemOpen
  \bibfield  {author} {\bibinfo {author} {\bibfnamefont {A.~V.}\ \bibnamefont
  {Afanasjev}}\ and\ \bibinfo {author} {\bibfnamefont {E.}~\bibnamefont
  {Litvinova}},\ }\bibfield  {title} {\bibinfo {title} {{Impact of collective
  vibrations on quasiparticle states of open-shell odd-mass nuclei and possible
  interference with the tensor force}},\ }\href
  {https://doi.org/10.1103/PhysRevC.92.044317} {\bibfield  {journal} {\bibinfo
  {journal} {Phys. Rev. C}\ }\textbf {\bibinfo {volume} {92}},\ \bibinfo
  {pages} {044317} (\bibinfo {year} {2015})}\BibitemShut {NoStop}%
\bibitem [{\citenamefont {Kubis}\ and\ \citenamefont
  {Kutschera}(1997)}]{Kubis1997Phys.Lett.B399_191}%
  \BibitemOpen
  \bibfield  {author} {\bibinfo {author} {\bibfnamefont {S.}~\bibnamefont
  {Kubis}}\ and\ \bibinfo {author} {\bibfnamefont {M.}~\bibnamefont
  {Kutschera}},\ }\bibfield  {title} {\bibinfo {title} {{Nuclear matter in
  relativistic mean field theory with isovector scalar meson}},\ }\href
  {https://doi.org/10.1016/S0370-2693(97)00306-7} {\bibfield  {journal}
  {\bibinfo  {journal} {Phys. Lett. B}\ }\textbf {\bibinfo {volume} {399}},\
  \bibinfo {pages} {191} (\bibinfo {year} {1997})}\BibitemShut {NoStop}%
\bibitem [{\citenamefont {Liu}\ \emph {et~al.}(2002)\citenamefont {Liu},
  \citenamefont {Greco}, \citenamefont {Baran}, \citenamefont {Colonna},\ and\
  \citenamefont {Di~Toro}}]{Liu2002Phys.Rev.C65_045201}%
  \BibitemOpen
  \bibfield  {author} {\bibinfo {author} {\bibfnamefont {B.}~\bibnamefont
  {Liu}}, \bibinfo {author} {\bibfnamefont {V.}~\bibnamefont {Greco}}, \bibinfo
  {author} {\bibfnamefont {V.}~\bibnamefont {Baran}}, \bibinfo {author}
  {\bibfnamefont {M.}~\bibnamefont {Colonna}},\ and\ \bibinfo {author}
  {\bibfnamefont {M.}~\bibnamefont {Di~Toro}},\ }\bibfield  {title} {\bibinfo
  {title} {{Asymmetric nuclear matter: The role of the isovector scalar
  channel}},\ }\href {https://doi.org/10.1103/PhysRevC.65.045201} {\bibfield
  {journal} {\bibinfo  {journal} {Phys. Rev. C}\ }\textbf {\bibinfo {volume}
  {65}},\ \bibinfo {pages} {045201} (\bibinfo {year} {2002})}\BibitemShut
  {NoStop}%
\bibitem [{\citenamefont {Roca-Maza}\ \emph {et~al.}(2011)\citenamefont
  {Roca-Maza}, \citenamefont {Vi\~{n}as}, \citenamefont {Centelles},
  \citenamefont {Ring},\ and\ \citenamefont
  {Schuck}}]{Roca-Maza2011Phys.Rev.C84_054309}%
  \BibitemOpen
  \bibfield  {author} {\bibinfo {author} {\bibfnamefont {X.}~\bibnamefont
  {Roca-Maza}}, \bibinfo {author} {\bibfnamefont {X.}~\bibnamefont
  {Vi\~{n}as}}, \bibinfo {author} {\bibfnamefont {M.}~\bibnamefont
  {Centelles}}, \bibinfo {author} {\bibfnamefont {P.}~\bibnamefont {Ring}},\
  and\ \bibinfo {author} {\bibfnamefont {P.}~\bibnamefont {Schuck}},\
  }\bibfield  {title} {\bibinfo {title} {{Relativistic mean-field interaction
  with density-dependent meson-nucleon vertices based on microscopical
  calculations}},\ }\href {https://doi.org/10.1103/PhysRevC.84.054309}
  {\bibfield  {journal} {\bibinfo  {journal} {Phys. Rev. C}\ }\textbf {\bibinfo
  {volume} {84}},\ \bibinfo {pages} {054309} (\bibinfo {year}
  {2011})}\BibitemShut {NoStop}%
\bibitem [{\citenamefont {Fayans}(1998)}]{Fayans1998JETPLett.68_169}%
  \BibitemOpen
  \bibfield  {author} {\bibinfo {author} {\bibfnamefont {S.~A.}\ \bibnamefont
  {Fayans}},\ }\bibfield  {title} {\bibinfo {title} {{Towards a universal
  nuclear density functional}},\ }\href {https://doi.org/10.1134/1.567841}
  {\bibfield  {journal} {\bibinfo  {journal} {JETP Lett.}\ }\textbf {\bibinfo
  {volume} {68}},\ \bibinfo {pages} {169} (\bibinfo {year} {1998})}\BibitemShut
  {NoStop}%
\bibitem [{\citenamefont {Reinhard}\ and\ \citenamefont
  {Nazarewicz}(2017)}]{Reinhard2017Phys.Rev.C95_064328}%
  \BibitemOpen
  \bibfield  {author} {\bibinfo {author} {\bibfnamefont {P.-G.}\ \bibnamefont
  {Reinhard}}\ and\ \bibinfo {author} {\bibfnamefont {W.}~\bibnamefont
  {Nazarewicz}},\ }\bibfield  {title} {\bibinfo {title} {{Toward a global
  description of nuclear charge radii: Exploring the Fayans energy density
  functional}},\ }\href {https://doi.org/10.1103/PhysRevC.95.064328} {\bibfield
   {journal} {\bibinfo  {journal} {Phys. Rev. C}\ }\textbf {\bibinfo {volume}
  {95}},\ \bibinfo {pages} {064328} (\bibinfo {year} {2017})}\BibitemShut
  {NoStop}%
\bibitem [{\citenamefont {Bertulani}\ \emph {et~al.}(2009)\citenamefont
  {Bertulani}, \citenamefont {L\"u},\ and\ \citenamefont
  {Sagawa}}]{Bertulani2009Phys.Rev.C80_027303}%
  \BibitemOpen
  \bibfield  {author} {\bibinfo {author} {\bibfnamefont {C.~A.}\ \bibnamefont
  {Bertulani}}, \bibinfo {author} {\bibfnamefont {H.~F.}\ \bibnamefont
  {L\"u}},\ and\ \bibinfo {author} {\bibfnamefont {H.}~\bibnamefont {Sagawa}},\
  }\bibfield  {title} {\bibinfo {title} {{Odd-even mass difference and isospin
  dependent pairing interaction}},\ }\href
  {https://doi.org/10.1103/PhysRevC.80.027303} {\bibfield  {journal} {\bibinfo
  {journal} {Phys. Rev. C}\ }\textbf {\bibinfo {volume} {80}},\ \bibinfo
  {pages} {027303} (\bibinfo {year} {2009})}\BibitemShut {NoStop}%
\bibitem [{\citenamefont {Bertulani}\ \emph {et~al.}(2012)\citenamefont
  {Bertulani}, \citenamefont {Liu},\ and\ \citenamefont
  {Sagawa}}]{Bertulani2012Phys.Rev.C85_014321}%
  \BibitemOpen
  \bibfield  {author} {\bibinfo {author} {\bibfnamefont {C.~A.}\ \bibnamefont
  {Bertulani}}, \bibinfo {author} {\bibfnamefont {H.}~\bibnamefont {Liu}},\
  and\ \bibinfo {author} {\bibfnamefont {H.}~\bibnamefont {Sagawa}},\
  }\bibfield  {title} {\bibinfo {title} {{Global investigation of odd-even mass
  differences and radii with isospin-dependent pairing interactions}},\ }\href
  {https://doi.org/10.1103/PhysRevC.85.014321} {\bibfield  {journal} {\bibinfo
  {journal} {Phys. Rev. C}\ }\textbf {\bibinfo {volume} {85}},\ \bibinfo
  {pages} {014321} (\bibinfo {year} {2012})}\BibitemShut {NoStop}%
\bibitem [{\citenamefont {Yamagami}\ \emph {et~al.}(2012)\citenamefont
  {Yamagami}, \citenamefont {Margueron}, \citenamefont {Sagawa},\ and\
  \citenamefont {Hagino}}]{Yamagami2012Phys.Rev.C86_034333}%
  \BibitemOpen
  \bibfield  {author} {\bibinfo {author} {\bibfnamefont {M.}~\bibnamefont
  {Yamagami}}, \bibinfo {author} {\bibfnamefont {J.}~\bibnamefont {Margueron}},
  \bibinfo {author} {\bibfnamefont {H.}~\bibnamefont {Sagawa}},\ and\ \bibinfo
  {author} {\bibfnamefont {K.}~\bibnamefont {Hagino}},\ }\bibfield  {title}
  {\bibinfo {title} {{Isoscalar and isovector density dependence of the pairing
  functional determined from global fitting}},\ }\href
  {https://doi.org/10.1103/PhysRevC.86.034333} {\bibfield  {journal} {\bibinfo
  {journal} {Phys. Rev. C}\ }\textbf {\bibinfo {volume} {86}},\ \bibinfo
  {pages} {034333} (\bibinfo {year} {2012})}\BibitemShut {NoStop}%
\bibitem [{\citenamefont {Sagawa}\ \emph {et~al.}(2013)\citenamefont {Sagawa},
  \citenamefont {Tanimura},\ and\ \citenamefont
  {Hagino}}]{Sagawa2013Phys.Rev.C87_034310}%
  \BibitemOpen
  \bibfield  {author} {\bibinfo {author} {\bibfnamefont {H.}~\bibnamefont
  {Sagawa}}, \bibinfo {author} {\bibfnamefont {Y.}~\bibnamefont {Tanimura}},\
  and\ \bibinfo {author} {\bibfnamefont {K.}~\bibnamefont {Hagino}},\
  }\bibfield  {title} {\bibinfo {title} {{Competition between $T=1$ and $T=0$
  pairing in $pf$-shell nuclei with $N=Z$}},\ }\href
  {https://doi.org/10.1103/PhysRevC.87.034310} {\bibfield  {journal} {\bibinfo
  {journal} {Phys. Rev. C}\ }\textbf {\bibinfo {volume} {87}},\ \bibinfo
  {pages} {034310} (\bibinfo {year} {2013})}\BibitemShut {NoStop}%
\bibitem [{\citenamefont
  {Yoshida}(2013)}]{Yoshida2013Prog.Theor.Exp.Phys.2013_113D02}%
  \BibitemOpen
  \bibfield  {author} {\bibinfo {author} {\bibfnamefont {K.}~\bibnamefont
  {Yoshida}},\ }\bibfield  {title} {\bibinfo {title} {{Spin--Isospin response
  of deformed neutron-rich nuclei in a self-consistent Skyrme
  energy-density-functional approach}},\ }\href
  {https://doi.org/10.1093/ptep/ptt091} {\bibfield  {journal} {\bibinfo
  {journal} {Prog. Theor. Exp. Phys.}\ }\textbf {\bibinfo {volume} {2013}},\
  \bibinfo {pages} {113D02} (\bibinfo {year} {2013})}\BibitemShut {NoStop}%
\bibitem [{\citenamefont {Sagawa}\ \emph {et~al.}(2016)\citenamefont {Sagawa},
  \citenamefont {Bai},\ and\ \citenamefont
  {Col\`{o}}}]{Sagawa2016Phys.Scr.91_083011}%
  \BibitemOpen
  \bibfield  {author} {\bibinfo {author} {\bibfnamefont {H.}~\bibnamefont
  {Sagawa}}, \bibinfo {author} {\bibfnamefont {C.~L.}\ \bibnamefont {Bai}},\
  and\ \bibinfo {author} {\bibfnamefont {G.}~\bibnamefont {Col\`{o}}},\
  }\bibfield  {title} {\bibinfo {title} {{Isovector spin-singlet ($ T = 1$, $ S
  = 0 $) and isoscalar spin-triplet ($T = 0 $, $ S = 1 $) pairing interactions
  and spin-isospin response}},\ }\href
  {https://doi.org/10.1088/0031-8949/91/8/083011} {\bibfield  {journal}
  {\bibinfo  {journal} {Phys. Scr.}\ }\textbf {\bibinfo {volume} {91}},\
  \bibinfo {pages} {083011} (\bibinfo {year} {2016})}\BibitemShut {NoStop}%
\bibitem [{\citenamefont {Teeti}\ and\ \citenamefont
  {Afanasjev}(2021)}]{Teeti2021Phys.Rev.C103_034310}%
  \BibitemOpen
  \bibfield  {author} {\bibinfo {author} {\bibfnamefont {S.}~\bibnamefont
  {Teeti}}\ and\ \bibinfo {author} {\bibfnamefont {A.~V.}\ \bibnamefont
  {Afanasjev}},\ }\bibfield  {title} {\bibinfo {title} {{Global study of
  separable pairing interaction in covariant density functional theory}},\
  }\href {https://doi.org/10.1103/PhysRevC.103.034310} {\bibfield  {journal}
  {\bibinfo  {journal} {Phys. Rev. C}\ }\textbf {\bibinfo {volume} {103}},\
  \bibinfo {pages} {034310} (\bibinfo {year} {2021})}\BibitemShut {NoStop}%
\bibitem [{\citenamefont {Oishi}\ \emph {et~al.}(2020)\citenamefont {Oishi},
  \citenamefont {Kru{\v{z}}i{\'{c}}},\ and\ \citenamefont
  {Paar}}]{Oishi2020J.Phys.G47_115106}%
  \BibitemOpen
  \bibfield  {author} {\bibinfo {author} {\bibfnamefont {T.}~\bibnamefont
  {Oishi}}, \bibinfo {author} {\bibfnamefont {G.}~\bibnamefont
  {Kru{\v{z}}i{\'{c}}}},\ and\ \bibinfo {author} {\bibfnamefont
  {N.}~\bibnamefont {Paar}},\ }\bibfield  {title} {\bibinfo {title} {{Role of
  residual interaction in the relativistic description of M1 excitation}},\
  }\href {https://doi.org/10.1088/1361-6471/abaeb1} {\bibfield  {journal}
  {\bibinfo  {journal} {J. Phys. G}\ }\textbf {\bibinfo {volume} {47}},\
  \bibinfo {pages} {115106} (\bibinfo {year} {2020})}\BibitemShut {NoStop}%
\bibitem [{\citenamefont {Oishi}\ \emph {et~al.}(2021)\citenamefont {Oishi},
  \citenamefont {Kru\v{z}i\'{c}},\ and\ \citenamefont
  {Paar}}]{Oishi2021Eur.Phys.J.A57_180}%
  \BibitemOpen
  \bibfield  {author} {\bibinfo {author} {\bibfnamefont {T.}~\bibnamefont
  {Oishi}}, \bibinfo {author} {\bibfnamefont {G.}~\bibnamefont
  {Kru\v{z}i\'{c}}},\ and\ \bibinfo {author} {\bibfnamefont {N.}~\bibnamefont
  {Paar}},\ }\bibfield  {title} {\bibinfo {title} {{Discerning nuclear pairing
  properties from magnetic dipole excitation}},\ }\href
  {https://doi.org/10.1140/epja/s10050-021-00488-7} {\bibfield  {journal}
  {\bibinfo  {journal} {Eur. Phys. J. A}\ }\textbf {\bibinfo {volume} {57}},\
  \bibinfo {pages} {180} (\bibinfo {year} {2021})}\BibitemShut {NoStop}%
\bibitem [{\citenamefont {Yoshida}\ and\ \citenamefont
  {Tanimura}(2021)}]{Yoshida2021Phys.Rev.C104_014319}%
  \BibitemOpen
  \bibfield  {author} {\bibinfo {author} {\bibfnamefont {K.}~\bibnamefont
  {Yoshida}}\ and\ \bibinfo {author} {\bibfnamefont {Y.}~\bibnamefont
  {Tanimura}},\ }\bibfield  {title} {\bibinfo {title} {{Spin-triplet
  proton-neutron pair in spin-dipole excitations}},\ }\href
  {https://doi.org/10.1103/PhysRevC.104.014319} {\bibfield  {journal} {\bibinfo
   {journal} {Phys. Rev. C}\ }\textbf {\bibinfo {volume} {104}},\ \bibinfo
  {pages} {014319} (\bibinfo {year} {2021})}\BibitemShut {NoStop}%
\bibitem [{\citenamefont {Bertsch}\ \emph {et~al.}(2009)\citenamefont
  {Bertsch}, \citenamefont {Bertulani}, \citenamefont {Nazarewicz},
  \citenamefont {Schunck},\ and\ \citenamefont
  {Stoitsov}}]{Bertsch2009Phys.Rev.C79_034306}%
  \BibitemOpen
  \bibfield  {author} {\bibinfo {author} {\bibfnamefont {G.~F.}\ \bibnamefont
  {Bertsch}}, \bibinfo {author} {\bibfnamefont {C.~A.}\ \bibnamefont
  {Bertulani}}, \bibinfo {author} {\bibfnamefont {W.}~\bibnamefont
  {Nazarewicz}}, \bibinfo {author} {\bibfnamefont {N.}~\bibnamefont
  {Schunck}},\ and\ \bibinfo {author} {\bibfnamefont {M.~V.}\ \bibnamefont
  {Stoitsov}},\ }\bibfield  {title} {\bibinfo {title} {{Odd-even mass
  differences from self-consistent mean field theory}},\ }\href
  {https://doi.org/10.1103/PhysRevC.79.034306} {\bibfield  {journal} {\bibinfo
  {journal} {Phys. Rev. C}\ }\textbf {\bibinfo {volume} {79}},\ \bibinfo
  {pages} {034306} (\bibinfo {year} {2009})}\BibitemShut {NoStop}%
\bibitem [{\citenamefont {Robledo}\ \emph {et~al.}(2012)\citenamefont
  {Robledo}, \citenamefont {Bernard},\ and\ \citenamefont
  {Bertsch}}]{Robledo2012Phys.Rev.C86_064313}%
  \BibitemOpen
  \bibfield  {author} {\bibinfo {author} {\bibfnamefont {L.~M.}\ \bibnamefont
  {Robledo}}, \bibinfo {author} {\bibfnamefont {R.}~\bibnamefont {Bernard}},\
  and\ \bibinfo {author} {\bibfnamefont {G.~F.}\ \bibnamefont {Bertsch}},\
  }\bibfield  {title} {\bibinfo {title} {{Pairing gaps in the
  Hartree-Fock-Bogoliubov theory with the Gogny D1S interaction}},\ }\href
  {https://doi.org/10.1103/PhysRevC.86.064313} {\bibfield  {journal} {\bibinfo
  {journal} {Phys. Rev. C}\ }\textbf {\bibinfo {volume} {86}},\ \bibinfo
  {pages} {064313} (\bibinfo {year} {2012})}\BibitemShut {NoStop}%
\bibitem [{\citenamefont {Horowitz}\ and\ \citenamefont
  {Piekarewicz}(2012)}]{Horowitz2012Phys.Rev.C86_045503}%
  \BibitemOpen
  \bibfield  {author} {\bibinfo {author} {\bibfnamefont {C.~J.}\ \bibnamefont
  {Horowitz}}\ and\ \bibinfo {author} {\bibfnamefont {J.}~\bibnamefont
  {Piekarewicz}},\ }\bibfield  {title} {\bibinfo {title} {{Impact of spin-orbit
  currents on the electroweak skin of neutron-rich nuclei}},\ }\href
  {https://doi.org/10.1103/PhysRevC.86.045503} {\bibfield  {journal} {\bibinfo
  {journal} {Phys. Rev. C}\ }\textbf {\bibinfo {volume} {86}},\ \bibinfo
  {pages} {045503} (\bibinfo {year} {2012})}\BibitemShut {NoStop}%
\bibitem [{\citenamefont {Reinhard}\ and\ \citenamefont
  {Nazarewicz}(2021)}]{Reinhard2021Phys.Rev.C103_054310}%
  \BibitemOpen
  \bibfield  {author} {\bibinfo {author} {\bibfnamefont {P.-G.}\ \bibnamefont
  {Reinhard}}\ and\ \bibinfo {author} {\bibfnamefont {W.}~\bibnamefont
  {Nazarewicz}},\ }\bibfield  {title} {\bibinfo {title} {{Nuclear charge
  densities in spherical and deformed nuclei: Toward precise calculations of
  charge radii}},\ }\href {https://doi.org/10.1103/PhysRevC.103.054310}
  {\bibfield  {journal} {\bibinfo  {journal} {Phys. Rev. C}\ }\textbf {\bibinfo
  {volume} {103}},\ \bibinfo {pages} {054310} (\bibinfo {year}
  {2021})}\BibitemShut {NoStop}%
\end{thebibliography}
\end{document}